\pdfoutput=1
\documentclass[iop, apj, twocolappendix, numberedappendix]{emulateapj}

\newif\iflocal
\localfalse

\usepackage{xspace}
\usepackage{amsmath}
\usepackage{framed} 
\usepackage{txfonts}
\usepackage{epstopdf}
\usepackage{color}
\usepackage{rotating}
\usepackage{natbib}
\usepackage{ulem}
\usepackage{xspace}
\usepackage[colorlinks=true,urlcolor=black,linkcolor=blue,citecolor=blue]{hyperref}
\usepackage{sistyle}
\SIthousandsep{,}

\iflocal
\def\includedir{/Users/benedito/University/docs/latex}
\def\figdir{figs}
\else
\def\includedir{.}
\def\figdir{.}
\fi

\usepackage{etoolbox}
\makeatletter
\patchcmd{\NAT@citex}
  {\@citea\NAT@hyper@{\NAT@nmfmt{\NAT@nm}\NAT@date}}
  {\@citea\NAT@nmfmt{\NAT@nm}\NAT@hyper@{\NAT@date}}
  {}
  {}
\patchcmd{\NAT@citex}
  {\@citea\NAT@hyper@{%
     \NAT@nmfmt{\NAT@nm}%
     \hyper@natlinkbreak{\NAT@aysep\NAT@spacechar}{\@citeb\@extra@b@citeb}%
     \NAT@date}}
  {\@citea\NAT@nmfmt{\NAT@nm}%
   \NAT@aysep\NAT@spacechar%
   \NAT@hyper@{\NAT@date}}
  {}
  {}
\patchcmd{\NAT@citex}
  {\@citea\NAT@hyper@{%
     \NAT@nmfmt{\NAT@nm}%
     \hyper@natlinkbreak{\NAT@spacechar\NAT@@open\if*#1*\else#1\NAT@spacechar\fi}%
       {\@citeb\@extra@b@citeb}%
     \NAT@date}}
  {\@citea\NAT@nmfmt{\NAT@nm}%
   \NAT@spacechar\NAT@@open\if*#1*\else#1\NAT@spacechar\fi%
   \NAT@hyper@{\NAT@date}}
  {}
  {}
\makeatother

\newcommand{\kpch}{\>{h^{-1}{\rm kpc}}}
\newcommand{\mpch}{\>h^{-1}{\rm {Mpc}}}

\newcommand{\msun}{\>{M_{\odot}}}

\newcommand{\msunh}{\>h^{-1} M_\odot}

\def\gcm3{\mathrm{g} / \mathrm{cm}^3}

\def\tdyn{t_{\rm dyn}}

\def\LCDM{$\Lambda$CDM\xspace}
\def\rhoc{\rho_{\rm c}}
\def\rhom{\rho_{\rm m}}

\def\gtsima{$\; \buildrel > \over \sim \;$}
\def\ltsima{$\; \buildrel < \over \sim \;$}
\def\prosima{$\; \buildrel \propto \over \sim \;$}
\def\gsim{\lower.7ex\hbox{\gtsima}}
\def\lsim{\lower.7ex\hbox{\ltsima}}
\def\simgt{\lower.7ex\hbox{\gtsima}}
\def\simlt{\lower.7ex\hbox{\ltsima}}
\def\simpr{\lower.7ex\hbox{\prosima}}

\def\sparta{\textsc{Sparta}\xspace}
\def\moria{\textsc{Moria}\xspace}
\def\shellfish{\textsc{Shellfish}\xspace}

\def\colossus{\textsc{Colossus}\xspace}
\def\rockstar{\textsc{Rockstar}\xspace}
\def\consistenttrees{\textsc{Consistent-Trees}\xspace}
\def\subfind{\textsc{SubFind}\xspace}

\def\planck{Planck\xspace}
\def\wmap{WMAP7\xspace}

\def\erebos{Erebos\xspace}

\def\vmax{V_{\rm max}}
\def\vpeak{V_{\rm peak}}

\def\gammadk{\Gamma_{\rm DK14}}
\def\gammadyn{\Gamma_{\rm dyn}}

\def\mvir{M_{\rm vir}}
\def\rvir{R_{\rm vir}}

\def\mtom{M_{\rm 200m}}
\def\rtom{R_{\rm 200m}}

\def\ntom{N_{\rm 200m}}
\def\nutom{\nu_{\rm 200m}}

\def\mtoc{M_{\rm 200c}}
\def\rtoc{R_{\rm 200c}}

\def\mfoc{M_{\rm 500c}}
\def\rfoc{R_{\rm 500c}}

\def\mdelta{M_{\Delta}}
\def\rdelta{R_{\Delta}}

\def\mtombnd{M_{\rm 200m,bnd}}

\def\rsp{R_{\rm sp}}

\def\msp{M_{\rm sp}}

\def\deltasp{\Delta_{\rm sp}}

\def\xsp{X_{\rm sp}}

\shorttitle{Splashback Catalogs}
\shortauthors{Splashback Catalogs}

\journalinfo{The Astrophysical Journal Supplement Series, {\rm 251:17 (20pp), 2020 December}}
\submitted{Received 2020 July 18; revised 2020 September 20; accepted 2020 October 6; published 2020 November 23}

\begin{document}


\iflocal
\def\figdir{figs}
\else
\def\figdir{.}
\fi


\defcitealias{diemer_13_scalingrel}{DKM13}
\defcitealias{diemer_14}{DK14}
\defcitealias{diemer_15}{DK15}
\defcitealias{diemer_17_sparta}{Paper I}
\defcitealias{diemer_17_rsp}{Paper II}


\title{The splashback radius of halos from particle dynamics: \\III. Halo catalogs, merger trees, and host--subhalo relations}
\author{Benedikt Diemer$^{1,2}$}
\altaffiliation{$^2$NHFP Einstein Fellow}
\affil{
$^1$Department of Astronomy, University of Maryland, College Park, MD 20742, USA; \href{mailto:diemer@umd.edu}{diemer@umd.edu}
}


\begin{abstract}
Virtually any investigation involving dark matter halos relies on a definition of their radius, of their mass, and of whether they are a subhalo. The halo boundary is most commonly defined to include a spherical overdensity contrast (such as $\rtoc$, $\rvir$, and $\rtom$), but different thresholds lead to significant differences in radius and mass. The splashback radius has recently been suggested as a more physically motivated (and generally larger) halo boundary, adding to the range of definitions. It is often difficult to assess the impact of a particular choice because most halo catalogs contain only one or a few definitions and generally only one set of host--subhalo relations. To alleviate this issue, we present halo catalogs and merger trees for $14$ $N$-body simulations of \LCDM and self-similar universes. Based on \rockstar catalogs, we compute additional halo properties using the \sparta code and recombine them with the original catalogs. The new catalogs contain numerous variants of spherical overdensity and splashback radii and masses and, most critically, host--subhalo relations for each definition. We also present a new merger tree format where the data are stored as a compressed, two-dimensional matrix. We perform basic tests of the relation between different definitions and present an updated model for the splashback--spherical overdensity connection. The \sparta code, as well as our catalogs and merger trees, are publicly available.
\end{abstract}



\section{Introduction}
\label{sec:intro}

In the widely accepted \LCDM framework, dark matter halos are the building blocks of structure, with galaxies growing at their centers \citep{rees_77, silk_77, white_78}. As soon as we develop this basic picture into a more quantitative understanding, we face the need to assign halos some measure of size and mass. For example, the mass function of halos is predicted to follow a relatively simple form \citep{press_74, bond_91}, but this statement is naturally predicated on there being a meaningful way to measure a halo's mass. Similarly, we wish to connect the properties of halos to their galaxies via techniques such as abundance matching, occupation distributions, or semianalytical modeling, all of which rely on a definition of the halo boundary \citep[see][for a review]{wechsler_18}. 

The most common solution is the spherical overdensity (SO) definition of halo properties, where the radius $\rdelta$ is defined to enclose a fixed overdensity $\Delta$ with respect to either the critical or mean density of the universe \citep[e.g.,][]{lacey_94}. This definition leads to variants such as $\rfoc$, $\rtoc$, $\rvir$, and $\rtom$ and the corresponding masses $\mdelta$ (in descending order of overdensity; see Table~\ref{table:defs} for the detailed definitions and Figure~\ref{fig:viz_radii} for a visualization). The main issue with SO radii is how to choose the density threshold. One popular definition is the so-called virial overdensity, $\Delta_{\rm vir} = 18 \pi^2 \approx 178$, derived from the collapse of an isolated top-hat overdensity in an Einstein--de Sitter universe \citep{gunn_72, peebles_80, lacey_93}. In a \LCDM universe, the overdensity evolves with time \citep{lahav_91, bryan_98}. However, the assumptions of the top-hat collapse model are not well justified, chiefly because halos do not form in isolation and because the initial peaks are not top-hat in shape \citep[e.g.,][]{dalal_10}. Nevertheless, SO definitions are simple to compute and easy to understand, and various density thresholds are adopted for mostly practical reasons. For example, $\rtoc$ is often used for historical reasons, whereas the X-ray cluster community prefers small radii such as $\rfoc$ or even $R_{\rm 2500c}$ because the X-ray signal is measurable only within a small aperture. 

Given the variety of SO definitions, one should always compare halo-related simulation results using multiple definitions. In practice, however, such checks are often unrealistic because halo catalogs are available in only one definition. In principle, it is possible to convert between definitions with a fitting function or by assuming a particular form of the density profile \citep{white_01_mass, white_02, hu_03, lukic_09, diemer_18_colossus}, but these conversions can be fairly inaccurate owing to deviations from the assumed profile or scatter in the concentration--mass relation \citep[e.g., Appendix C of][]{diemer_15}. Thus, it is often best to choose one definition and understand the consequences of that choice.

The situation is even more complicated for subhalos, whose center lies within another, larger halo. There, we cannot always define SO masses because the density profile includes a large, possibly dominant contribution from the host. A common solution is to remove gravitationally unbound particles \citep[e.g.,][]{springel_01_subfind, han_12_hbt, behroozi_13_rockstar}, but the results depend on the exact algorithm. Moreover, subhalos tend to undergo unphysical stripping due to numerical effects --- the classic overmerging problem that still plagues modern simulations \citep[e.g.,][]{vankampen_95, moore_96,  klypin_99_overmerging, vandenbosch_17}. Both issues can be avoided by defining subhalo masses at their peak or at infall, neglecting the subsequent mass evolution. The complex density structure around subhalos can also be circumvented by using $\vmax$, the maximum of the circular velocity curve. However, this definition measures the mass within a rather small radius and is still prone to spikes during mergers \citep{behroozi_14}. The chosen definition has a sizable impact on the galaxy--halo connection \citep{reddick_13}, highlighting that there is no one correct choice. 

\begin{deluxetable}{ll}
\tablecaption{Definitions of the Symbols Used in this Paper
\label{table:defs}}
\tablewidth{0.47\textwidth}
\tablehead{
\colhead{Symbol} &
\colhead{Meaning}
}
\startdata
$\rhom$ & Mean matter density of the universe \\
$\rhoc$ & Critical density of the universe \\
$\Delta$ & An overdensity with respect to either $\rhom$ or $\rhoc$ \\
$R_{\rm X}$ & A particular definition $X$ of the halo boundary \\
$M_{\rm X}$ & Mass inside $R_{\rm X}$ \\
$N_{\rm X}$ & Number of particles inside $R_{\rm X}$ \\
$\rtom$ & Radius enclosing an overdensity of $200 \times \rhom$ \\
$\rtoc$ & The radius enclosing an overdensity of $200 \times \rhoc$ \\
$\rvir$ & $\rdelta$ with varying overdensity \citep{bryan_98} \\
$M_{\rm X,peak}$ & Peak mass attained during halo history \\
$M_{\rm X,acc}$ & Mass of subhalo at accretion onto host \\
$M_{\rm \Delta,all}$ & SO mass computed from all particles \\
$M_{\rm \Delta,bnd}$ & SO mass computed from bound particles only \\
$M_{\rm \Delta,tcr}$ & SO mass computed from only tracer particles in subhalos \\
$V_{\Delta}$ & Circular velocity, $V_{\Delta} \equiv \sqrt{G\mdelta / \rdelta}$ \\
$\vmax$ & Maximum of circular velocity at any radius \\
$\vpeak$ & Maximum $\vmax$ attained during halo history \\
$\nu$ &  Peak height, $\nu \equiv \nutom = \delta_{\rm c} / \sigma(\mtom, z)$ \\
$\rsp$ & Splashback radius of a halo \\
$\msp$ & Splashback mass of a halo (estimated separately from $\rsp$) \\
$\deltasp$ & Splashback overdensity wrt. $\rhom$, $\deltasp \equiv 3 \msp / (4 \pi \rsp^3) / \rho_{\rm m}$ \\
$R_{\rm sp,mn}$ & $\rsp$ defined as the mean of the particle apocenters \\
$R_{\rm sp,50\%}$ & $\rsp$ defined as the median of the particle apocenters \\
$R_{\rm sp,75\%}$ & $\rsp$ defined as the 75th percentile of the particle apocenters \\
$t_{\rm dyn}$ & Dynamical time or crossing time, $t_{\rm dyn} \equiv 2 \rtom / V_{\rm 200m}$ \\
$\gammadyn$ & Logarithmic mass accretion rate  over $t_{\rm dyn}$, $\Delta \log(M) / \Delta \log(a)$
\enddata
\end{deluxetable}

In addition to these practical considerations, we may question whether SO definitions are the physically correct choice for the halo boundary. Besides their somewhat indecisive theoretical foundation, SO definitions lead to a sizable number of ``backsplash'' or ``ejected'' halos that reside outside $\rvir$ but eventually merge \citep[e.g.,][]{balogh_00, mamon_04, gill_05}. The sphere of influence of halos also extends beyond $\rvir$ in that infalling subhalos begin to lose mass at around two ``virial'' radii \citep{behroozi_14}. Finally, SO radii and masses suffer from pseudo-evolution, arguably unphysical changes due to the evolution of the threshold density with the critical or mean density of the universe \citep{diemand_05, cuesta_08, diemer_13_pe, zemp_14, more_15}. All of these problems indicate that SO radii do not typically encompass the entire halo.

\def\figsize{0.84}
\begin{figure*}
\centering
\includegraphics[trim =  31mm 2mm 30mm 2mm, clip, scale=\figsize]{\figdir/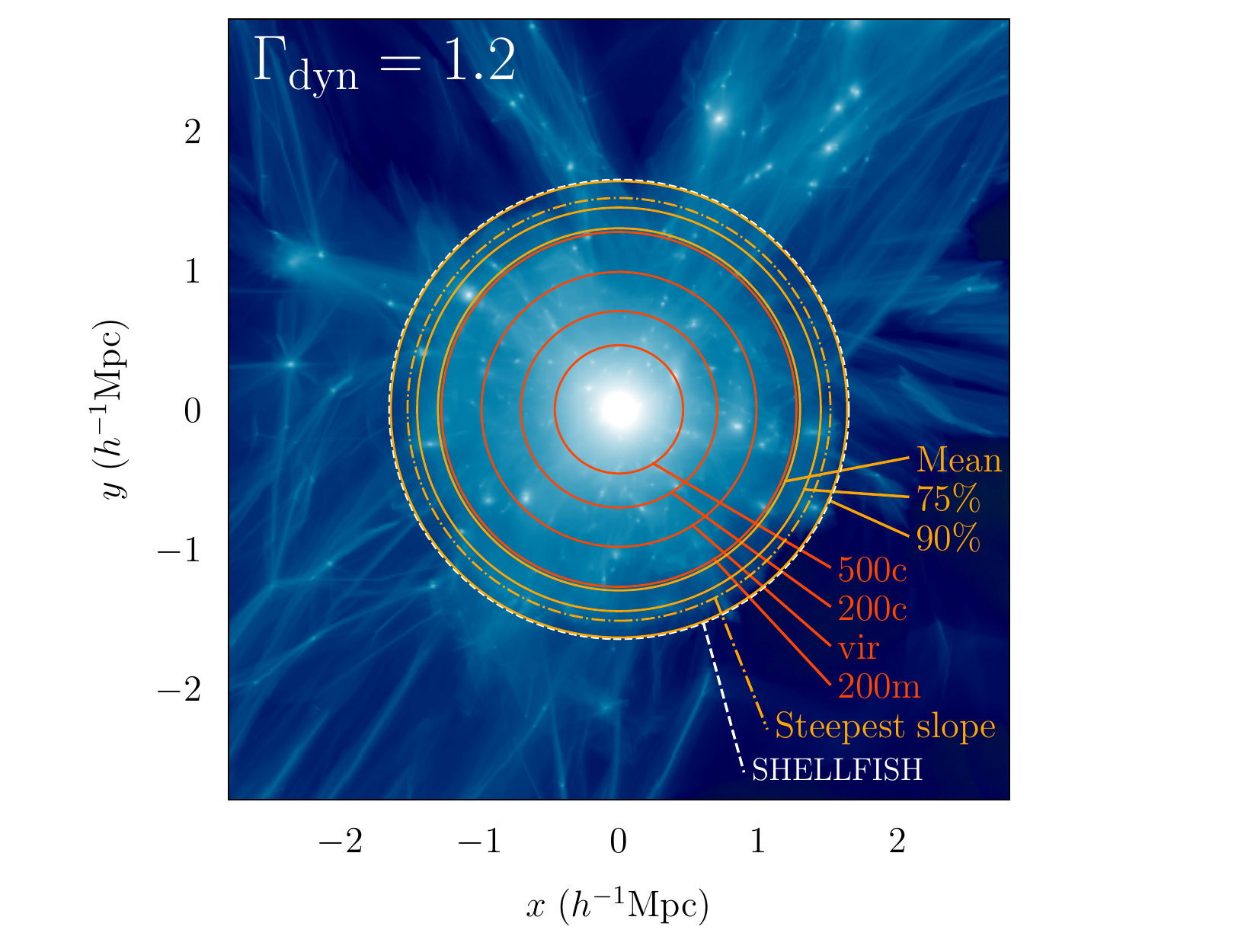}
\includegraphics[trim =  28mm 2mm 30mm 2mm, clip, scale=\figsize]{\figdir/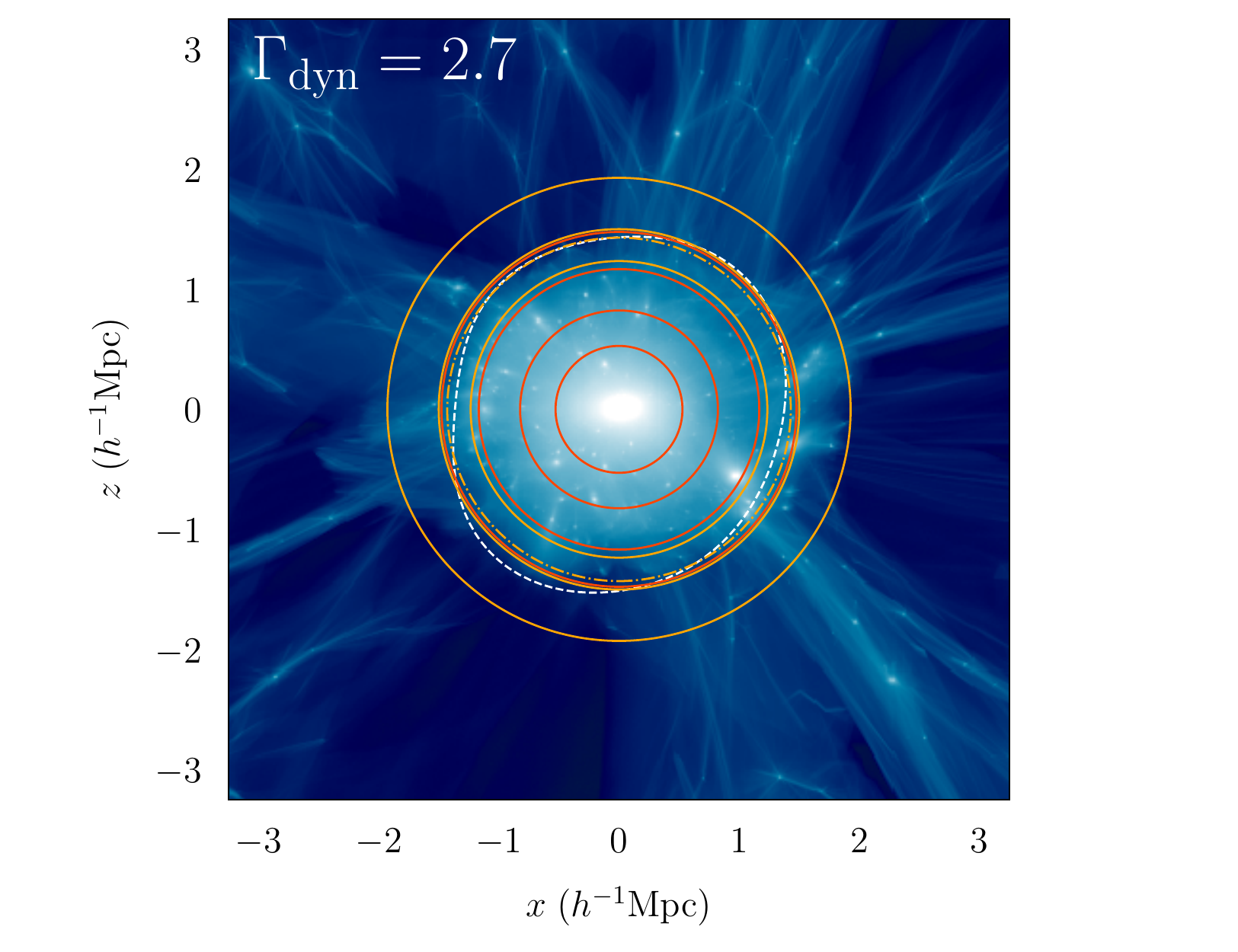}
\caption{Visualization of radius definitions for two representative halos with low and high accretion rates (left and right panels, respectively). Both halos have masses of about $1.5 \times 10^{14} \msunh$ and were selected from the L0125-WMAP7 simulation at $z = 0$. The logarithmic color map shows projected density through slabs of thickness $0.3 \rtom$. Each line shows one of the radius definitions discussed in this paper. The red circles indicate, from the center outward, $\rfoc$, $\rtoc$, $\rvir$, and $\rtom$. The yellow circles show different estimates of the splashback radius, namely $R_{\rm sp,mn}$, $R_{\rm sp,75\%}$, and $R_{\rm sp,90\%}$. In the left panel, all splashback radii are larger than $\rtom$ owing to the relatively low accretion rate of the halo, whereas $\rtom \approx R_{\rm sp,75\%}$ for the fast-accreting halo in the right panel. The dot-dashed yellow lines show the \citet{more_15} prediction for the radius where the slope of the density profile is steepest, which corresponds to a visible drop in the projected density. The relationship of this radius to the dynamically calculated radii from \sparta also depends on the accretion rate: in the left panel, this radius is greater than $R_{\rm sp,75\%}$; in the right panel, it is smaller. The white dashed lines show slices through the three-dimensional splashback shells computed by \shellfish \citep{mansfield_17}. For the slowly accreting halo on the left, the shell is almost spherical and coincides with $R_{\rm sp,90\%}$. The fast-accreting halo on the right is less spherical, and the \shellfish shell visibly traces the density drop toward the lower left of the halo. The background visualizations were created using the gotetra code by P. Mansfield (\href{https://github.com/phil-mansfield/gotetra}{https://github.com/phil-mansfield/gotetra}), which uses a tetrahedron-based estimate of the density field \citep{kaehler_12, abel_12, hahn_13}. The same halos are shown in Figure~1 of \citet{more_15}, although with a thinner projection depth and a different color scheme.}
\label{fig:viz_radii}
\end{figure*}

To remedy these issues, the splashback radius, $\rsp$, has been suggested as a more physically motivated alternative \citep{diemer_14, adhikari_14, more_15}. This radius corresponds to the apocenter of the first orbits of particles and subhalos and is inspired by the spherical collapse model, where it represents the boundary between infalling and orbiting material \citep{fillmore_84, bertschinger_85, lithwick_11, vogelsberger_11_similarity, adhikari_14, shi_16_rsp}. Even though the boundary is not as uniquely defined in realistic \LCDM halos \citep[e.g.,][]{aung_20_phasespace}, the sharp drop due to particles piling up at their apocenter has been detected in simulations \citep{diemer_14, more_15, okumura_18, fong_18, banerjee_20, xhakaj_20, mansfield_20_ab} and observations \citep[][see also \citealt{tully_15}, \citealt{patej_16}, \citealt{umetsu_17}, \citealt{contigiani_19_wl}, \citealt{tomooka_20}, \citealt{zu_17}, \citealt{busch_17}]{more_16, baxter_17, chang_18, nishizawa_18, shin_19_rsp, zuercher_19, murata_20}.

These detections have provided powerful motivation to further explore the splashback radius in simulations. While the initial investigations and observational detections have relied on the steepening feature in stacked density profiles, the profiles of individual halos are often too noisy and influenced by substructure to detect it. Two other techniques to measure $\rsp$ have been suggested thus far. First, the \shellfish code of \citet{mansfield_17} considers the full 3D density field around halos and finds the density drop in infinitesimally thin lines of sight, combining those measurements into a non--spherical splashback shell. This method relies only on density data at the current time and resolves the full shape of the splashback boundary, but it works reliably only for very well--resolved halos. Second, the \sparta code introduced in \citealt{diemer_17_sparta} (hereafter \citetalias{diemer_17_sparta}) tracks the trajectories of dark matter particles to find their first apocenter and defines the halo's splashback radius as an average of those apocenters. First results from this technique were shown in \citealt{diemer_17_rsp} (hereafter \citetalias{diemer_17_rsp}). While the initial results from \shellfish and \sparta have given us a sense of the relation between $\rsp$, SO radii, and mass accretion rates, they did not allow for a systematic exploration of the impact of mass definitions on structure formation in general. For such investigations, we need halo catalogs, that is, lists of halos at each snapshot including information about the relation of host and subhalos. The latter is critical: without knowing the subhalo status, we cannot separate host halos (e.g., to compute the mass function or assembly bias) or assign galaxies to (sub)halos.

In the present work, we fill this gap by providing the first halo catalogs and merger trees that contain a large range of SO and splashback definitions, as well as the corresponding host--subhalo relations. We have slightly improved the splashback calculations of \citetalias{diemer_17_sparta} and have added a number of SO-related calculations to \sparta. We present a new tool called \moria that combines results from \sparta with the original halo catalogs to create enhanced catalogs and merger trees (see Figure~\ref{fig:schematic_moria} for a schematic of the workflow). Our catalogs are publicly available in multiple formats and will enable the community to easily switch between mass definitions. In this paper, we present the new algorithms and some numerical tests. In a number of follow-up papers, we will analyze the impact of the halo boundary definition on subhalo occupation and mass functions.

The paper is structured as follows. We begin by describing the $N$-body simulations and halo catalogs on which we base our enhanced catalogs in Section~\ref{sec:simulations}. We describe the new algorithms in the \sparta and \moria codes in Sections~\ref{sec:sparta} and \ref{sec:moria}, respectively. We test our data products and present some basic results in Section~\ref{sec:results}. We discuss the strengths, weaknesses, and potential applications of our catalogs in Section~\ref{sec:discussion} before concluding in Section~\ref{sec:conclusion}.

Throughout the paper, we follow the notation established in \citetalias{diemer_17_sparta} and \citetalias{diemer_17_rsp} with minor changes (Table~\ref{table:defs}). The variance of the power spectrum, $\sigma$, is computed based on the transfer function approximation of \citet{eisenstein_98}. The dynamical time is calculated as the time to cross $\rtom$ at velocity $V_{\rm 200m}$, which is independent of halo mass and approaches $1/5$ of the Hubble time at high redshift \citepalias{diemer_17_sparta}. The accretion rate, $\gammadyn$, is then defined as the logarithmic mass growth over this time interval. All cosmological calculations are performed using the \colossus code \citep{diemer_18_colossus}. The \sparta code and its documentation are publicly available at \href{http://www.benediktdiemer.com/code}{benediktdiemer.com/code}, as are our catalogs and merger trees at \href{http://www.benediktdiemer.com/data}{benediktdiemer.com/data}.

 
\section{Simulation Data}
\label{sec:simulations}

\begin{deluxetable*}{lcccccccccccc}
\tablecaption{$N$-body Simulations
\label{table:sims}}
\tablewidth{\textwidth}
\tablehead{
\colhead{Name} &
\colhead{$L\, (\mpch)$} &
\colhead{$N^3$} &
\colhead{$m_{\rm p}\, (\msunh)$} &
\colhead{$\epsilon\, (\kpch)$} &
\colhead{$\epsilon / (L / N)$} &
\colhead{$z_{\rm initial}$} &
\colhead{$z_{\rm final}$} &
\colhead{$N_{\rm snaps}$} &
\colhead{$z_{\rm f-snap}$} &
\colhead{$z_{\rm f-cat}$} &
\colhead{Cosmology} &
\colhead{Reference}
}
\startdata
L2000-WMAP7 & $2000$ & $1024^3$ & $5.6 \times 10^{11}$  & $65$  & $1/30$ & $49$ & $0$ & $100$ & $20$ & $4.2$ & \wmap & \citetalias{diemer_15} \\
L1000-WMAP7 & $1000$ & $1024^3$ & $7.0 \times 10^{10}$ & $33$ & $1/30$ & $49$ & $0$ &  $100$ & $20$ & $6.2$ & \wmap & \citetalias{diemer_13_scalingrel} \\
L0500-WMAP7 & $500$  & $1024^3$ & $8.7 \times 10^{9}$  & $14$ & $1/35$  & $49$ & $0$ &  $100$ & $20$ & $8.8$ & \wmap & \citetalias{diemer_14} \\
L0250-WMAP7 & $250$  & $1024^3$ & $1.1 \times 10^{9}$  & $5.8$  & $1/42$  & $49$ & $0$ &  $100$ & $20$ & $11.5$ & \wmap & \citetalias{diemer_14} \\
L0125-WMAP7 & $125$  & $1024^3$ & $1.4 \times 10^{8}$  & $2.4$  & $1/51$  & $49$ & $0$ &  $100$ & $20$ & $14.5$ & \wmap & \citetalias{diemer_14} \\
L0063-WMAP7 & $62.5$ & $1024^3$ & $1.7 \times 10^{7}$  & $1.0$  & $1/60$ & $49$ & $0$ &  $100$ & $20$ & $17.6$ & \wmap & \citetalias{diemer_14} \\
L0031-WMAP7 & $31.25$ & $1024^3$ & $2.1 \times 10^{6}$  & $0.25$  & $1/122$ & $49$ & $2$ &  $64$ & $20$ & $20$ & \wmap & \citetalias{diemer_15} \\
L0500-Planck & $500$  & $1024^3$ & $1.0 \times 10^{10}$  & $14$ & $1/35$  & $49$ & $0$ &  $100$ & $20$ & $9.1$ & \planck & \citetalias{diemer_15} \\
L0250-Planck & $250$  & $1024^3$ & $1.3 \times 10^{9}$  & $5.8$  & $1/42$  & $49$ & $0$ &  $100$ & $20$ & $12.3$ & \planck & \citetalias{diemer_15} \\
L0125-Planck & $125$  & $1024^3$ & $1.6 \times 10^{8}$  & $2.4$  & $1/51$  & $49$ & $0$ &  $100$ & $20$ & $15.5$ & \planck & \citetalias{diemer_15} \\
L0100-PL-1.0 & $100$  & $1024^3$ & $2.6 \times 10^{8}$  & $0.5$  & $1/195$  & $119$ & $2$ & $64$ & $20$ & $20$ & PL, $n=-1.0$ & \citetalias{diemer_15} \\
L0100-PL-1.5 & $100$  & $1024^3$ & $2.6 \times 10^{8}$  & $0.5$  & $1/195$  & $99$ & $1$ & $78$ & $20$ & $20$ & PL, $n=-1.5$ & \citetalias{diemer_15} \\
L0100-PL-2.0 & $100$  & $1024^3$ & $2.6 \times 10^{8}$  & $1.0$  & $1/98$  & $49$ & $0.5$ & $100$ & $20$ & $15.5$ & PL, $n=-2.0$ & \citetalias{diemer_15} \\
L0100-PL-2.5 & $100$  & $1024^3$ & $2.6 \times 10^{8}$  & $1.0$  & $1/98$  & $49$ & $0$ & $100$ & $20$ & $5.4$ & PL, $n=-2.5$ & \citetalias{diemer_15} \\
TestSim200 & $62.5$  & $256^3$  & $1.1 \times 10^{9}$  & $5.8$  & $1/42$  & $49$ & $-0.1$ & $193$ & $9$  & $9$ & \wmap & \citetalias{diemer_17_sparta} \\
TestSim100 & $62.5$  & $256^3$  & $1.1 \times 10^{9}$  & $5.8$  & $1/42$  & $49$ & $-0.1$ & $96$  & $9$  & $9$ & \wmap & \citetalias{diemer_17_sparta} \\
TestSim50  & $62.5$  & $256^3$  & $1.1 \times 10^{9}$  & $5.8$  & $1/42$  & $49$ & $-0.1$ & $48$  & $9$  & $9$ & \wmap & \citetalias{diemer_17_sparta}
\enddata
\tablecomments{The \erebos suite of $N$-body simulations. $L$ denotes the box size in comoving units, $N^3$ the number of particles, $m_{\rm p}$ the particle mass, and $\epsilon$ the force softening length in physical units. The redshift range of each simulation is determined by the first and last redshifts $z_{\rm initial}$ and $z_{\rm final}$, but snapshots were output only between $z_{\rm f-snap}$ and $z_{\rm final}$. The earliest snapshots of some simulations do not yet contain any halos, and the first catalog with halos is output at $z_{\rm f-cat}$; the \sparta and \moria data also begin at that redshift. The cosmological parameters are given in Section~\ref{sec:simulations}, ``PL'' indicates self-similar cosmologies with a power-law initial spectrum with slope $n$. The references correspond to \citet[][\citetalias{diemer_13_scalingrel}]{diemer_13_scalingrel}, \citet[][\citetalias{diemer_14}]{diemer_14}, and \citet[][\citetalias{diemer_15}]{diemer_15}. Our system for choosing force resolutions is discussed in \citetalias{diemer_14}.}
\end{deluxetable*}

Our catalogs are based on the \erebos suite of $N$-body simulations, essentially the same as in \citetalias{diemer_17_sparta} and \citetalias{diemer_17_rsp} (Table~\ref{table:sims}). This collection contains two \LCDM cosmologies, scale-free universes, and a lower-resolution simulation for testing. The first \LCDM cosmology is that of the {\it Bolshoi} simulation \citep{klypin_11}, which is consistent with \wmap \citep[][$\Omega_{\rm m} = 0.27$, $\Omega_{\rm b} = 0.0469$, $h = 0.7$, $\sigma_8 = 0.82$, and $n_{\rm s} = 0.95$]{komatsu_11}. For this cosmology, we have run seven boxes with side lengths decreasing by factors of two from $2000$ down to $31.25 \mpch$ and corresponding particle masses that span more than five orders of magnitude (Table~\ref{table:sims}). The second cosmology is similar to the \citet{planck_14} cosmology ($\Omega_{\rm m} = 0.32$, $\Omega_{\rm b} = 0.0491$, $h = 0.67$, $\sigma_8 = 0.834$, and $n_{\rm s} = 0.9624$). For this cosmology, we use three boxes of $500$, $250$, and $125 \mpch$. We will denote these cosmologies as \wmap and \planck wherever we refer to the respective simulations collectively. The self-similar simulations represent Einstein--de Sitter universes with power-law initial power spectra of slopes $-1$, $-1.5$, $-2$, and $-2.5$. The parameters of such simulations can be expressed without explicit reference to length or time scales \citep[e.g.,][]{knollmann_08}, but we adjusted them to \LCDM-like parameters of $L = 100 \mpch$, $h = 0.7$, and $\sigma_8 = 0.82$ for historical reasons. Finally, we use three versions of the same, smaller test simulation, where we have removed none, half, and three-quarters of the snapshots in order to test the convergence of our algorithms with snapshot spacing.

The initial power spectra for the \LCDM simulations were generated using \textsc{Camb} \citep{lewis_00}, and the initial conditions for all simulations were created using the \textsc{2LPTic} code \citep{crocce_06}. The simulations were run with \textsc{Gadget2} \citep{springel_05_gadget2}. We use \rockstar \citep[][version 0.99.9-RC3+]{behroozi_13_rockstar} and \consistenttrees \citep[][version 1.01]{behroozi_13_trees} to construct halo catalogs and merger trees. \rockstar finds the particles in friends-of-friends groups in six-dimensional phase space. As our new \moria catalogs are directly built on top of the \rockstar catalogs, we will discuss details of the \rockstar algorithm throughout the paper. Our catalogs use $\rtom$ as the main mass definition, where we use only gravitationally bound particles (although we will discuss other definitions at length).

Our \LCDM cosmologies bracket the currently favored range of those cosmological parameters that have a large impact on dark matter structure formation, most notably $\Omega_{\rm m}$. Power-law cosmologies are less commonly studied, but their self-similarity offers a unique opportunity to study halo structure in a simplified manner with only one free parameter (the power spectrum slope). Any meaningful result must be independent of redshift, allowing us to establish resolution limits by comparing different redshifts \citep{joyce_20}. We further discuss the strengths and weaknesses of the \erebos simulation suite in Section~\ref{sec:discussion}.

\section{New Algorithms in SPARTA}
\label{sec:sparta}

In this section, we briefly review the splashback algorithm presented in \citetalias{diemer_17_sparta} and \citetalias{diemer_17_rsp}, referring the reader to those works for details. We describe a number of improvements and additions to \sparta, namely, an updated splashback algorithm (Section~\ref{sec:sparta:rsp}), SO calculations with and without gravitational unbinding (Sections~\ref{sec:sparta:defs_so_all} and \ref{sec:sparta:defs_so_bnd}), and the concept of tracer masses (Section~\ref{sec:sparta:defs_so_tcr}).

\subsection{Computing the Splashback Radius}
\label{sec:sparta:rsp}

The original purpose of the \sparta code was to calculate the splashback radius from the dynamics of individual particles. \sparta tracks particles as they approach a halo and records the time and location of their infall and their first apocenter (or ``splashback event''). We tag particles at infall if they are deemed to belong to a subhalo. Subhalos whose mass is greater than $1/100$ of the host mass may suffer from significant dynamical friction and are thus biased in their splashback radii. After removing particles from such subhalos from consideration, we smooth the distribution of the remaining particle splashbacks in time and define the halo's splashback radius as their mean, median, or higher percentiles (Figure~\ref{fig:viz_radii}). The same procedure is repeated for the mass enclosed within the radii of the particle splashbacks (meaning that the splashback mass can slightly deviate from the mass within the splashback radius). 

Since the publication of \citetalias{diemer_17_rsp}, we have made a number of improvements to the \sparta algorithm. First, we have fixed a bug due to which the Hubble drag term was underestimated by a factor of $h$ \citep{croton_13_h}, leading to earlier splashback times and splashback radii that were a few percent too large (because halos grow over time). Furthermore, the \citetalias{diemer_17_sparta} version of \sparta could not compute $\rsp$ for up to 5\% of halos depending on the mass range. We have increased the completeness (the fraction of halos with valid splashback measurement) of the $\rsp$ catalogs in a number of ways. First, the previous version of \sparta stopped tracking all particles as soon as a halo became a subhalo. The new version keeps tracking particles if the subhalo epoch lasts for only one snapshot because such fast flyby events do not necessarily mean that the particle trajectories are interrupted. Similarly, we do not categorically prohibit computing $\rsp$ and $\msp$ for subhalos any more, as long as there are sufficient particle splashback events near the time in question  \citepalias{diemer_17_sparta}. We do, however, still abort all particle trajectories when a halo becomes a subhalo for more than one snapshot. Subsequently, the subhalo will quickly run out of past splashback events, and a determination of $\rsp$ will not be possible any longer. This behavior is physically sensible because the splashback radius is ill-defined for subhalos. We have also updated our algorithm to determine which particles belong to a subhalo at infall (Section~\ref{sec:sparta:defs_so_tcr}). Altogether, the algorithmic changes shift the average $\rsp$ and $\msp$ by a few percent. To incorporate these differences, we recalibrate the model of \citetalias{diemer_17_rsp} in Section~\ref{sec:results:model}.

\subsection{SO Masses (All Particles)}
\label{sec:sparta:defs_so_all}

SO radii and masses are routinely calculated by halo finders such as \rockstar, but a number of subtleties can make it difficult to obtain exactly the desired definition. First, many halo finders offer only a limited number of definitions or particularly common ones such as $\rtoc$ or $\rvir$. Second, we can include all particles (strict SO) or only gravitationally bound particles. The resulting masses tend to be close for host halos, but there are some exceptions. For subhalos, strict SO masses are often ill-defined, as they can include the entire host halo in some cases. To facilitate a thorough exploration of mass definitions, \sparta allows the user to compute any number of SO radii and masses with arbitrary density thresholds. 

\sparta computes SO masses by considering a sorted list of particle radii within a user-defined factor times $\rtom$. This factor generally needs to be somewhat larger than unity because the common $\Delta_{\rm vir}$ definition tends to overdensities of about $178$ at high redshift, rendering $\rvir$ larger than $\rtom$. For subhalos, we start at $4 R_{\rm 200m,tcr}$, where the halo radius is computed from tracked particles only (Section~\ref{sec:sparta:defs_so_tcr} and Diemer \& Behroozi 2020, in preparation). Given a list of particles inside the search radius, sorted by radius, we start from the outermost particle and move inward until a density threshold is achieved for the first time. In some rare cases, the threshold can be crossed multiple times, and we wish to obtain the outermost occurrence. We set $\mdelta = i \times m_{\rm p}$ where $i$ is the index of the particle where the threshold was crossed. We note that the internally used $\rtom$ definition in \sparta interpolates the density between particles, but this difference is negligible in all but the smallest halos.

The calculation of an SO radius can fail for two reasons. First, the density may not reach the required threshold at any radius because even the very core of the halo is not dense enough. This issue occurs mostly for high-density definitions such as $\rfoc$ and can arise even in hosts halos. Second, the density may never fall below the threshold within the available particle distribution or within any reasonable radius. This issue cannot occur for hosts in \sparta because halos are defined via their $\rtom$ internally. For subhalos, however, the entire host halo may be included if the subhalo is close enough to its center. This issue predominantly affects low-density definitions such as $\rvir$ and $\rtom$. If the calculation fails, we record an error code in the catalogs. We investigate the completeness of our SO measurements in Section~\ref{sec:results:completeness}.

\subsection{SO Masses (Bound Particles)}
\label{sec:sparta:defs_so_bnd}

In addition to strict SO masses, we also wish to include bound-only masses in our catalogs, particularly for subhalos. Our base catalogs from \rockstar already contain bound-only masses that are computed as follows. \rockstar considers the particles in a friends-of-friends (FOF) group (or subgroup for subhalos) and unbinds particles if their kinetic energy exceeds their potential energy. The process is not repeated iteratively. If a halo contains more than some fraction (half, in our case) of unbound particles, it is abandoned as a transient feature and not included in the catalog.

While this unbinding procedure is sensible in theory and gives excellent results in practice, all unbinding techniques turn out to be highly subjective and code dependent because they strongly depend on the initial particle distribution considered. For example, if we wanted to be conservative and be sure to include all bound halo particles, we could start from all particles inside $2 \rtom$ --- but with this much mass in the overall distribution, virtually all particles within $\rtom$ would be bound. Conversely, if we considered only particles within $\rfoc$, very large fractions would be unbound. Since no SO definition is special, there is no ``correct'' distribution of initial particles. FOF-based halo finders such as \rockstar or \subfind use FOF groups as the initial particle set, which is sensible but depends on the linking length and the algorithm for discerning subclumps \citep[e.g.,][]{springel_01_subfind, han_12_hbt, behroozi_13_rockstar}. This issue is exacerbated in subhalos, where the position of the subhalo within the host will influence its overall density and thus how many particles would be considered bound. After much experimentation, we provide a fairly general unbinding algorithm in \sparta. We check the inequality 
\begin{equation}
\label{eq:bound}
-\Phi_{\rm i} = G \sum_{j \neq i}^{N_{\rm ptl}} \frac{1}{r_{\rm ij} + \epsilon} \geq f_{\rm bnd} \frac{v_{\rm i}^2}{2}  \,,
\end{equation}
where $N_{\rm ptl}$ is the number of particles to consider, $f_{\rm bnd}$ is a threshold for the binding-to-kinetic energy ratio, and $\epsilon$ is the force softening scale of the simulation. To compute the sum over many particles efficiently, we have adapted the tree algorithm of \rockstar for \sparta (see Diemer \& Behroozi 2020, in preparation, for details). The user can set the initial radius for particles, the boundness factor $f_{\rm bnd}$ in Equation~\ref{eq:bound}, and whether the procedure is iterated until it converges. We set the initial radius for host halos to $\rtom$ and for subhalos to the tracer-only radius (Section~\ref{sec:sparta:defs_so_tcr}), which eliminates the issue of unphysical all-particle radii in subhalos; however, it does not eliminate the issue of host material contributing to the potential.

We compare our bound-only masses to those from \rockstar in Section~\ref{sec:results:bnd}. Broadly speaking, we find that our results agree reasonably well for the majority of subhalos, although with strong outliers where the host contributes a significant amount of material. These tests demonstrate that the potential calculation in \sparta works as expected, but, given the issue of host contamination, it is not clear how useful the bound-only masses computed by \sparta would be. We have thus left them out of the catalogs and instead included the \rockstar bound-only masses for four common overdensity thresholds (Section~\ref{sec:results:cats}).

\subsection{Ghosts and Tracer Masses}
\label{sec:sparta:defs_so_tcr}

The algorithms described in the previous sections apply mostly to host halos, where we can meaningfully compute splashback and SO radii and masses without appealing to FOF algorithms or other ways to decide on the membership of particles. We have, however, also introduced entirely new algorithms for dealing with subhalos in \sparta. These methods will be described in detail in Diemer \& Behroozi (2020, in preparation), but we briefly review them here because the results are included in our catalogs. 

Unlike the \sparta version of \citetalias{diemer_17_sparta}, the code now tracks all particles in subhalos. When the halo first becomes a subhalo (when it crosses $\rtom$), we identify the particles that truly belong to the subhalo using an updated algorithm. Specifically, we apply user-defined thresholds to exclude particles that first joined the subhalo close to the host (as they were probably swept up during infall) and that are not strongly gravitationally bound. Thereafter, we assume that subhalos only ever lose particles and never accrete new ones. When particles stray more than $2 R_{\rm 200m,tcr}$ from the subhalo, they are removed. The ``tracer masses,'' including $R_{\rm 200m,tcr}$, are computed like normal SO masses but only including the remaining distribution of tracked particles. This definition gives similar but distinct results to \rockstar's unbinding algorithm (Diemer \& Behroozi 2020, in preparation).

We also use the tracking algorithm to extend the lives of subhalos after they are lost by the halo finder. When a subhalo disappears from the input catalogs, we keep following its particles, at which point we term the halo a ``ghost.'' We track the ghost's particles until there are only $10$ particles left or until it has sunk to the center of the host halo. We compute the position and velocity of the ghost based on its most bound particles. Conceptually, ghosts are similar to ``orphans'' and ``cores,'' where subhalos are represented as a single particle or a number of the most bound particles \citep[e.g.,][]{wang_06_orphans, guo_10, heitmann_19, heitmann_20}. Tracking all particles, however, allows us to compute physically meaningful tracer masses for ghosts. In Diemer \& Behroozi (2020, in preparation), we show that tracking ghosts significantly increases the completeness of our merger trees.

\section{The MORIA Extension}
\label{sec:moria}

\sparta creates an output file with information about halos throughout a simulation's range of cosmic time, including their splashback radii and masses. This output file, however, does not constitute a halo catalog because it is structured as a series of halo histories, because it does not duplicate the majority of the halo finder data, and because \sparta's splashback algorithm is complete to only about 95\% \citepalias{diemer_17_sparta}. Moreover, a halo catalog should contain information about the relationships between halos, namely, their host--subhalo relationships. 

In this section, we introduce a new extension to \sparta called \moria\footnote{In ancient Greece, {\it moriai} (plural of {\it moria}, or $\mu o \rho \acute{\iota} \alpha$) were sacred olive trees that belonged to the state rather than an individual, a fitting name given that the purpose of \moria is to create publicly available merger trees from \sparta data.}, which creates halo catalogs and merger trees that contain both the halo finder's original output and \sparta's results. We describe the general code design (Section~\ref{sec:moria:general}), algorithms to increase the completeness of the splashback data (Section~\ref{sec:moria:completeness}), calculations of the mass accretion rate (Section~\ref{sec:moria:accrate}), the computation of host--subhalo assignments (Section~\ref{sec:moria:hostsub}), our new merger tree format (Section~\ref{sec:moria:trees}), and \moria's system for setting resolution limits (Section~\ref{sec:moria:limits}). The procedure for running \sparta and \moria on a simulation is summarized in Figure~\ref{fig:schematic_moria}; we refer the reader to the \sparta code documentation for detailed instructions.

\subsection{General Code Design}
\label{sec:moria:general}

\begin{figure}
\centering
\vspace{0.4cm}
\includegraphics[trim =  10mm 120mm 18mm 9mm, clip, scale=0.25]{\figdir/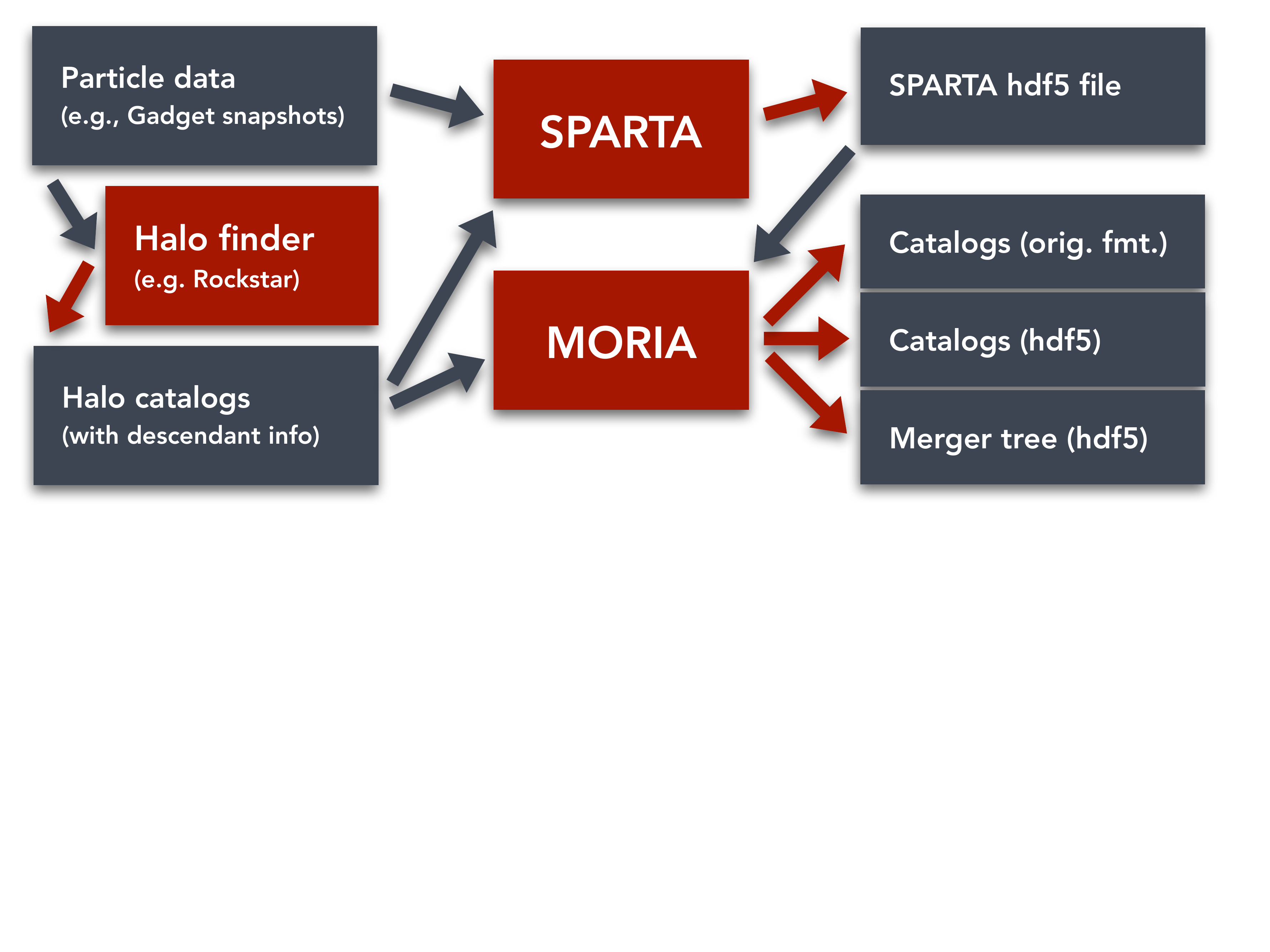}
\caption{Flow chart for the \sparta and \moria codes. Red fields represent codes; blue fields represent data products. From the codes' perspective, blue arrows signify inputs, whereas red arrows signify outputs. Particle data, the original output from the simulation, are fed into a halo finder to create catalogs. \sparta needs the halos to be connected between snapshots, that is, it needs progenitor/descendant information. Based on the information in the snapshot and catalog files, \sparta creates an hdf5 output file. \moria combines the information in that file with the original halo catalogs to create enhanced catalogs and merger trees. The simulation data (the left-hand side of the schematic) are discussed in Section~\ref{sec:simulations}, the \sparta and \moria codes in Sections~\ref{sec:sparta} and \ref{sec:moria}, respectively.}
\label{fig:schematic_moria}
\end{figure}

\moria works in post-processing, meaning that it can be run multiple times on the same \sparta output (Figure~\ref{fig:schematic_moria}). The main reason for this choice is that the CPU time consumed by \moria is orders of magnitude smaller than that of \sparta because the former processes only halo data but not particles. For example, it would be highly inconvenient to rerun \sparta only to create a new halo catalog with a few extra fields from the original catalogs. Nevertheless, \moria performs I/O operations on large catalog and hdf5 files and uses tree searches to find host--subhalo relations, making it too computationally demanding for scripting languages. Thus, \moria is, like \sparta, written in pure C, but it runs on a single process.

The inputs to \moria are a \sparta output file and the original catalogs used by \sparta to create that file. The output format is flexible: the user can choose to write catalogs in \moria's native hdf5 format, the original catalog format (e.g., ASCII files for \rockstar), or a merger tree format that is essentially the same hdf5 format as the catalogs. The content of those files is entirely up to the user, who can choose any number of SO and splashback mass definitions that are present in the \sparta output file, as well as any number of fields from the original catalog. Mass definitions that exist in both the halo finder and \sparta outputs will be named accordingly, allowing for comparisons between different algorithms. Finally, the user can choose to compute an arbitrary number of host--subhalo relations based on any of the available radius definitions (Section~\ref{sec:moria:hostsub}). Virtually all parameters mentioned in the following sections can be changed by the user at run-time.

\subsection{Increasing the Completeness of Splashback Data}
\label{sec:moria:completeness}

In \citetalias{diemer_17_sparta}, we showed that the \sparta algorithm for determining $\rsp$ is at least about 95\% complete for host halos with $\ntom \geq 1000$ (and often better depending on halo mass and redshift). This completeness is somewhat unsatisfactory for a number of reasons. First, we would like to push below $1000$ particles to include the large fraction of halos that are smaller than this limit. Second, even 5\% of halos without splashback values can lead to inconvenient systematics, e.g., in mass functions. Third, many of the interrupted splashback histories occur because the halo temporarily becomes a subhalo, meaning that it enters within $R_{\rm 200m,bnd}$ of a larger halo. When we compute host--subhalo relations for the splashback definitions, some of those subhalos may become hosts but not have $\rsp$ measurements. This case occurs relatively rarely because the splashback radius is typically larger than $\rtom$, but, depending on the exact definition, there is a significant fraction of halos for which it is slightly smaller \citepalias{diemer_17_rsp}. We have partially addressed this issue in Section~\ref{sec:sparta:rsp}, but we would ideally like to achieve a completeness of 100\% for host halos.

\moria addresses this issue by interpolating $\rsp$ and $\msp$ across time. In many cases, splashback histories are interrupted by flyby events with more than one snapshot as a subhalo or a temporary lack of particles at first apocenter. Such gaps can be bridged because the overall splashback radius and mass histories tend to be relatively smooth, particularly when expressed as $\rsp / \rtom$ and $\msp / \mtom$. As a first guess, \moria looks for an earlier and later epoch within one dynamical time where the halo was a host and where $\rsp$ and $\msp$ were successfully measured by \sparta. Furthermore, we check that $\rsp/\rtom$ and $\msp/\mtom$ (for all splashback definitions) are not too dissimilar from our expectation for a halo of the given mass and mass accretion rate, as quantified by the model from \citetalias{diemer_17_rsp} (see Section~\ref{sec:results:model} and Equation~\ref{eq:fit1}). We require the value to be within a factor of two or a logarithmic factor of $5 \sigma$, where the standard deviation is quantified by Equation~\ref{eq:sigma}. These limits may seem overly generous, but the nominal standard deviation from the model can be as small as $0.02$ dex or 5\%. This deviation does not capture the significant tails in the distribution, which we do want to allow to avoid biasing our results. If at least one valid past or future epoch is found, we assume that $\rsp/\rtom$ and $\msp/\mtom$ vary slowly over a dynamical time and accept their values for the missing epoch. If both a past and future epoch are found, we linearly interpolate $\rsp/\rtom$ and $\msp/\mtom$ in time. Another potential source of error are the values of $\rtom$ and $\mtom$ used to convert the ratios to $\rsp$ and $\msp$. If $\rtom$ includes a large amount of material from a nearby halo (Section~\ref{sec:sparta:defs_so_all}), keeping the ratio fixed can lead to extreme values of $\rsp$. Thus, we check the ratio between $R_{\rm 200m,all}$ and $R_{\rm 200m,bnd}$ and use the latter if the ratio exceeds two. This choice is consistent with the fitting function on Equation~\ref{eq:fit1}, where we exclude halos with extreme all-to-bound ratios. 

The accuracy of the interpolation depends on a number of factors, including the time spacing of snapshots, how many snapshots lack $\rsp$ determinations, and on the lowest particle number considered. We have tested the interpolation from past snapshots at $z = 0$ in TestSim100. For halos with $\ntom > 200$, we find that the guess based on previous snapshots is biased low by about 3\% in radius and 5\% in mass on average, with about 15\% scatter. The low estimate is expected as $\rsp/\rtom$ typically increases with time. These results are likely to be a worst-case scenario because $\rsp/\rtom$ increases rapidly at $z = 0$ as mass accretion rates are falling quickly; this is particularly relevant in a small box such as TestSim100, where the majority of halos are low-mass galaxy halos. We find similar values for guesses based on future snapshots and much better results for interpolated values, depending on the time interval over which we interpolate.  For short intervals of one or a few snapshots, the typical error is less than 1\%. We have also experimented with extrapolating the evolution of $\rsp/\rtom$ from past snapshots but found that, while it can slightly reduce the bias, it leads to somewhat erratic results and large scatter.

If no epoch with valid $\rsp$ determination is found within a dynamical time, we use the mass and mass accretion rate of the halo to compute the fitting model of Section~\ref{sec:results:model}. This procedure is not circular because the model is calibrated using only halos where $\rsp$ and $\msp$ were computed by \sparta. The fitting function is problematic in that we are imposing a median relation with significant scatter, but this guess is preferable over having no estimate at all (which could lead to significant differences in the subhalo statistics; see Section~\ref{sec:moria:hostsub}). The catalogs contain a status value that indicates how the splashback data for each halo and epoch were computed, allowing the user to easily exclude interpolated or predicted values. We test the completeness of our catalogs in Section~\ref{sec:results:completeness}.

\subsection{Mass Accretion Rates}
\label{sec:moria:accrate}

Over the past few years, it has become clear that the mass accretion rate of halos is an important property that determines their density profiles, splashback radius, and certain baryonic properties \citep{diemer_14, more_15, lau_15, green_20}. However, the optimal definition of the accretion rate is not obvious: while we would ideally like to define an infinitesimal derivative of mass akin to theoretical models \citep{adhikari_14, shi_16_rsp}, this quantity has no meaning in simulations where it is entirely dominated by shot noise and mergers. 

In \moria, we keep the definition of \citetalias{diemer_17_sparta}, the logarithmic change in all-particle $\mtom$ per scale factor over one dynamical time, or $\gammadyn$. However, \citet{xhakaj_19_accrate} recently reported that our calculation of $\gammadyn$ differs from \rockstar's owing to three seemingly subtle differences: the mass definition (bound-only $\mvir$ in \rockstar, all-particle $\mtom$ in \moria), interpolating between two past snapshots (which is done in \rockstar but not in \moria), and the definition of the dynamical time ($t_{\rm dyn,vir}$ in \rockstar, $t_{\rm dyn,200m}$ in \moria). The latter is by far the most important factor, highlighting how sensitive accretion rates can be to the time interval over which they are computed. For comparison, the \rockstar mass accretion rates are included in our catalogs (Section~\ref{sec:results:cats}). We note that the \citet{behroozi_13_rockstar} definition of dynamical time is a factor of two shorter than ours, meaning that their rate measured over $2 \tdyn$ corresponds most closely to our definition.

Once again, a complication arises for subhalos and backsplash halos that were subhalos one dynamical time ago. We could replace the unphysical all-particle masses of subhalos with bound-only masses from \rockstar, but this mixture would lead to an apples-to-oranges comparison and ignore the physical reality that there is no clear equivalency between the SO masses of hosts and subhalos. Moreover, the splashback radius is influenced by the total change of mass within the orbits of particles, which includes all matter regardless of whether it ``truly'' belongs to the halo. Thus, we use $M_{\rm 200m,all}$ but adaptively reduce the interval over which we compute $\gammadyn$ to avoid subhalo epochs. We stop if the interval becomes shorter than $0.25 \tdyn$, which occurs if the halo was a subhalo until very recently. In this rare case, we use the median value of $\gammadyn$ as predicted by the fitting function of Section~\ref{sec:results:model}, ignoring the substantial scatter of $(0.41 - 0.07 \nu)$ dex. Physically speaking, we have to accept that there are halos for which it is difficult to define a meaningful accretion rate. For example, the evolution of backsplash halos can be dominated by stripping due to the flyby event. For subhalos, there is no sensible way to define an accretion rate at all. For completeness, the catalogs give $\gammadyn$ at the epoch when the subhalo was accreted, which we compute as for host halos. The catalogs contain a flag that indicates how the accretion rate was computed. 

At $z \approx 0$, $\gammadyn$ is computed by the default procedure for about 99\% of host halos, over a reduced interval for 1\%, and from the fitting function for a negligible fraction. In the smallest box at $z = 0$, L0063-WMAP7, those numbers rise to 4\% and 0.2\%. For subhalos, between 1\% and 17\% have reduced-interval accretion rates (at their infall redshift), and between 1\% and 4\% have only the fitting function estimate. These numbers vary only slightly with redshift. Moreover, \citet{mansfield_20_resolution} find that mass accretion rates in the \erebos simulations are well converged even at $100$ particles per halo. In conclusion, we are able to measure mass accretion rates for virtually all halos for which this quantity is physically sensible.

\subsection{Host--subhalo Relations}
\label{sec:moria:hostsub}

\begin{figure}
\centering
\includegraphics[trim =  43mm 8mm 40mm 12mm, clip, scale=0.28]{\figdir/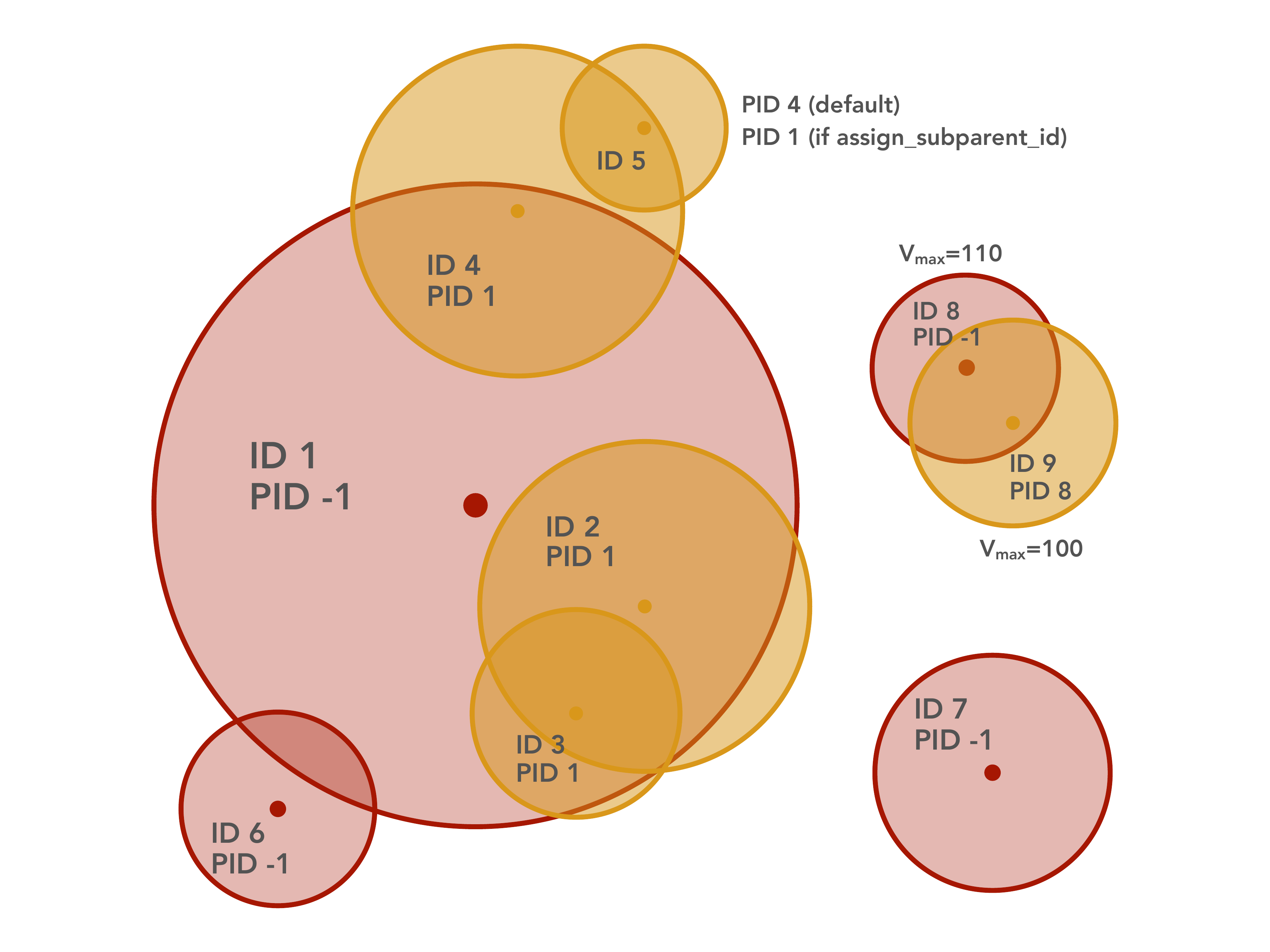}
\caption{Schematic illustration of the subhalo assignment in \moria, which essentially follows the \consistenttrees method. Each circle indicates a host halo (red) or subhalo (orange) radius according to some radius definition. Halos are subhalos if their center lies within the radius of another halo with higher $\vmax$. In this drawing, we assume that $\vmax \propto R$ with the exception of halos 8 and 9, where the halo with the smaller radius becomes the host because it has a larger $\vmax$. Halos 3 and 5 illustrate choices we need to make about sub-subhalos. In the case of halo 3, we assign halo 1 as the parent (rather than halo 2, which is also a parent). In the case of halo 5, the user can choose whether to assign the direct parent or parent-parent. In our catalogs, we have chosen to assign parent ID 4.}
\label{fig:schematic_hostsub}
\end{figure}

Our process of assigning host--subhalo relations is illustrated in Figure~\ref{fig:schematic_hostsub}. We find all halos whose centers lie within another, larger halo \citep[see][for alternative percolation algorithms]{garcia_19}. Here we need to define what ``larger'' means, since we will be dealing with a range of mass definitions. This ordering definition is a user-defined parameter in \moria. Given that none of the mass definitions discussed so far can be computed and are meaningful for all halos, we follow \citet{behroozi_13_rockstar} in using $\vmax$ (as measured by \rockstar) to sort the halos. Starting with the largest halo, we use a tree search algorithm to look for all halos within its radius in a given mass definition (respecting the periodic boundary conditions). We mark all such halos as subhalos of the current halo unless they are already a subhalo.

Some subhalos are inside multiple hosts, such as halo 3 in Figure~\ref{fig:schematic_hostsub}. In this case, we assign the largest parent's ID (i.e., 1 in our example; this logic is equivalent to the ``upid'' field in \consistenttrees catalogs). Another choice arises for subhalos whose only host is also a subhalo (halo 5 in Figure~\ref{fig:schematic_hostsub}). \moria lets the user decide which parent ID to use; in the catalogs presented in this paper we assign the direct parent's ID (4 in this example). We have tested our assignment against \consistenttrees and find perfect agreement when using the same radius definition.

One subtle question is how to handle missing radius estimates, for instance, when SO definitions cannot be computed (Section~\ref{sec:sparta:defs_so_all}). We set those radii to zero, meaning that those halos cannot have subhalos. This solution seems sensible regardless of whether the SO could not be computed because the the halo never reaches the required central density at the center or because it is so close to a host that it includes its mass. Both issues apply predominantly to small halos that would not contain many subhalos regardless. Finally, we note that not all parent halos are necessarily part of the catalogs, depending on the chosen resolution cut (Section~\ref{sec:results:cats}).

\subsection{Merger Tree Format}
\label{sec:moria:trees}

\begin{figure}
\centering
\includegraphics[trim =  5mm 0mm 0mm 0mm, clip, scale=0.24]{\figdir/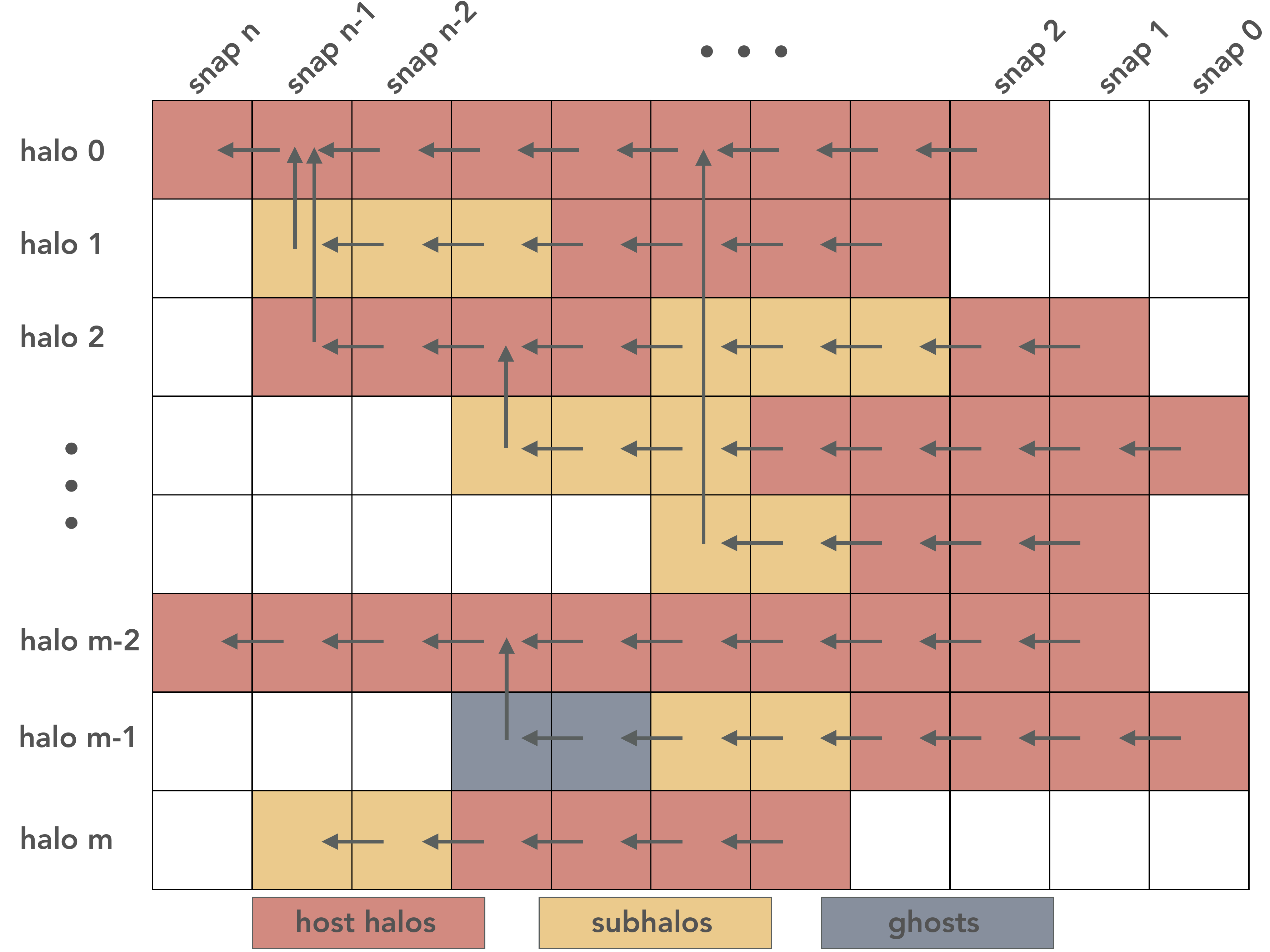}
\caption{Schematic depiction of the \moria merger tree format. Unlike conventional formats, a \moria tree is represented by two-dimensional arrays spanning the number of halos and the number of snapshots. White fields represent times when the given halo did not exist. Time flows from right to left, with gray arrows indicating the descendant of each halo. When a halo ends, it either has a descendant at the final snapshot or it merges into another halo. The ordering of the tree is determined by those mergers (see Section~\ref{sec:moria:trees} for details).}
\label{fig:schematic_tree}
\end{figure}

\begin{deluxetable*}{lcl}
\tablecaption{Partial list of fields included in the \moria catalogs and merger trees
\label{table:fields}}
\tablewidth{\textwidth}
\tablehead{
\colhead{Field} &
\colhead{From} &
\colhead{Explanation}
}
\startdata
\multicolumn{3}{l}{\rule{-7pt}{3ex} {\bf Organizational and merger information:}} \\
\rule{-2pt}{2ex} 
\rule{1pt}{0ex} \verb|id| & C,S & Halo ID from the original catalogs (except ghosts, where the ID is determined by \sparta) \\
\rule{1pt}{0ex} \verb|descendant_id| & C,S & ID of the halo at the next snapshot, or of the halo it merged into if it is ending (tree only) \\
\rule{1pt}{0ex} \verb|descendant_index| & M & Tree index of descendant; same as current halo unless merging (tree only) \\
\rule{1pt}{0ex} \verb|mask_alive| & M & True if there is an alive halo at this index and snapshot, false if not (tree only)\\
\rule{1pt}{0ex} \verb|mask_cut| & M & True if halo is part of catalog as well as tree (tree only) \\
\rule{1pt}{0ex} \verb|num_prog| & C & Number of progenitors (not counting ghosts) \\
\multicolumn{3}{l}{\rule{-7pt}{3ex} {\bf Halo status:}} \\
\rule{-2pt}{2ex} 
\rule{1pt}{0ex} \verb|status_sparta|  & S & Indicates host/sub/ghost status, switches in host, becoming a subhalo, becoming host, etc. \\
\rule{1pt}{0ex} \verb|phantom| & C & Nonzero if halo was interpolated by \consistenttrees \\
\rule{1pt}{0ex} \verb|status_acc_rate| & M & Status indicating whether accretion rate was computed or guessed \\
\rule{1pt}{0ex} \verb|status_sparta_rsp| & S & Status of $\rsp$ analysis in \sparta at this snapshot (success, not enough particles, etc.) \\
\rule{1pt}{0ex} \verb|status_moria_rsp|  & M & Status of $\rsp$ analysis in \moria (taken from \sparta, interpolated, guessed from fitting function, etc.) \\
\rule{1pt}{0ex} \verb|status_moria_hps_M<definition>|  & M & Status of each SO mass (success, not dense enough, never below threshold, etc.) \\
\rule{1pt}{0ex} \verb|status_moria_hps_R<definition>|  & M & Status of each SO radius definition (probably the same as for the SO mass of the same definition) \\
\multicolumn{3}{l}{\rule{-7pt}{3ex} {\bf Important times in halo history and merger information:}} \\
\rule{-2pt}{2ex} 
\rule{1pt}{0ex} \verb|Mpeak_Scale| & C & Scale where highest $\mtombnd$ was reached during halo's history \\
\rule{1pt}{0ex} \verb|Acc_Scale| & C & Scale where subhalo last became a subhalo, if it ever did \\
\rule{1pt}{0ex} \verb|First_Acc_Scale| & C & Scale where this halo first became a subhalo, if ever \\
\rule{1pt}{0ex} \verb|scale_of_last_MM| & C & Scale where last merger with mass ratio greater than $0.3$ occurred, if ever \\
\rule{1pt}{0ex} \verb|Time_to_future_merger| & C & Time (in Gyr) until the halo merges into a larger halo, if ever \\
\rule{1pt}{0ex} \verb|Future_merger_MMP_ID| & C & ID of most massive progenitor into which halo merges (-1 if it does not exist at this time) \\
\multicolumn{3}{l}{\rule{-7pt}{3ex} {\bf Mass and radius definitions:}} \\
\rule{-2pt}{2ex} 
\rule{1pt}{0ex} \verb|R200m_all_spa_internal|  & S & Internally used $\rtom$ from \sparta ($R_{\rm 200m,all}$ for hosts, $R_{\rm 200m,tcr}$ for subhalos) \\
\rule{1pt}{0ex} \verb|M200m_all_spa_internal|  & S & Internally used $\mtom$ from \sparta (corresponding to \verb|R200m_all_spa_internal|) \\
\rule{1pt}{0ex} \verb|nu200m_internal|  & M & Peak height, $\nutom$, calculated from internal $\mtom$ or bound mass if $M_{\rm 200m,all} > 2 M_{\rm 200m,bnd}$ \\
\rule{1pt}{0ex} \verb|R<delta>_all_spa| & S & SO radius computed from all particles (with $\Delta = $ 500c, 200c, vir, and 200m) \\
\rule{1pt}{0ex} \verb|R<delta>_tcr_spa| & S & SO radius computed from tracked particles for subhalos, same as all-particle radius for hosts \\
\rule{1pt}{0ex} \verb|R<delta>_bnd_cat|  & R & SO radius computed from bound particles only \\
\rule{1pt}{0ex} \verb|M<delta>_all_spa| & S & SO mass computed from all particles \\
\rule{1pt}{0ex} \verb|M<delta>_tcr_spa| & S & SO mass computed from tracked particles for subhalos, same as all-particle mass for hosts \\
\rule{1pt}{0ex} \verb|M<delta>_bnd_cat| & R & SO mass computed from bound particles only \\
\rule{1pt}{0ex} \verb|Rsp-apr-mn| & S & $\rsp$ computed from mean of the particle splashback distribution \\
\rule{1pt}{0ex} \verb|Rsp-apr-p<percentile>| & S & $\rsp$ computed from percentiles of the particle splashback distribution (50, 70, 75, 80, 85, 90) \\
\rule{1pt}{0ex} \verb|Msp-apr-mn| & S & $\msp$ computed from mean of the particle splashback distribution \\
\rule{1pt}{0ex} \verb|Msp-apr-p<percentile>| & S & $\msp$ computed from percentiles of the particle splashback distribution (50, 70, 75, 80, 85, 90) \\
\rule{1pt}{0ex} \verb|Vmax| & R & Maximum circular velocity \\
\rule{1pt}{0ex} \verb|Macc| & C & $\mtombnd$ at accretion (for subhalos) \\
\rule{1pt}{0ex} \verb|Vacc| & C & $\vmax$ at accretion (for subhalos) \\
\rule{1pt}{0ex} \verb|M200m_peak_cat| & C & Highest $\mtombnd$ attained during halo's history \\
\rule{1pt}{0ex} \verb|Vpeak| & C & Highest $\vmax$ attained during halo's history \\
\rule{1pt}{0ex} \verb|Vmax@Mpeak| & C & $\vmax$ at scale \verb|Mpeak_Scale| \\
\rule{1pt}{0ex} \verb|First_Acc_Mvir| & C & $\mvir$ when first becoming subhalo (if ever) \\
\rule{1pt}{0ex} \verb|First_Acc_Vmax| & C & $\vmax$ when first becoming subhalo (if ever) \\
\multicolumn{3}{l}{\rule{-7pt}{3ex} {\bf host--subhalo relations:}} \\
\rule{-2pt}{2ex} 
\rule{1pt}{0ex} \verb|parent_id_cat|  & C & Parent ID according to original catalog; equivalent to \verb|upid| (most massive host) \\
\rule{1pt}{0ex} \verb|parent_id_R<delta>_all_spa| & M & Parent ID for all-particle SO definitions \\
\rule{1pt}{0ex} \verb|parent_id_R<delta>_bnd_cat| & M & Parent ID for bound-only SO definitions \\
\rule{1pt}{0ex} \verb|parent_id_R<delta>_tcr_spa| & M & Parent ID for tracer SO definitions (experimental and very similar to all-particle assignment) \\
\rule{1pt}{0ex} \verb|parent_id_Rsp-apr-mn| & M & Parent ID for $\rsp$ from mean of the particle splashback distribution \\
\rule{1pt}{0ex} \verb|parent_id_Rsp-apr-p<percentile>| & M & Parent ID for $\rsp$ from percentiles of the particle splashback distribution (50, 70, 75, 80, 85, 90) \\
\multicolumn{3}{l}{\rule{-7pt}{3ex} {\bf Accretion rates:}} \\
\rule{-2pt}{2ex} 
\rule{1pt}{0ex} \verb|acc_rate_200m_dyn| & M & Fiducial accretion rate $\gammadyn$ as computed by \moria, see \verb|status_acc_rate|; dimensionless units \\
\rule{1pt}{0ex} \verb|Acc_Rate_1*Tdyn| & C & Acc. rate over $0.5 \tdyn$ as defined here but using $t_{\rm dyn,vir}$; in absolute units of $\msun / h / {\rm yr}$ \\
\rule{1pt}{0ex} \verb|Acc_Rate_2*Tdyn| & C & Acc. rate in $\mtombnd$ over $\tdyn$ as defined here but using $t_{\rm dyn,vir}$ \\
\rule{1pt}{0ex} \verb|Acc_Rate_Inst| & C & Acc. rate in $\mtombnd$ over one snapshot (very noisy) \\
\rule{1pt}{0ex} \verb|Acc_Rate_100Myr| & C & Acc. rate in $\mtombnd$ over 100 Myr (typically short compared to $\tdyn$) \\
\rule{1pt}{0ex} \verb|Acc_Rate_Mpeak| & C & Acc. rate in $M_{\rm 200m,bnd,peak}$ from current $z$ to $z + 0.5$ \\
\rule{1pt}{0ex} \verb|Acc_Log_Vmax_1*Tdyn| & C & Difference in $\log_{10} \vmax$ over $0.5 t_{\rm dyn,vir}$ \\
\rule{1pt}{0ex} \verb|Acc_Log_Vmax_Inst| & C & Difference in $\log_{10} \vmax$ over one snapshot \\
\rule{1pt}{0ex} \verb|Log_(VmaxVmax_max(Tdyn;Tmpeak))| & C & $\log_{10} \vmax$ over $\vmax(t-\tdyn)$ or $\vmax(t_{\rm peak})$ (the latter if $M_{\rm peak}$ happened more than $\tdyn$ ago) \\
\multicolumn{3}{l}{\rule{-7pt}{3ex} {\bf Other halo properties:}} \\
\rule{-2pt}{2ex} 
\rule{1pt}{0ex} \verb|x| & R,S & Halo position from \rockstar, except for ghosts (and optionally phantoms) where computed by \sparta \\
\rule{1pt}{0ex} \verb|v| & R,S & Halo velocity from \rockstar, except for ghosts (and optionally phantoms) where computed by \sparta 
\enddata
\tablecomments{The letters in the ``From'' column indicate the possible sources of a field, namely, R (\rockstar), C (\consistenttrees), S (\sparta), or M (\moria). This list is incomplete; see the \sparta documentation for additional fields and units. In identifiers, ``delta'' can be replaced by ``200m,'' ``vir,'' ``200c,'' or ``500c.'' The ``percentile'' for $\rsp$ definitions can be 50, 70, 75, 80, 85, or 90.}
\end{deluxetable*}

The term ``merger tree'' can refer to a format of outputting time-ordered halo information or to the history of one halo and its progenitors; here, we mean the former. The \moria merger trees contain basically the same information as the collection of catalogs from all snapshots, but the information is ordered differently. A typical ordering is ``depth-first,'' where the subsequent entries represent the most massive progenitor branch of one halo going back in time. Once that branch ends, the next progenitor (starting at some earlier time) is listed, and so on. In this way, one can construct a logical scheme where halos are listed sequentially as in the catalog entries, except that a redshift or snapshot number must be assigned to each entry \citep[e.g.,][see \citealt{srisawat_13} for a code comparison]{springel_01_subfind, behroozi_13_trees, rodriguezgomez_15, elahi_19_treefrog}.

The \moria merger tree format is designed in an entirely different fashion as illustrated in Figure~\ref{fig:schematic_tree}. We store halos in a 2D array spanning all halo histories and the number of snapshots in the simulation (a similar system is used in the \textsc{HBT+} halo finder of \citealt{han_18_hbt}). This format has the obvious advantage that it is easy to extract a catalog-like dataset (by selecting a time slice) or a history-like dataset (by selecting a halo index). At first sight, the format seems wasteful since about one-third of the array is occupied by halos that did not yet or do no longer exist. These empty fields, however, occupy very little additional disk space owing to the hdf5 compression algorithm. The only difference between our catalogs and a time slice of the tree data is that the tree includes the entire histories of all halos that are included in the catalogs at any snapshot, meaning that, at a given redshift, the tree will include some halos that would not pass the threshold in the corresponding catalog (Section~\ref{sec:moria:limits}). If so, a flag in the tree data indicates that a halo is present only in the tree file. We provide Python code to load the catalog and tree data, which takes care of this cut automatically if desired.

The halo histories in the trees correspond to progenitor--descendant relations that we take directly from the original \consistenttrees catalogs. These relations uniquely determine the next epoch in a halo history (horizontal gray arrows in Figure~\ref{fig:schematic_tree}) or the tree branch into which a halo merges at the end of its life if it does not survive until the end of the simulation (vertical arrows). \moria does not attempt to improve the progenitor--descendant relations except for ghosts (blue fields in Figure~\ref{fig:schematic_tree}). 

We are now free to sort the tree file along the halo axis in any convenient fashion. We split the halos into subtrees, where each subtree contains a host halo at the final snapshot, its subhalos, and all histories that merged into them at previous redshifts. The sub-trees are sorted by the same quantity that is used to impose a mass cut on the catalogs and trees, the peak $\mtom$ along a halo history in our case. Thus, the first ``line'' in a tree file is the halo with the largest $M_{\rm 200m,peak}$ that survived until the final snapshot. This row in the 2D array will be followed by the halos that merged into the top halo at the second-to-last snapshot, again ordered by their $M_{\rm 200m,peak}$. Thus, a halo history is not necessarily followed by its own progenitors, leading to upward-pointing gray arrows that cross multiple lines in Figure~\ref{fig:schematic_tree}.

\subsection{Limits on Mass and Halo Status}
\label{sec:moria:limits}

\moria makes it a priority to apply a well-defined lower-mass cutoff to the catalogs, with the twofold purpose of dissuading the user from accidentally using poorly resolved halos and saving disk space. The latter effect is substantial because the steep mass function of halos means that halos near the resolution limit account for a significant fraction of a catalog's size. The cutoff can be given in units of halo mass according to some definition, as a number of particles corresponding to that mass, $\vmax$, or $\vmax$ corresponding to a number of particles. In the latter case, we estimate the limiting value of $\vmax$ using the empirical formula of \citet{klypin_11},
\begin{equation}
\vmax \approx 2.8 \times 10^{-2} \mvir^{0.316} \,.
\end{equation}
Generally, it is preferable to set limits in units of particle number since they directly connect to the mass resolution of each simulation. We discuss the mass limits chosen for our catalogs in Section~\ref{sec:results:cats}.

Furthermore, the user can decide how to treat phantoms and ghosts. Phantoms are rare cases where \consistenttrees infers the existence of a halo from past and future snapshots even though \rockstar's FOF algorithm did not detect it. This can happen to both host and subhalos, although the latter make up for the vast majority of phantoms. \sparta uses the tracked particles in phantoms to compute their positions and velocities, which will not exactly agree with those from the halo finder (Section~\ref{sec:sparta:defs_so_tcr}). The user can choose to replace the \rockstar values with those from \sparta, but we have not used this feature in our catalogs because it causes the host--subhalo relations to slightly differ from \consistenttrees even if the same mass definition is used. We do not include ghosts in our catalogs because they do not have any catalog-defined properties, by definition; they are, however, included in the merger trees.


\begin{figure*}
\centering
\includegraphics[trim =  0mm 19mm 2mm 0mm, clip, scale=0.66]{\figdir/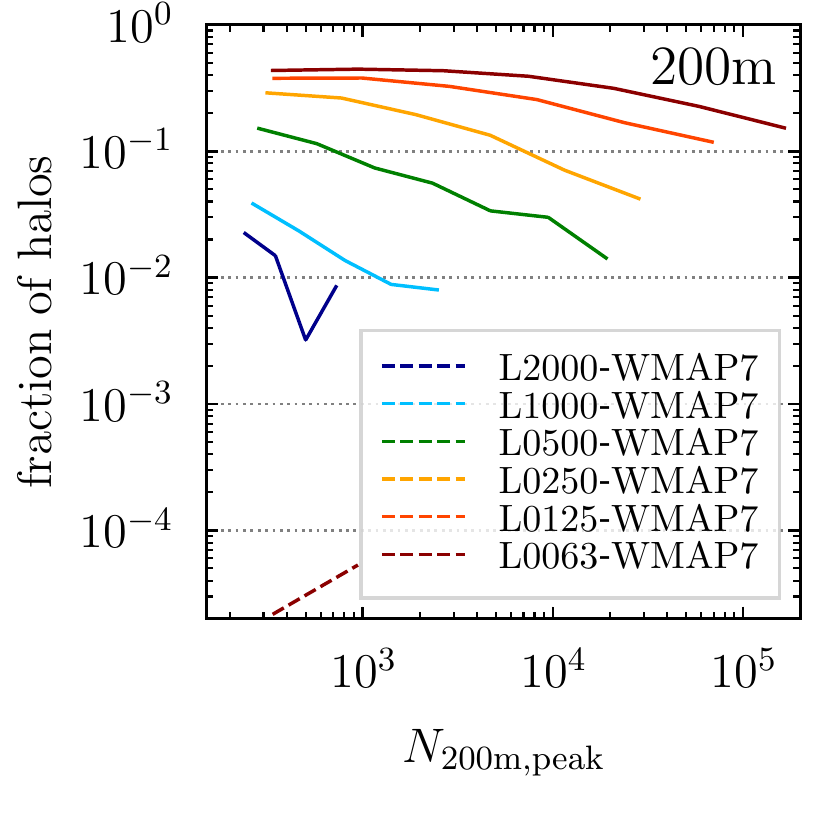}
\includegraphics[trim =  20mm 19mm 2mm 0mm, clip, scale=0.66]{\figdir/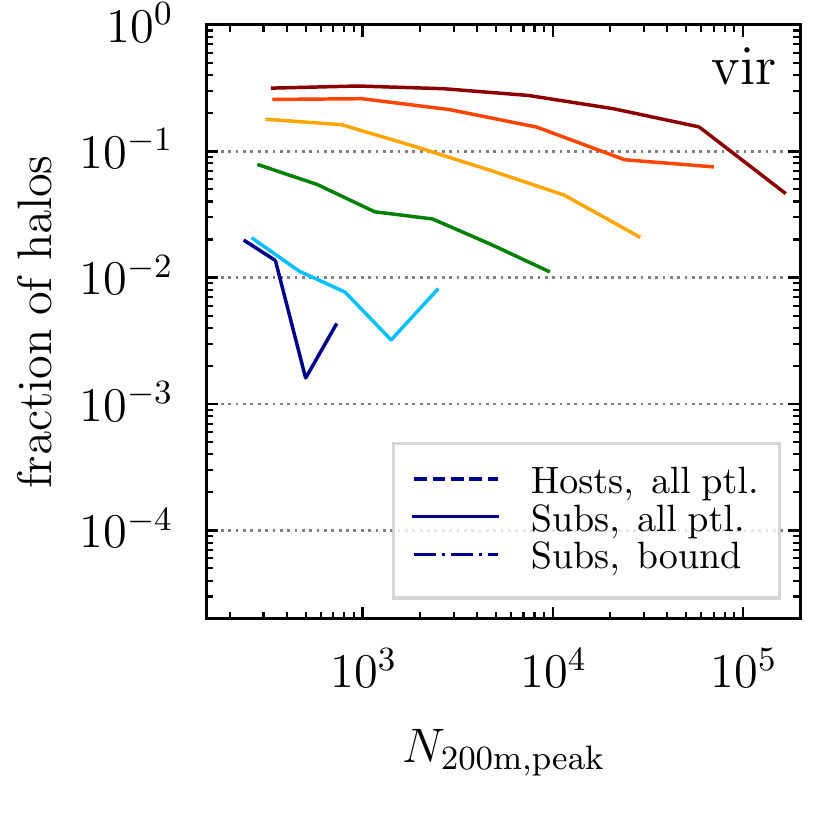}
\includegraphics[trim =  20mm 19mm 2mm 0mm, clip, scale=0.66]{\figdir/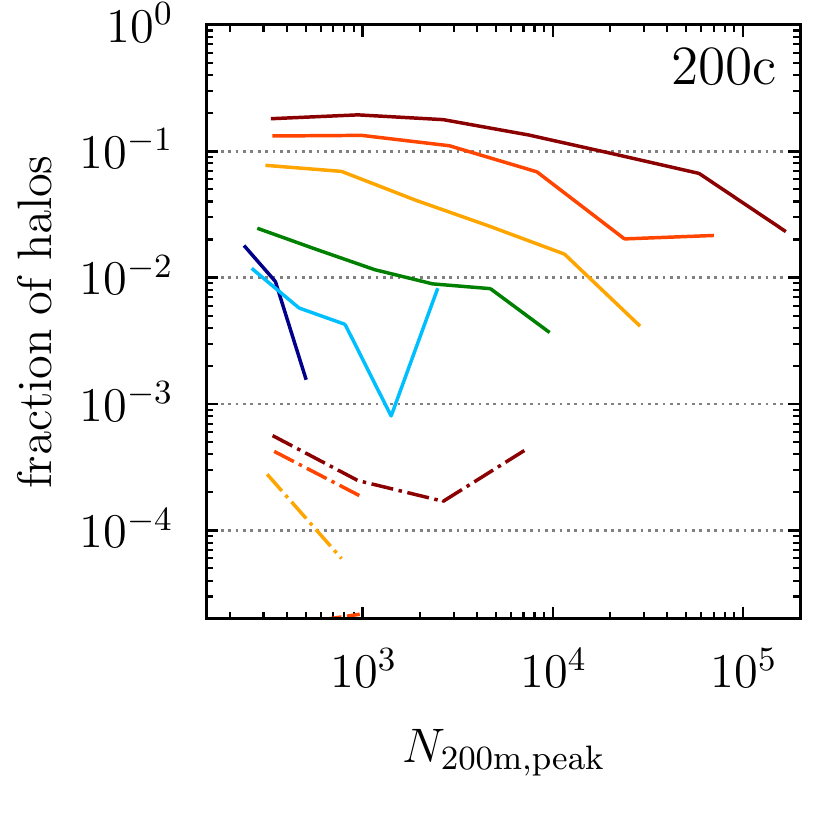}
\includegraphics[trim =  20mm 19mm 2mm 0mm, clip, scale=0.66]{\figdir/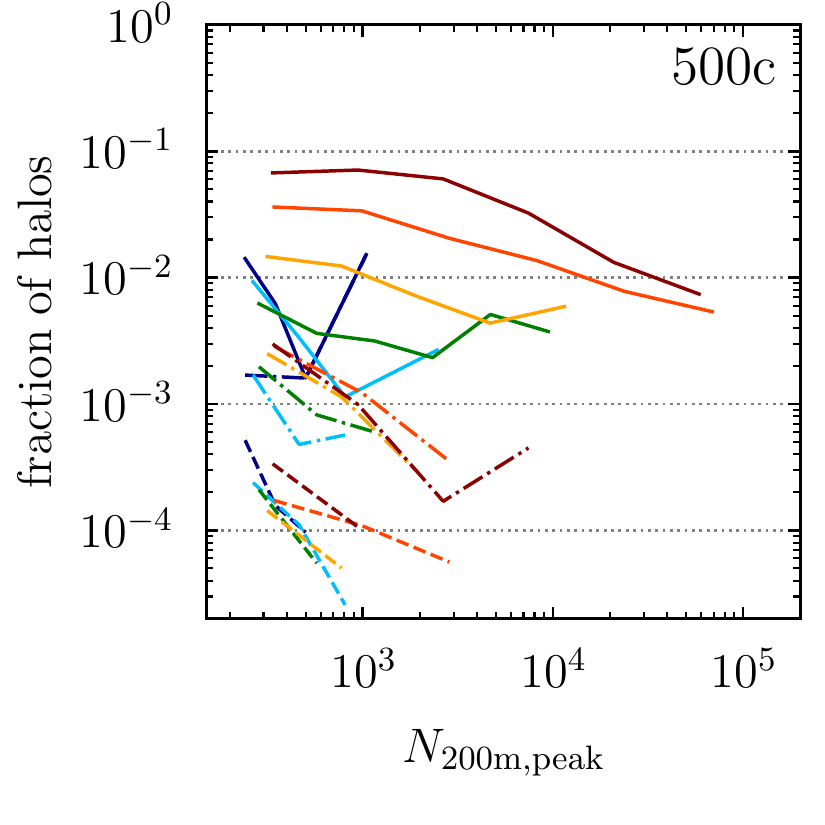}
\includegraphics[trim =  0mm 3mm 2mm 0mm, clip, scale=0.66]{\figdir/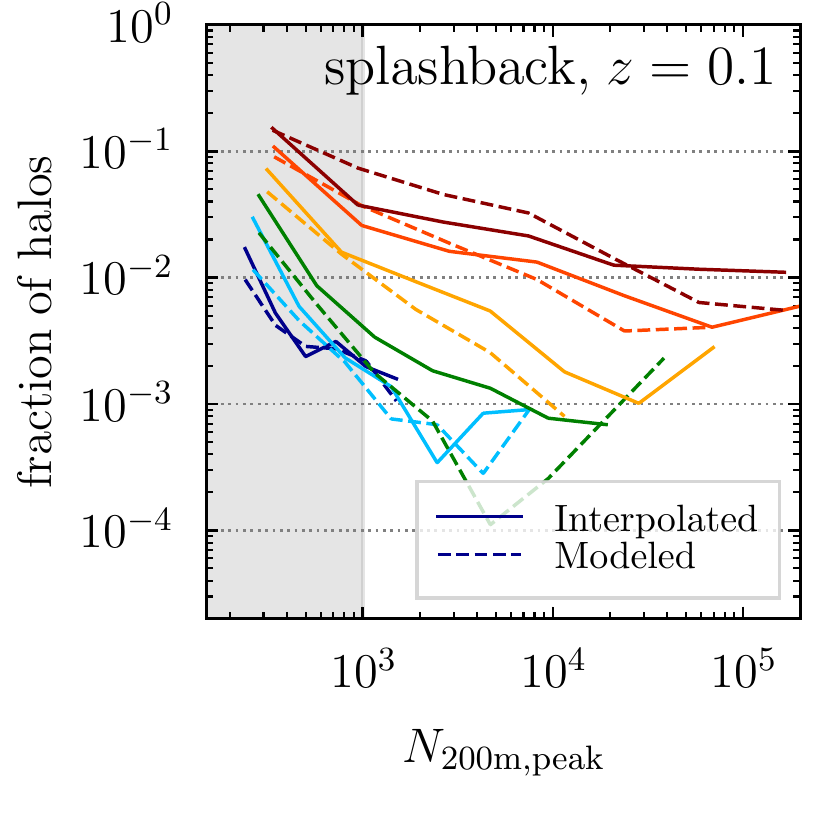}
\includegraphics[trim =  20mm 3mm 2mm 0mm, clip, scale=0.66]{\figdir/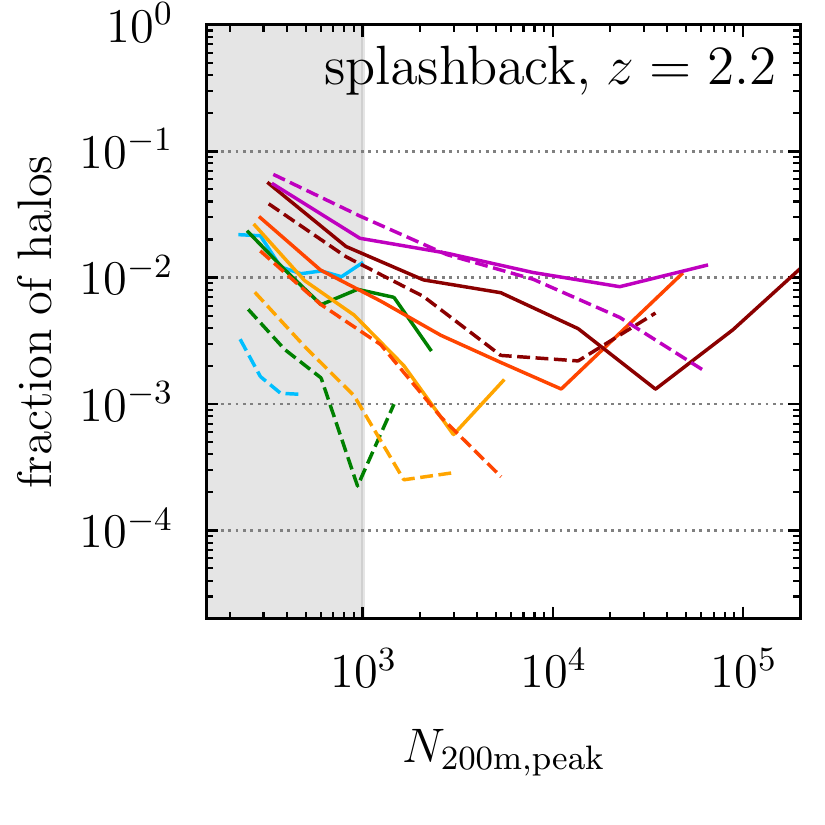}
\includegraphics[trim =  20mm 3mm 2mm 0mm, clip, scale=0.66]{\figdir/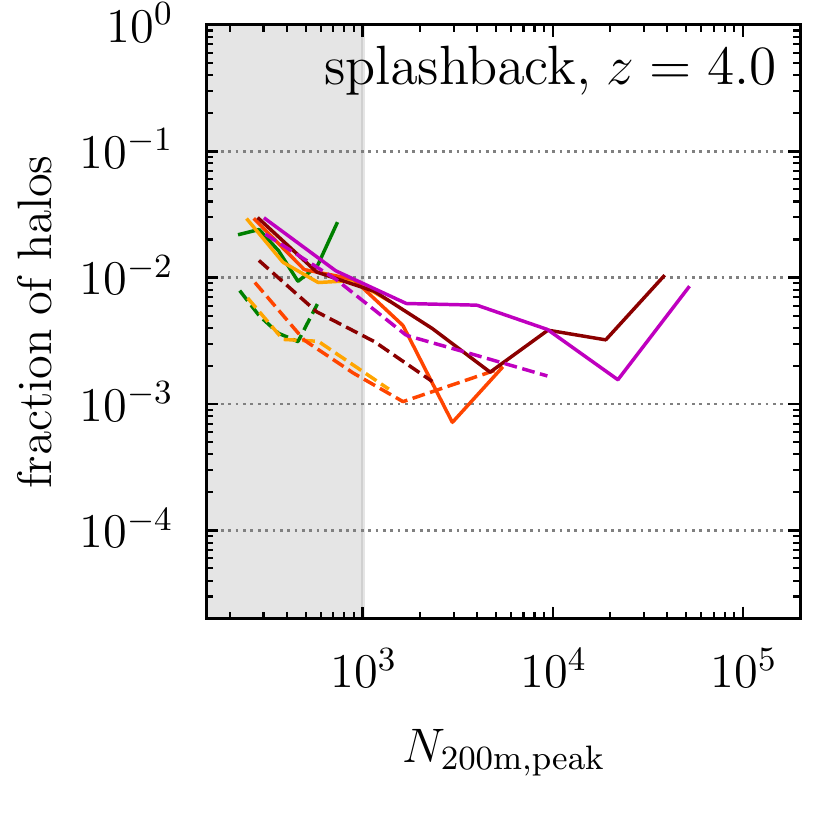}
\includegraphics[trim =  20mm 3mm 2mm 0mm, clip, scale=0.66]{\figdir/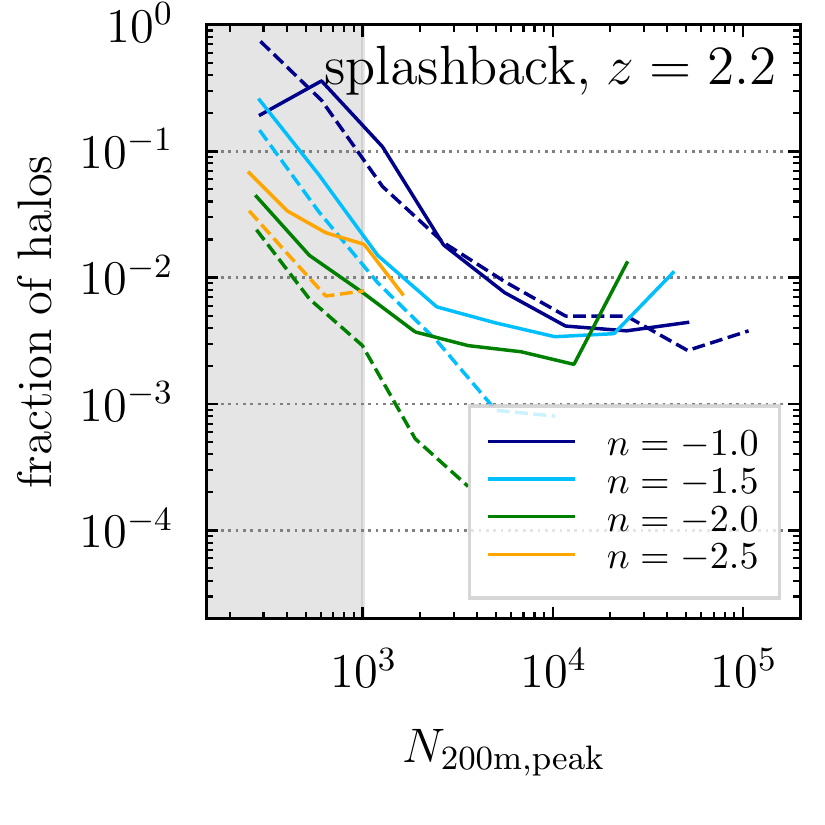}
\caption{Incompleteness of the catalogs as a function of the peak particle number. The colors refer to the different simulations in the \wmap sample, except in the bottom right panel where they indicate the self-similar simulations. {\it Top row:} Each panel shows an SO definition at $z = 0.1$. The lines show the fraction of halos for which an SO mass could not be computed, either because the threshold density was not reached at the center (common for high-density definitions such as $\rtoc$) or because the density of the halo did not fall below the threshold even at large radii (common for low-density definitions such as $\rtom$). Ending lines mean that that the corresponding fraction falls to zero. Dashed lines refer to all-particle masses for hosts, solid lines to all-particle masses for subhalos, and dotted--dashed lines to bound-only masses for subhalos. The all-particle definitions are frequently ill-defined for subhalos, leading to large incompleteness. The fractions depend on the box size because the small boxes contain more resolved subhalos whose strict SO masses would include the entire host halo. All-particle masses for hosts and bound-only masses for subhalos can almost always be computed, with the exception of a small fraction of halos in high-contrast definitions such as $\rfoc$. The lines for bound-only host halos look very similar to the all-particle versions and are thus omitted. {\it Bottom row:} The panels show the completeness of the splashback definitions for host halos; all splashback definitions share the same status. Missing values are recovered by \moria by either interpolating using adjacent snapshots (solid lines) or by predicting a value from a fitting function (dashed lines). Both fractions are low, especially at $N>1000$, which is the suggested threshold for using the splashback data (gray area). At high redshift, the number of interpolated and modeled radii drops. The bottom right panel demonstrates that the completeness is somewhat worse in the self-similar simulations with shallow slopes, especially $n=-1$. See Section~\ref{sec:results:completeness} for details.}
\label{fig:completeness}
\end{figure*}

\section{Results}
\label{sec:results}

The main product of this work are publicly available halo catalogs and merger trees. We summarize their most important properties in Section~\ref{sec:results:cats} and analyze their completeness in Section~\ref{sec:results:completeness}. We compare different SO definitions in Sections~\ref{sec:results:so} and \ref{sec:results:bnd}. In Sections~\ref{sec:results:rsp} to \ref{sec:results:steepest}, we consider the relationship between SO and splashback definitions and present an updated model for the splashback--SO connection.

\subsection{Halo Catalogs and Merger Trees}
\label{sec:results:cats}

Our catalogs and trees are based on \rockstar and \consistenttrees catalogs run with $R_{\rm 200m,bnd}$ as the main definition, with the ``strict SO'' setting switched off and the ``bound-props'' setting on, meaning that all properties, including SO masses, are calculated from bound particles only. While the distance units used in the \rockstar catalogs are comoving $\mpch$ and $\kpch$, all halo radii are given in physical $\kpch$ in the \moria catalogs. 

We have chosen a mass limit of $N_{\rm 200m,bnd,peak} \geq 200$ in \moria, meaning that a halo is included in the catalogs if it had at least $200$ particles within $R_{\rm 200m,bnd}$ at some time in the past. The trees contain the entire history of a halo if it exceeds the limit at any snapshot. There is no one ideal limit for all purposes because different halo properties converge at different particle numbers \citep{mansfield_20_resolution}. SO masses tend to be relatively well converged down to a few hundred particles, even for subhalos \citep[e.g.,][]{leroy_20}, although their mass is subject to significant statistical variance \citep{trenti_10, benson_17}. For splashback radii, we would ideally enforce a higher limit but instead caution the user that splashback masses and radii below $1000$ particles become less complete (\citetalias{diemer_17_sparta} and Section~\ref{sec:results:completeness}).

We have included a large number of mass definitions, accretion rates, and status fields, some of which we summarize in Table~\ref{table:fields}. We refer the reader to the code documentation for units and further details. Most importantly, we include SO masses and radii in the $\mfoc$, $\mtoc$, $\mvir$, and $\mtom$ definitions, each calculated with all particles (from \sparta), bound-only (from \rockstar), and from tracers only (for subhalos, from \sparta). We also include splashback radii and masses corresponding to the mean, median, 70\%, 75\%, 80\%, 85\%, and 90\% of the particle splashback distribution. For each of these radius definitions, we provide host--subhalo relations, that is, an array of host IDs for each halo. We also include the original host-sub assignment from \consistenttrees. We include a number of flags that indicate the status of the halo in \sparta, which definitions were successfully computed, and whether fields were modified or added by \moria. In addition to the fields listed in  Table~\ref{table:fields}, we have added almost all fields available in the \rockstar\ / \consistenttrees catalogs, including scale radius, velocity dispersion, angular momentum, spin parameters, axis ratios, position and velocity offsets, half-mass radii and times, and tidal forces \citep[see][although some fields have been added since those original publications]{behroozi_13_rockstar, behroozi_13_trees}.

We have set the compression level of the hdf5 files to the lowest level, 1, which gives a good compromise between file size and speed. The files are readable with any up-to-date hdf5 distribution. In total, our catalogs for the $14$ \erebos simulations contain about $7$ million halo histories and a total of $342$ million halo epochs. On average, 64\% of the tree arrays contain halos that are alive; the remaining 36\% are filled with zeros. Out of the active halo epochs, about 88\% are occupied by hosts, 8.6\% by subhalos, and 3.8\% by ghosts. These fractions vary from simulation to simulation, with smaller box sizes containing higher fractions of subhalos and ghosts, as well as an overall greater number of halos due to the higher variance of the density field on smaller scales. 

\subsection{Completeness}
\label{sec:results:completeness}

\def\figsize{0.57}
\begin{figure*}
\centering
\includegraphics[trim =  4mm 22mm 3mm 0mm, clip, scale=\figsize]{\figdir/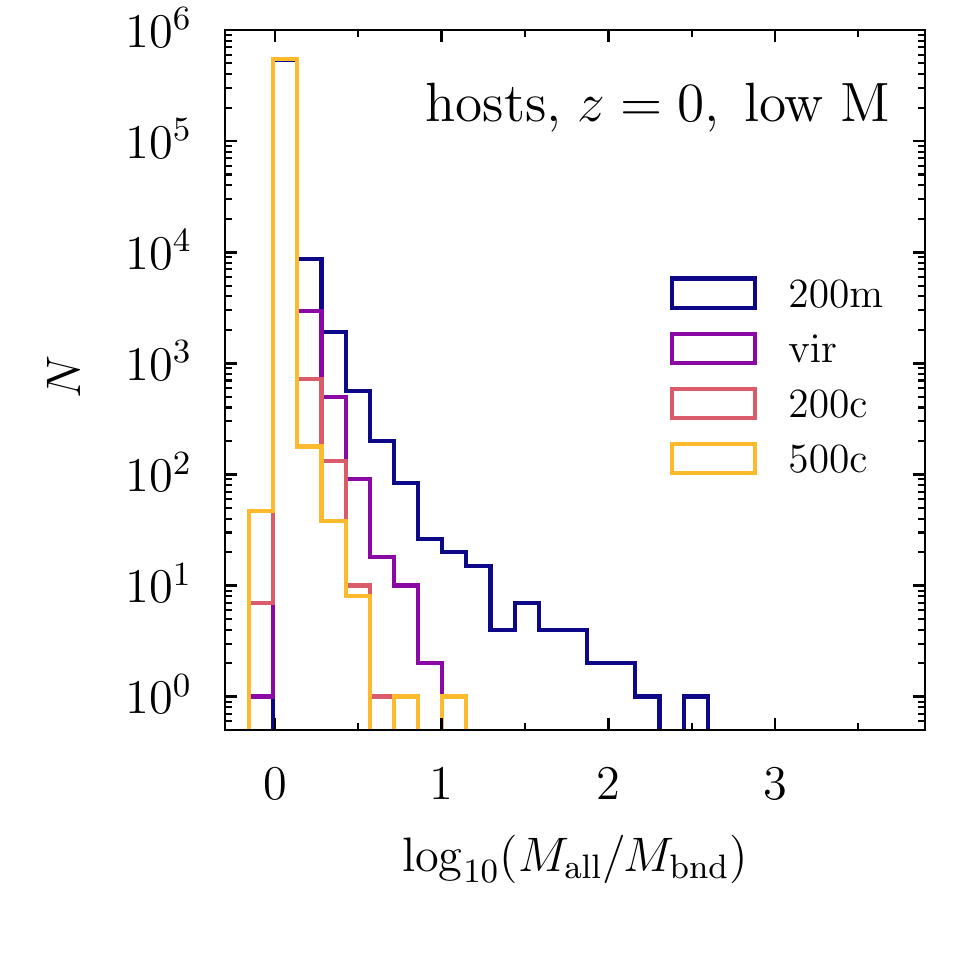}
\includegraphics[trim =  21mm 22mm 3mm 0mm, clip, scale=\figsize]{\figdir/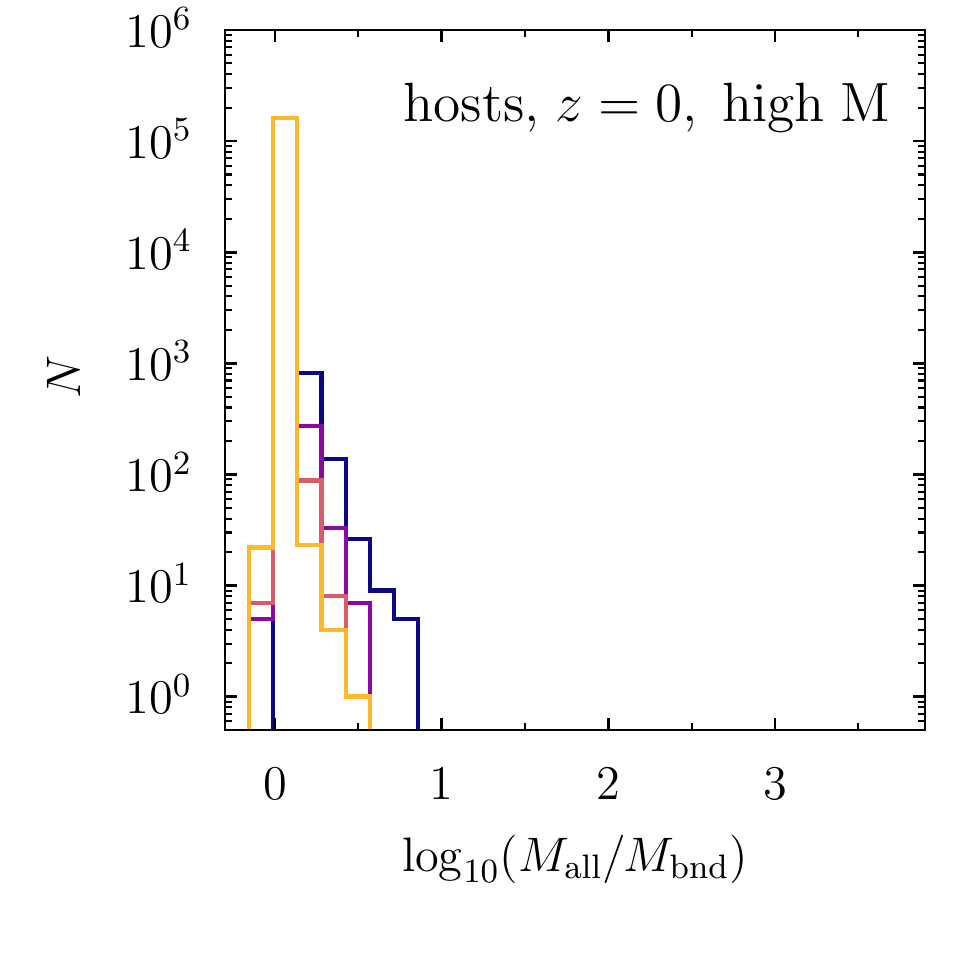}
\includegraphics[trim =  21mm 22mm 3mm 0mm, clip, scale=\figsize]{\figdir/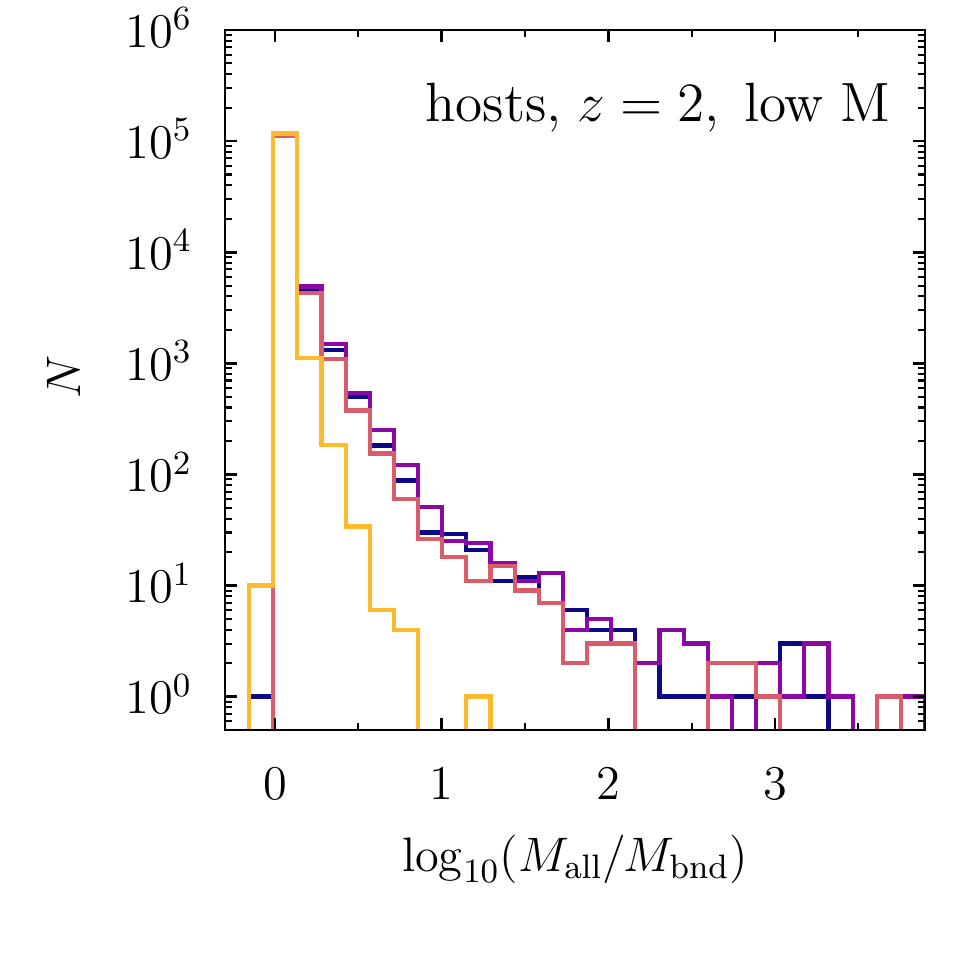}
\includegraphics[trim =  21mm 22mm 3mm 0mm, clip, scale=\figsize]{\figdir/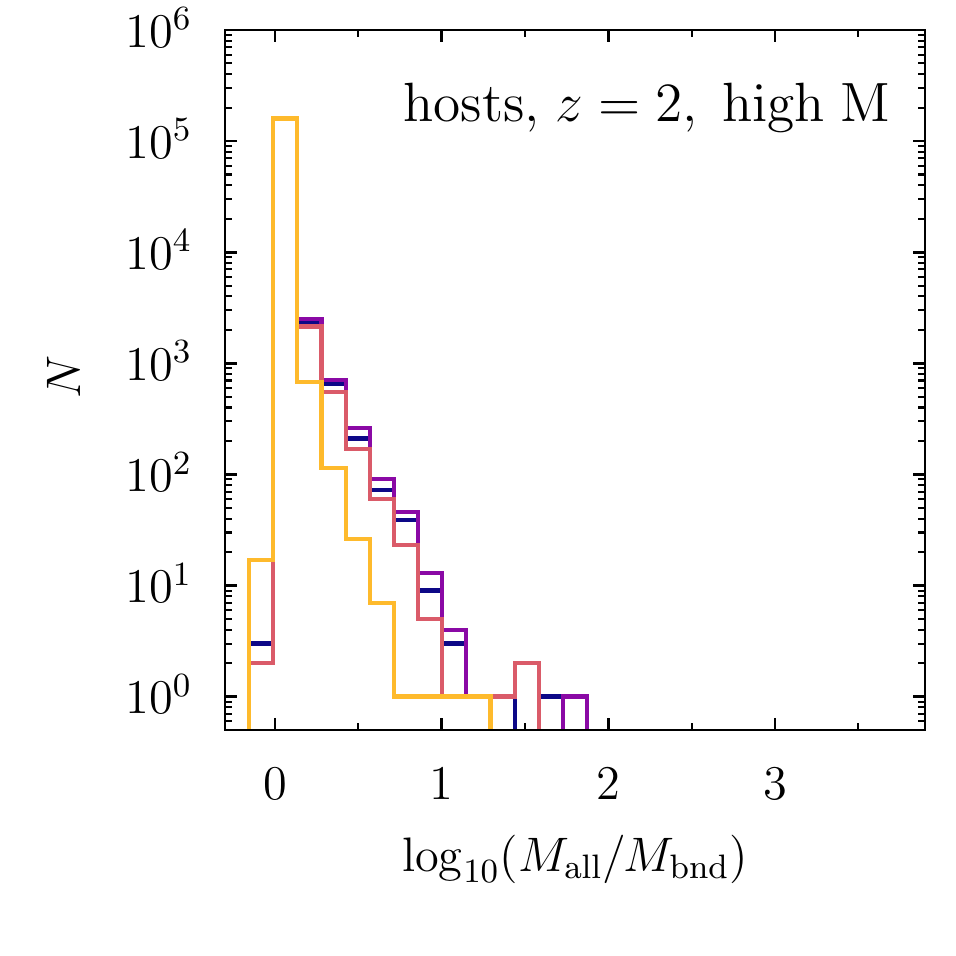}
\includegraphics[trim =  4mm 5mm 3mm 1mm, clip, scale=\figsize]{\figdir/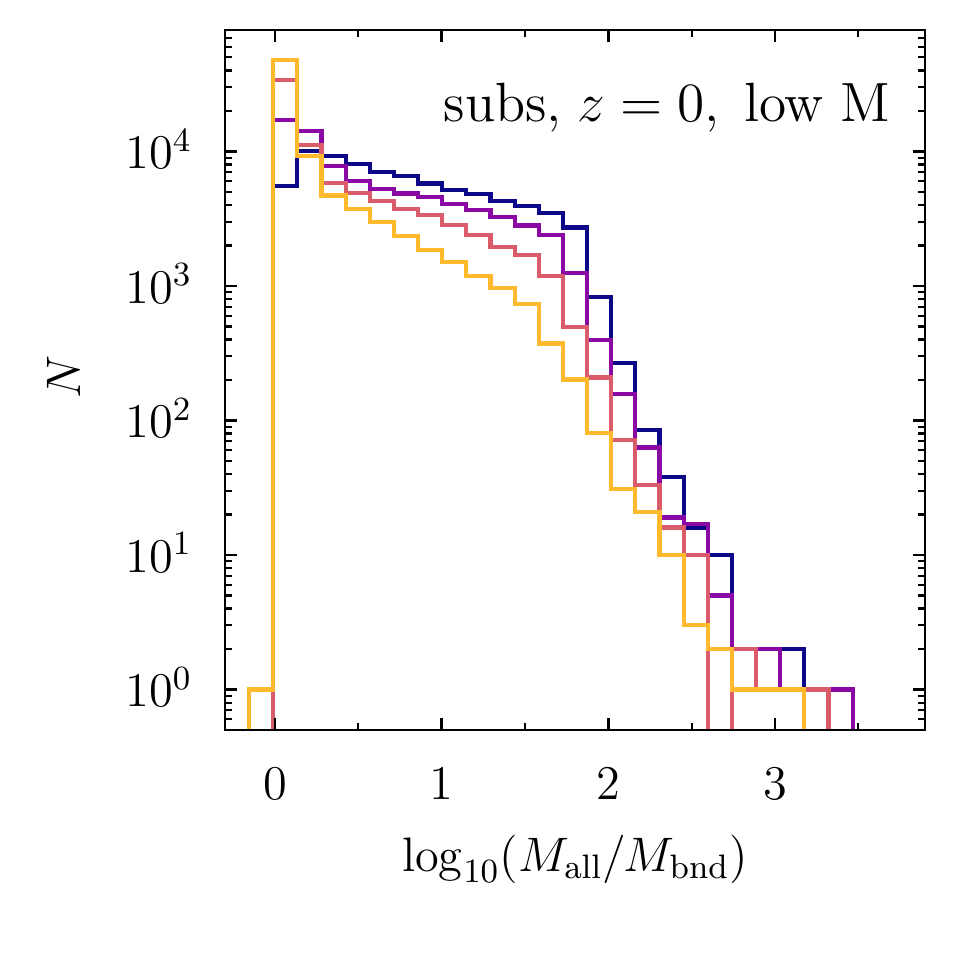}
\includegraphics[trim =  21mm 5mm 3mm 1mm, clip, scale=\figsize]{\figdir/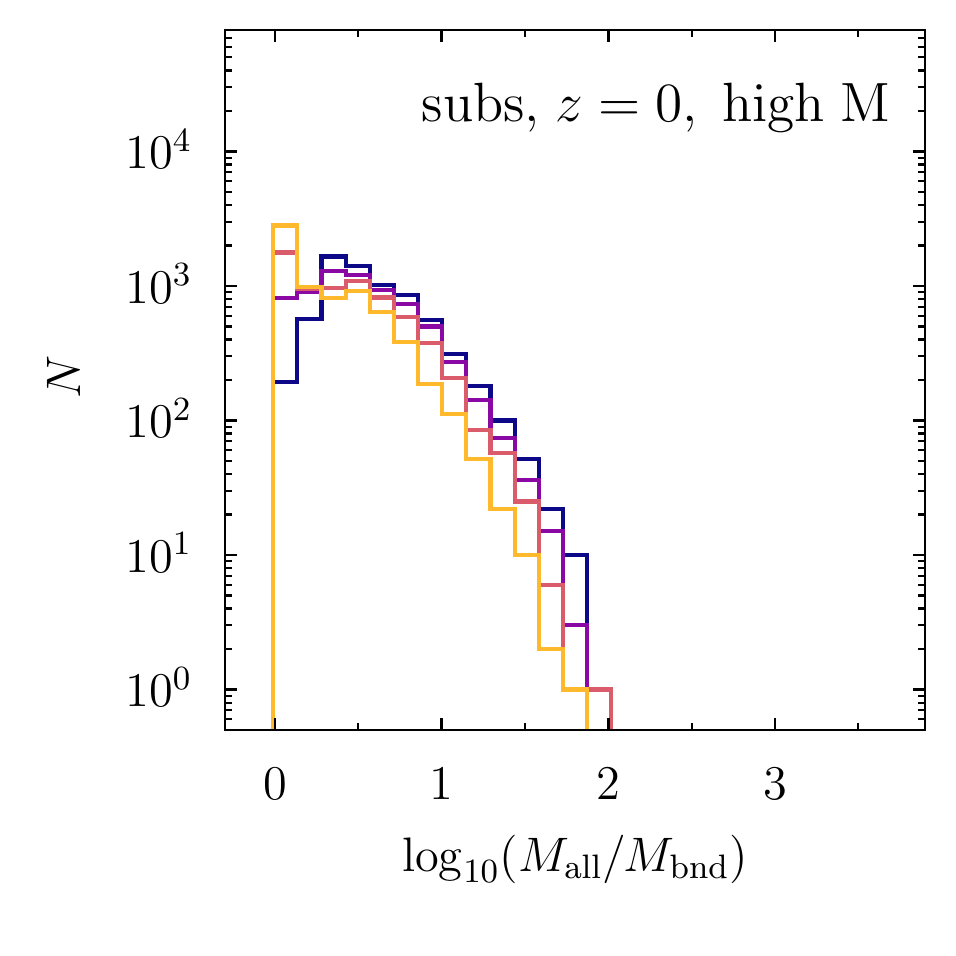}
\includegraphics[trim =  21mm 5mm 3mm 1mm, clip, scale=\figsize]{\figdir/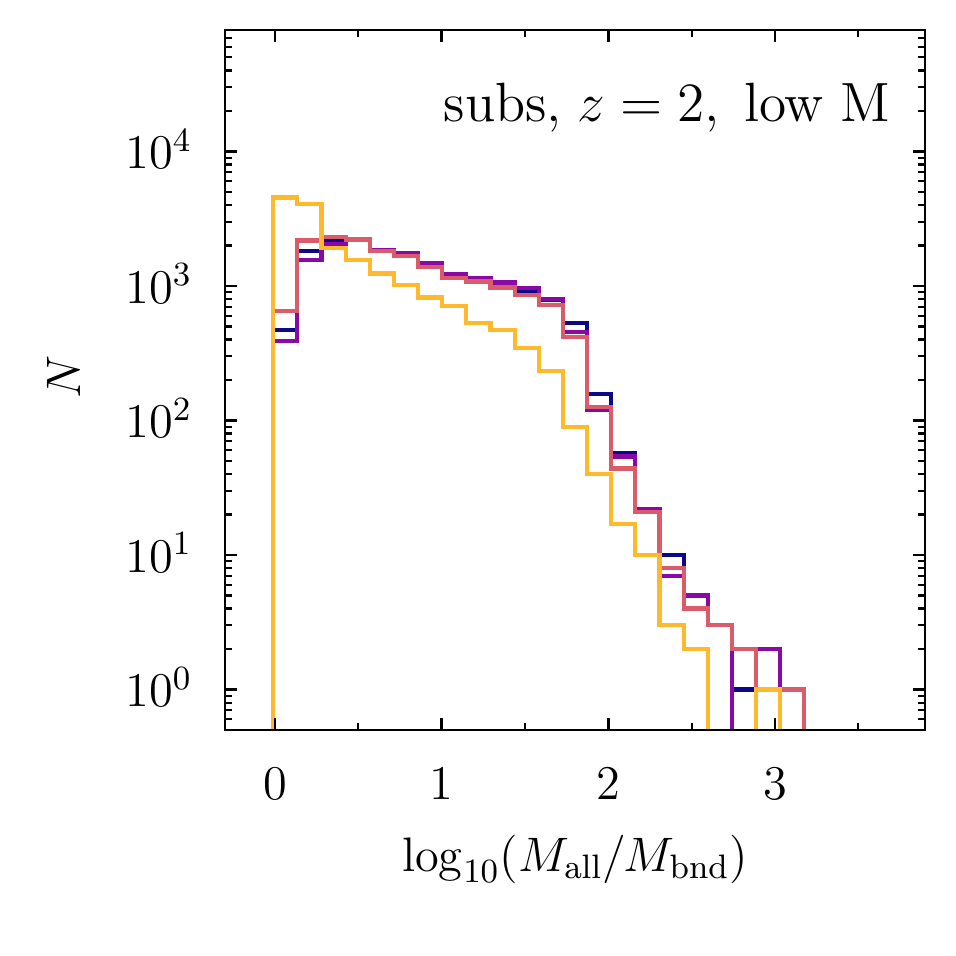}
\includegraphics[trim =  21mm 5mm 3mm 1mm, clip, scale=\figsize]{\figdir/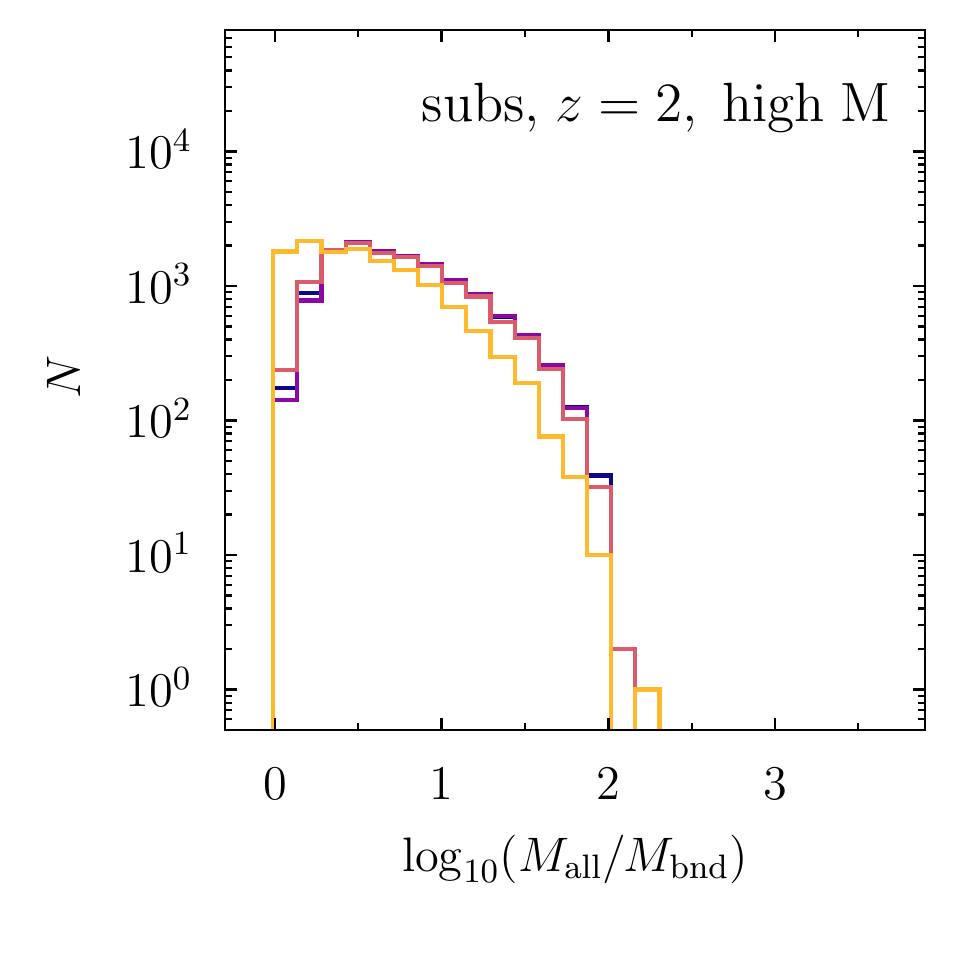}\caption{Ratio of all-particle and bound-only SO masses for host halos (top row) and subhalos (bottom row), low and high redshift, and low and high masses (defined as $0.5 < \nu < 1$ and $1.5 < \nu < 2$). In each panel, the colored lines show histograms of the radius ratio for four mass definitions. Since $M \propto R^3$ for SO definitions, the logarithmic differences in radius are three times smaller. For host halos, the definitions agree for the vast majority of halos, but there is a tail toward high ratios caused by halos that are close to another, larger halo. In such cases, the all-particle mass can encompass much or all of the other halo's mass. As expected, this effect is more common for lower-threshold definitions such as $\rtom$, at low mass, and at high redshift. The all-particle masses of a large fraction of subhalos are ill-defined; the cutoff of the distribution around a ratio of $1000$ is partly caused by \sparta's algorithm. The dependence on mass definition is even stronger for subhalos, with high-threshold definitions such as $\rfoc$ performing better. However, high thresholds also lead to more subhalos for which no mass can be found at all, which are excluded from this figure. See Section~\ref{sec:results:so} for details.}
\label{fig:ratio_so}
\end{figure*}

As discussed in Sections~\ref{sec:sparta:rsp} to \ref{sec:sparta:defs_so_bnd}, the calculation of SO and splashback radii can fail in certain situations. Figure~\ref{fig:completeness} shows the corresponding incompleteness of our catalogs for a variety of simulations and redshifts. The top four panels address SO definitions, where the completeness varies strongly between all-particle and bound-only definitions and between hosts and subhalos (different line styles in Figure~\ref{fig:completeness}). 

The calculation of bound-only radii can fail if the halo never reaches the threshold density at its center. By construction, there are no halos for which the halo finder fails to calculate the main definition, $R_{\rm 200m,bnd}$. As the threshold increases to 500c, an increasing number of subhalos fail to reach the threshold, but the number remains well below a percent (dotted--dashed lines). The same issue can occur even for host halos, although with a frequency of less than $10^{-3}$ (dashed lines). The results for bound-only definitions in host halos are similar to those for all particles, which is not surprising since virtually all particles near the halo center are bound. 

We also attempt to determine all-particle SO masses for subhalos even though they are often intrinsically ill-defined because the density never falls below the threshold (unless the entire host halo is included). The resulting incompleteness can reach up to 50\% for $\rtom$ but decreases significantly toward $\rfoc$, which can be computed for the vast majority of subhalos (solid lines in Figure~\ref{fig:completeness}). The incompleteness in all-particle subhalo masses is a function of mass rather than particle number, leading to the large differences between simulations. This is not the case for bound-only definitions, host halos, or splashback definitions though. Either way, the large incompleteness of all-particle subhalo masses is not important because they are given for comparison rather than as a physically meaningful mass definition.

In some very rare cases, a host halo is close enough to another halo that \sparta cannot find a radius where the density profile decreases below $200 \rhom$. We observe this case in the $10^{-4}$ of host halos in L0063-WMAP7 (dashed line in top left panel of Figure~\ref{fig:completeness}). We emphasize that this incompleteness captures only cases where $R_{\rm 200m,all}$ could not be computed; there are many more cases where it is computed but includes significant material from neighboring halos (Section~\ref{sec:results:so}).

We now turn to the splashback radius (bottom row of Figure~\ref{fig:completeness}). The issue of completeness was discussed at length in \citetalias{diemer_17_sparta}, with the conclusion that \sparta produces splashback data for about 95\% of halos with $\ntom \geq 1000$. Here, the situation has changed because we have included smaller halos down to $200$ particles, selected them by peak mass rather than current mass, and computed a value for $\rsp$ and $\msp$ regardless of whether they could be computed by \sparta. As discussed in Section~\ref{sec:moria:completeness}, we either interpolate the splashback radius from past and/or future values (solid lines) or estimate it using the fitting function (dashed lines). For most applications, the model estimates should not be used because the scatter in the true $\rsp / \rtom$ ratio is large \citepalias{diemer_17_rsp}. At $z \approx 0$, we find that model estimates account for just under 10\% of halos with $N_{\rm peak} = 1000$ in the smallest boxes and much less than a percent in the larger boxes. This trend is driven by the larger number of mergers and close flybys in the smaller boxes, which lead to interrupted splashback histories \citep{diemer_20_subs}. The second and third panels show redshifts $2$ and $4$ (including a pink line for L0031, which runs only up to $z = 2$). By $z = 4$, the fraction of estimated values falls below a percent in all simulations. Finally, we consider the self-similar simulations in the bottom right panel. At $N = 1000$, the modeled fraction varies between 10\% for $n = -1$ and around 1\% for the other simulations. 

In summary, Figure~\ref{fig:completeness} confirms that pushing far below the canonical value of $1000$ particles per halo leads to possibly problematic fractions of interpolated or estimated splashback values. We find very similar results for the {\it Planck} cosmology at the same box sizes. For SO definitions, we do not find a large evolution in the fractions with redshift, meaning that the general trends shown in Figure~\ref{fig:completeness} persist. Given that we will largely use bound-only and tracer masses for subhalos, we can think of the SO definitions as complete. The splashback definitions are nearly complete for hosts as long as we use reasonably well resolved halos. One exception are the last few snapshots of a simulation: due to the missing future particle splashback events, the completeness and reliability of \sparta's splashback results decrease for the last $\approx 0.2 t_{\rm dyn}$ \citepalias{diemer_17_sparta}. Thus, it is generally advisable to consider the splashback results at $z \approx 0.1$ rather than at $z = 0$. In the following sections, we combine data from all simulations of a given cosmology.

\subsection{SO Definitions: All-particle vs. Bound-only}
\label{sec:results:so}

We now consider the ratios between different mass and radius definitions quantitatively, starting with SO definitions. Both all-particle and bound-only definitions are commonly used, for instance, in studies of the halo mass function and merger trees, respectively. While it is widely known that unbinding and the density threshold can have a significant effect, these differences are rarely quantified. 

In Figure~\ref{fig:ratio_so}, we show histograms of the ratio between all-particle and bound-only SO definitions for halos from all simulations in our \wmap sample for which both were measured and for which the bound-only radius contains at least 200 particles. We use the $R_{\rm 200m,bnd}$ definition to split hosts and subhalos. Furthermore, we crudely split the sample by peak height into low-mass ($0.5 < \nu < 1$) and higher-mass halos ($1.5 < \nu < 2$). Similarly, we plot the ratio at $z = 0$ and $z = 2$ to highlight the general redshift trends. When referring to SO definitions, we use radii and masses interchangeably as they are uniquely coupled via $M \propto R^3$. The logarithmic mass ratios translate into three times smaller logarithmic radius ratios.

For host halos (top row of Figure~\ref{fig:ratio_so}), we notice a very small fraction where $R_{\rm all} < R_{\rm bnd}$ owing to numerical effects; in general, the bound-only mass is always smaller. We have checked that all-particle masses measured by \sparta and \rockstar agree perfectly (if \rockstar's strict SO option is activated), except in some rare cases where the FOF groups used by \rockstar are slightly incomplete at large radii. However, in TestSim100 we find that only about 1\% of halos exhibit more than 1\% difference in $\rtom$. Returning to the all-bound comparison, Figure~\ref{fig:ratio_so} shows that unbinding plays almost no role for the vast majority of host halos. The median ratios are unity in all definitions, and the mean ratios deviate from unity by no more than 3\%. The strength of the tail depends on the density threshold, with $\rfoc$ leading to fewer large ratios. For example, at $z = 0$, the all-bound ratio for $\mfoc$ is higher than 10\% for only 0.5\% of halos but for about 9\% when using $\mtom$. These differences are caused by halos that are close to another, larger halo whose material is partially included in the all-particle mass. High density thresholds are useful in avoiding this proximity issue, but they also lead to more failures of the calculation.

While it is generally safe to use all-particle and bound-only masses interchangeably for hosts, there are noticeable differences for the SO radii in the lower-threshold definitions. For example, there is a standard deviation of 14\% in $\rvir$ in the $z = 2$ low-mass example. The corresponding tail toward large radius ratios highlights the importance of unbinding in the context of subhalo relations: even few halos with spuriously large radii could, in principle, create a larger number of subhalos. We investigate this difference in \citet{diemer_20_subs} and find that subhalo relations based on all-particle definitions lead to up to 5\% more subhalos at the low-mass end.

\begin{figure}
\centering
\includegraphics[trim =  3mm 6mm 0mm 0mm, clip, scale=0.64]{\figdir/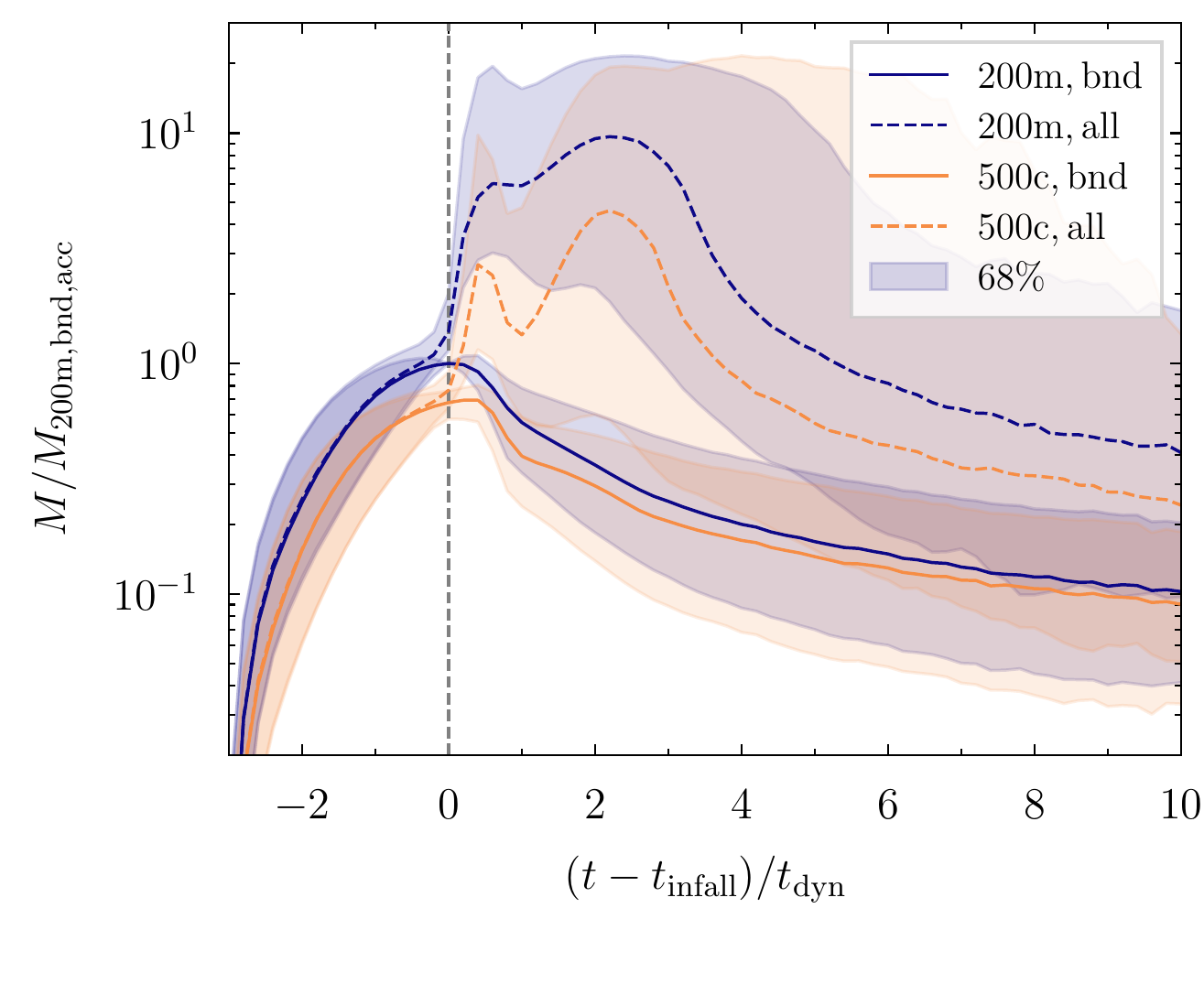}
\caption{Mass evolution of subhalos before and after infall (left and right of vertical gray line). The lines show the median evolution of $\mtom$ (blue) and $\mfoc$ (orange) in their bound-only (solid) and all-particle (dashed) versions, including only epochs where the respective mass could be measured. Time is measured in units of the dynamical time at infall. All masses are normalized to $M_{\rm 200m,bnd}$ at infall, so that the solid blue line is unity at $t_{\rm infall}$ by construction. Before infall, the all-particle and bound-only masses grow in unison and with modest scatter in their evolution. Just before infall, the all-particle masses sweep up additional mass from the future host, including roughly 40\% more mass on average for $\mtom$ and about 10\% for $\mfoc$. After infall, the bound-only mass begins to decrease whereas the all-particle mass rapidly grows. However, around one dynamical (or crossing) time, the mass decreases again as the subhalos reach their apocenter.}
\label{fig:subevo_200m}
\end{figure}

We now turn to subhalos (bottom panels of Figure~\ref{fig:ratio_so}). Here, the ratios are peaked near unity, but a significant fraction of halos have larger ratios. The distributions begin to fall off around a mass ratio of 100, but this cutoff is not necessarily physical because \sparta stops looking for a solution when the all-particle radius gets too large. On the one hand, the main conclusion from Figure~\ref{fig:ratio_so} is that all-particle masses are ill-defined for subhalos and should simply not be used. The median mass ratios vary between $1.2$ and $5$, increasing with redshift and decreasing with mass and overdensity threshold. Considering radii, on the other hand, a surprisingly large fraction of subhalos do have well-defined all-particle radii: the median radius ratio varies between $1.1$ and $1.8$. The large standard deviations of up to unity reflect that the result of the calculation will strongly depend on how close to the host center the subhalo is located. In summary, the bound-to-all ratio appears to increase in cases where halos are close to neighbors (for hosts) or to their host center (for subhalos). 

We now explicitly confirm this picture by considering the mass evolution of subhalos in the \wmap cosmology in Figure~\ref{fig:subevo_200m}. We use our new merger tree format to pick out all epochs where a halo becomes a subhalo and to extract its mass evolution before that time (while it was a host) and afterward (while it is a subhalo, as opposed to a ghost or backsplash halo). We define the time of infall as the last snapshot before each halo became a subhalo according to the $\rtom$ definition (vertical dashed line in Figure~\ref{fig:subevo_200m}). We require halos to contain at least $200$ particles at infall, but this limit does not qualitatively change the results. Similarly, the mass evolution does not change much if we limit the infall redshift or peak height; the figure includes infalls at any redshift. We omit epochs where a mass definition cannot be measured, meaning that the median all-particle masses would be even larger if we could include the most extreme cases. 

Figure~\ref{fig:subevo_200m} shows that, before infall, the all-particle and bound-only masses evolve more or less in unison. After infall, the median all-particle mass rapidly increases by a factor of up to 10. The large scatter highlights that the mass evolution strongly depends on the orbital parameters and on the internal structure of host and subhalo. Despite the scatter, we observe a dip in the median all-particle mass after one dynamical time, when many subhalos have reached their first apocenter in the less dense outskirts of their hosts. The corresponding dips from subsequent orbits are smoothed out by phase mixing. Focusing on the evolution around infall, the all-particle mass begins to deviate from the bound-only mass about half a dynamical time before infall. As expected, the increase over the bound-only mass is larger for low-threshold definitions such as $\mtom$ because their larger radius includes more host material. While $\mfoc$ increases by only about 10\% at infall, $\mtom$ has increased about 40\% on average. This mass increase is largely responsible for the tails in the host halo distribution in Figure~\ref{fig:ratio_so}.

\subsection{SO Definitions: Unbinding Algorithm}
\label{sec:results:bnd}

\def\figsize{0.57}
\begin{figure}
\centering
\includegraphics[trim =  4mm 4mm 3mm 1mm, clip, scale=\figsize]{\figdir/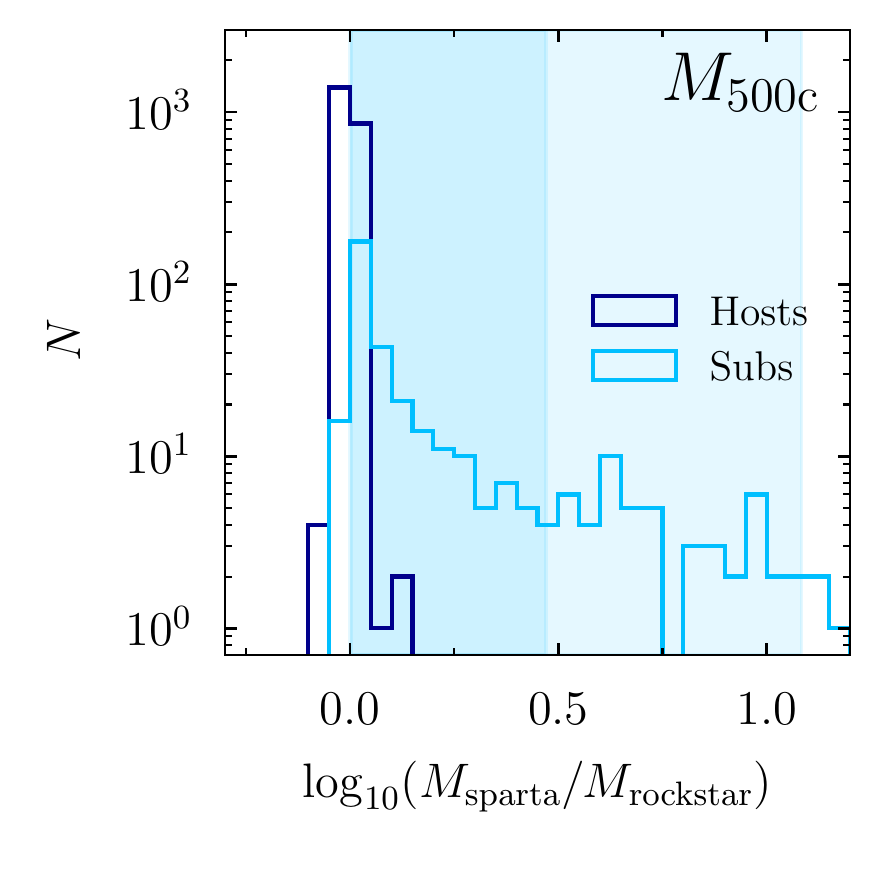}
\includegraphics[trim =  21mm 4mm 3mm 1mm, clip, scale=\figsize]{\figdir/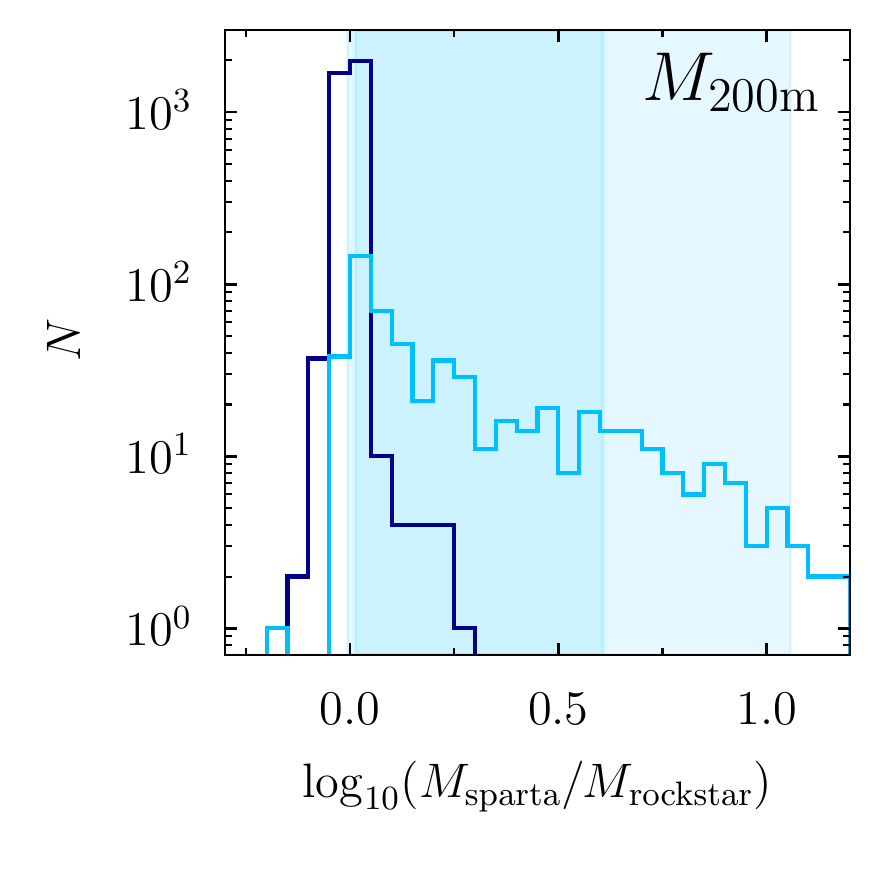}
\caption{Ratio of bound-only masses computed from an initial FOF group (by \rockstar) and from all particles within $\rtom$ (by \sparta), at $z = 0$ in TestSim100. The bound masses agree almost exactly for host halos (dark blue). For subhalos (light blue), there is a large fraction where the masses agree but also an extensive tail toward large values (the shaded areas show the 68\% and 95\% intervals for subhalos). The tail is caused by halos where $\rtom$ includes sufficient host particles to spuriously bind a lot of material. This case is more likely for $\rtom$ than for $\rfoc$ because the latter relies on more strongly bound particles.}
\label{fig:ratio_bnd}
\end{figure}

As mentioned in Section~\ref{sec:sparta:defs_so_bnd}, we found that bound-only masses depend strongly on the chosen algorithm, most notably on the initial set of particles. In Figure~\ref{fig:ratio_bnd}, we compare bound-only masses computed by \sparta to those from \rockstar. Here, we consider all particles within $\rtom$, defined as the all-particle radius for hosts and the tracer radius for subhalos. We do not iteratively unbind to match \rockstar's procedure and because iterating does not significantly alter our conclusions. The results in Figure~\ref{fig:ratio_bnd} refer to TestSim100 because we have not included \sparta's bound-only masses in our fiducial catalogs.

As expected, the results agree very well for host halos, with median ratios of unity in all definitions and standard deviations between 2\% in $\mfoc$ and 8\% in $\mtom$. This agreement reconfirms that we can safely neglect boundedness for host halos except in some extreme merger scenarios. For subhalos, the estimates can differ significantly, with a long tail toward large ratios. The shaded areas in Figure~\ref{fig:ratio_bnd} show the 68\% and 95\% intervals for subhalos. While the logarithmic scale of the figure highlights the tails, the median ratios vary between only $1.1$ for $\mfoc$ and $1.35$ for $\mtom$. For $\mfoc$, the estimates agree to 10\% or better for about half the halos; for $\mtom$ that fraction goes down to 30\%. However, the tails and standard deviations are extremely large in all definitions. The reason for these disagreements are halos where $\rtom$ includes numerous host particles, rendering the entire distribution more bound than it would be if we considered only ``true'' subhalo particles. The 6D-FOF groups used for subhalos in \rockstar do not constitute a ``correct'' answer either, but they contain less host material because the velocity of the particles is considered. We thus provide the \rockstar bound-only masses in our catalogs. We continue the search for physically motivated definitions of subhalo masses in Diemer \& Behroozi (2020, in preparation), where we further explore the concept of tracer masses.

\subsection{Splashback Definitions}
\label{sec:results:rsp}

In \citetalias{diemer_17_rsp}, we investigated the splashback radii and masses computed by \sparta as a function of $\rtom$ and $\mtom$, accretion rate, redshift, and cosmology. However, we did not show the distribution of ratios and did not consider subhalos. Figure~\ref{fig:ratio_sp} shows $\rsp / \rtom$ and $\msp / \mtom$ but using the bound-only SO radii and masses from \rockstar to make the ratios more comparable to Figure~\ref{fig:ratio_so} and to avoid the issue of spuriously large all-particle radii (Section~\ref{sec:results:so}). The distributions are clearly peaked around the median (shaded areas), but they do exhibit broad tails. Some of the most extreme ratios are caused by flyby encounters that can result in negative accretion rates and large splashback radii. The cutoff above a radius ratio of three is partly caused by \sparta's algorithm, which does not trace particles to arbitrary distances \citepalias{diemer_17_sparta}. The distribution of ratios does not strongly depend on redshift, although the average splashback radius and mass slightly decrease with $z$ (in agreement with \citetalias{diemer_17_rsp}). We choose $R_{\rm sp,75\%}$ as a representative definition in Figure~\ref{fig:ratio_sp}; the distributions for the other percentiles look similar (although slightly shifted in radius and mass). One important difference between \citetalias{diemer_17_rsp} and our new catalogs is that \moria reconstructs splashback quantities by interpolation, extrapolation, and estimates from our fitting function. Those results are shown as dashed lines in Figure~\ref{fig:ratio_sp}. The good agreement with the measured splashback quantities is somewhat by construction since we interpolate in $\rsp / \rtom$ and $\msp / \mtom$ space, but it demonstrates that the interpolation does not introduce any unexpected features. 

\def\figsize{0.57}
\begin{figure}
\centering
\includegraphics[trim =  4mm 21mm 3mm 1mm, clip, scale=\figsize]{\figdir/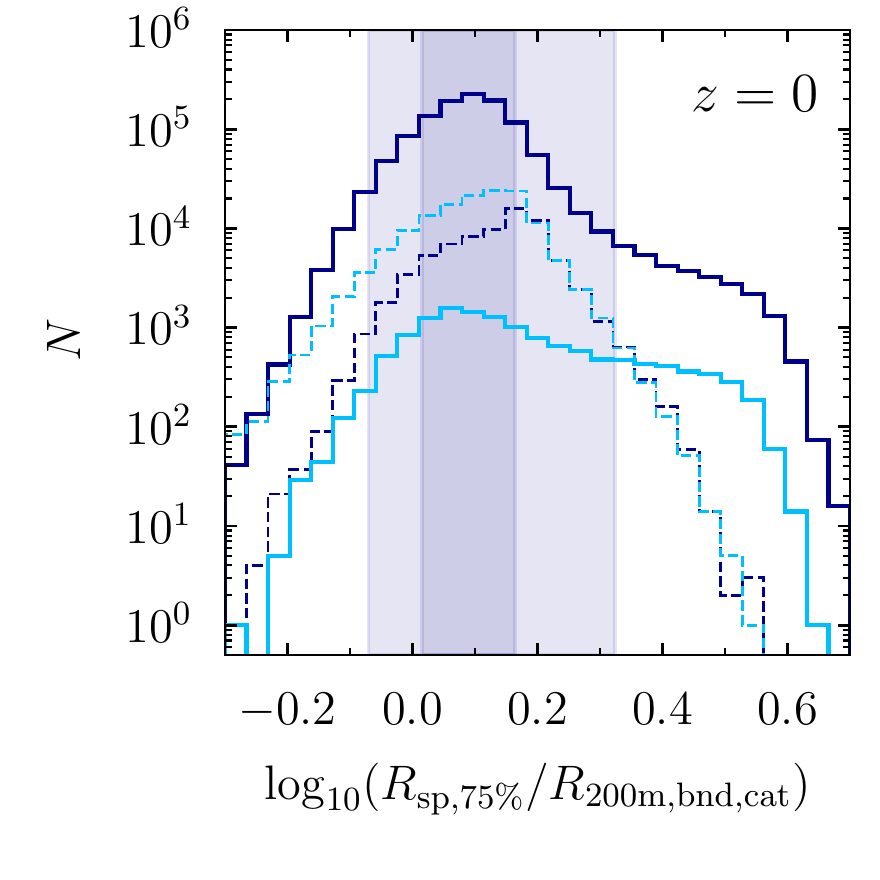}
\includegraphics[trim =  21mm 21mm 3mm 1mm, clip, scale=\figsize]{\figdir/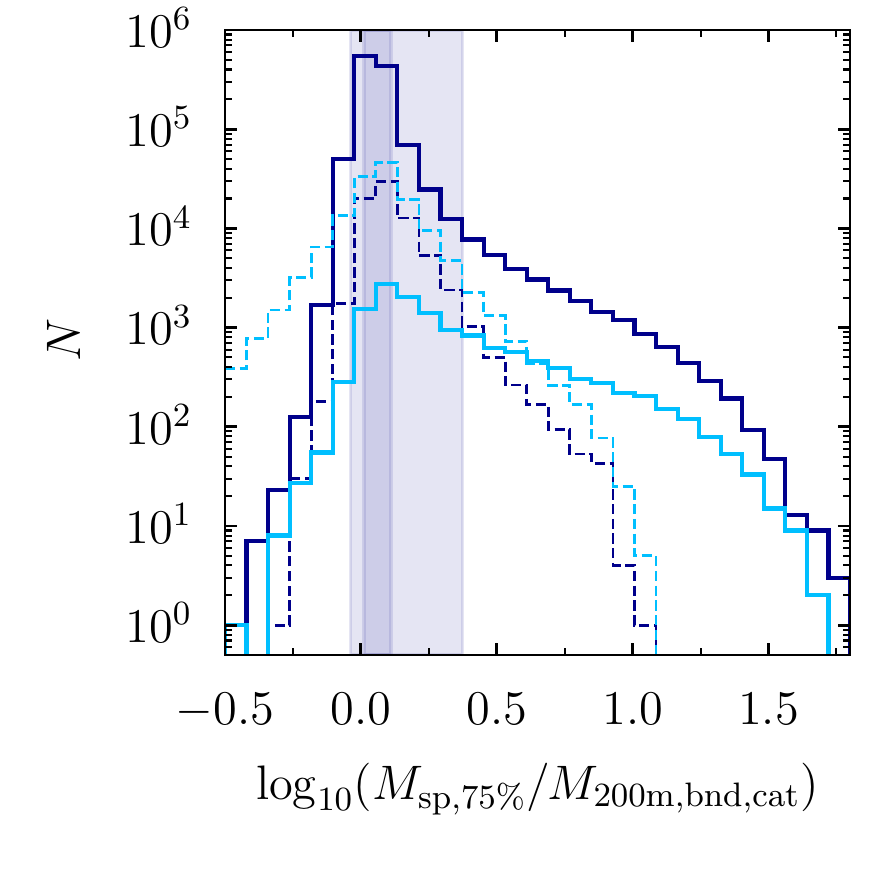}
\includegraphics[trim =  4mm 4mm 3mm 1mm, clip, scale=\figsize]{\figdir/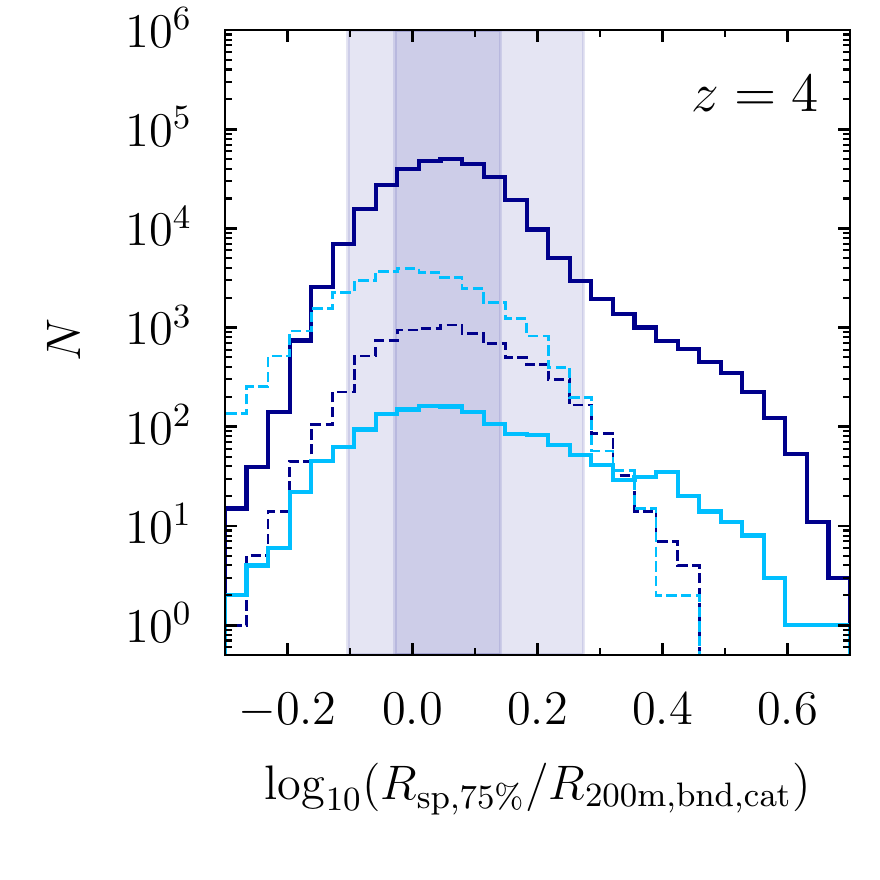}
\includegraphics[trim =  21mm 4mm 3mm 1mm, clip, scale=\figsize]{\figdir/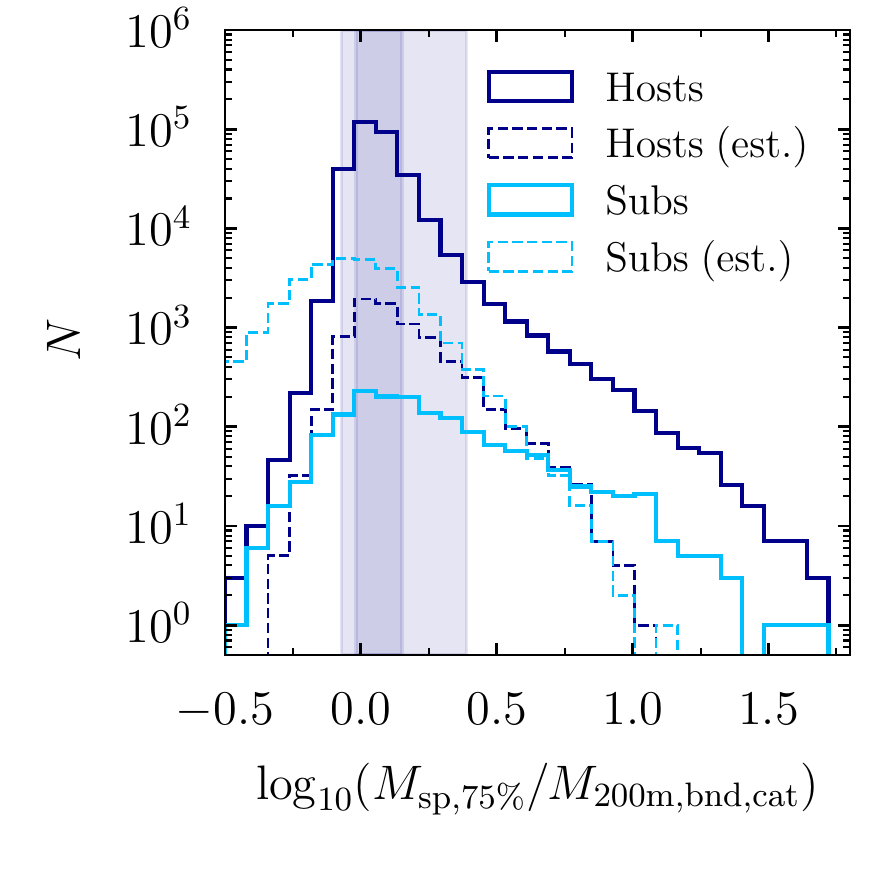}
\caption{Ratio of the 75\%-splashback radius and mass to $\rtom$ (left) and $\mtom$ (right), for $z \approx 0$ (top) and $z = 4$ (bottom). The histograms include all halos in the \wmap sample with $\ntom > 1000$. The splashback radius depends on the accretion rate and mass, which broadens the distribution. The solid lines show splashback quantities computed by \sparta, the dashed lines those interpolated or estimated by \moria. For subhalos, only a small fraction has \sparta estimates. While the distribution has large tails, the 68\% and 95\% intervals for hosts (shaded areas) are well confined. The tails for the estimated values are limited by construction as described in Section~\ref{sec:moria:completeness}. The tails toward high ratios are partially due to outliers with genuinely large splashback radii and partly due to differences between including all particles and the bound-only SO mass to which we are comparing. The broadness of the distributions also highlights that the splashback radius is a unique definition that cannot easily be guessed based on SO definitions. See Section~\ref{sec:results:rsp} for details.}
\label{fig:ratio_sp}
\end{figure}

\def\figsize{0.64}
\begin{figure}
\centering
\includegraphics[trim =  7mm 23mm 1mm 1mm, clip, scale=\figsize]{\figdir/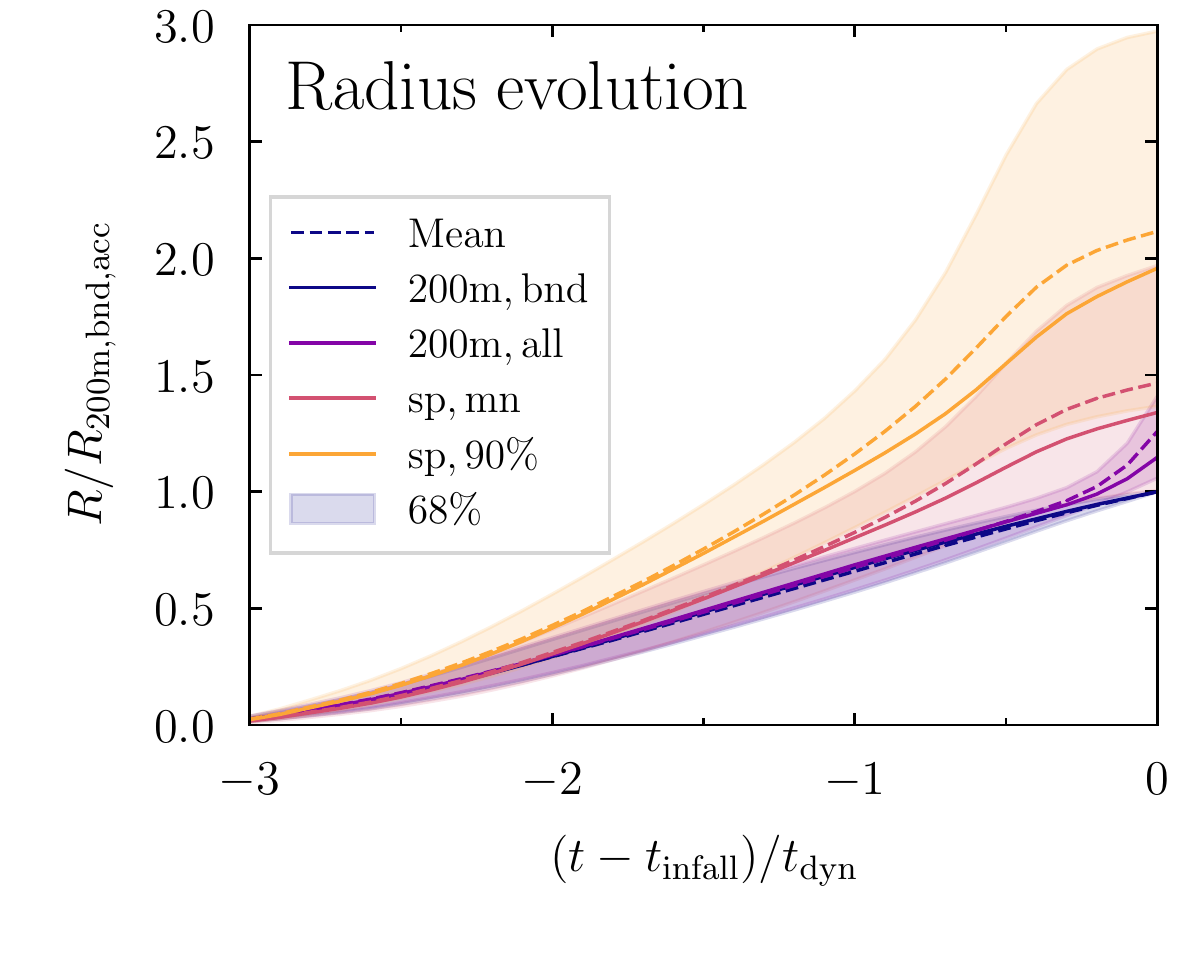}
\includegraphics[trim =  7mm 6mm 1mm 1mm, clip, scale=\figsize]{\figdir/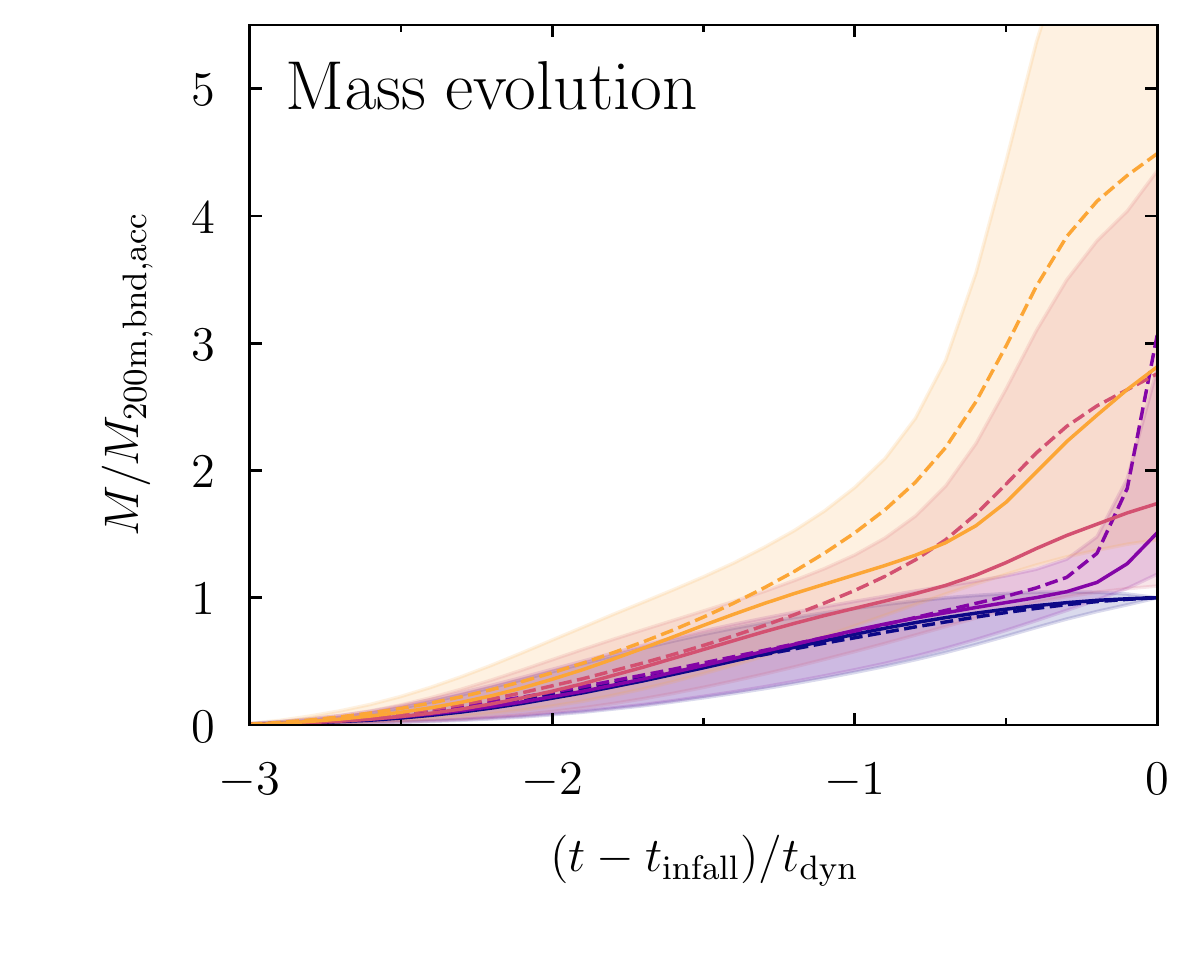}
\caption{Evolution of SO and splashback radii (top) and masses (bottom) prior to infall into a larger halo. The figure corresponds to the left-hand side of Figure~\ref{fig:subevo_200m} for $M_{\rm 200m,bnd}$ and $M_{\rm 200m,all}$ (blue and purple lines), but we are not showing the evolution after infall because splashback radii are not physically meaningful for subhalos. The solid lines show the median evolution, the dashed lines the mean. For SO definitions such as $M_{\rm 200m,all}$, the inclusion of material belonging to the future host leads to a growth in both radius and mass. For the splashback definitions (red and orange lines), the situation is quite different: while the mean mass does grow significantly, the splashback radius follows a smooth trajectory that does not seem to be impacted much by the future merger.}
\label{fig:evo_sp}
\end{figure}

For subhalos, \sparta does not compute particle splashbacks because orbits reflect the combined potential of host and subhalo and do not typically exhibit well-defined apocenters. However, if a halo becomes a subhalo only briefly, there may be enough past and future particle splashbacks to compute $\rsp$ and $\msp$. The corresponding distributions are similar to those of host halos (solid light-blue lines in Figure~\ref{fig:ratio_sp}). The estimated subhalo splashback radii (dashed light-blue lines) refer to the time of infall and are thus not directly comparable to the current-time radius and mass according to the halo finder; they are given mostly for completeness.

Based on the mass evolution of SO halos shown in Figure~\ref{fig:subevo_200m}, we might worry about the evolution of splashback radii and masses close to merger events. For example, if the radii were artificially inflated as a result of particles being stripped from the halo, that could lead to other halos in the vicinity falsely being identified as subhalos. However, we find no evidence for such an issue. Figure~\ref{fig:evo_sp} shows the average radius and mass evolution before infall for all (future) subhalos with more than $1000$ particles in the \wmap cosmology. As in Figure~\ref{fig:subevo_200m}, we observe a smooth mass evolution in bound-only $\mtom$ (blue lines) and a sudden growth of  all-particle $\mtom$ shortly before infall (purple lines); $R_{\rm 200m,all}$ follows this growth by construction. The splashback radii (orange and red lines), on the other hand, do not exhibit any sudden growth. The splashback mass is independent of the radius and does include material in the vicinity of the host, leading to a gradual increase in $\msp$ during the last dynamical time before the merger. The mean (dashed lines) grows much more sharply than the median, indicating a strongly asymmetric distribution with a tail toward large mass increases. We conclude that our splashback radii and masses behave as expected in the vicinity of mergers, at least on average.

\subsection{An Updated Model for the Splashback--SO Relation}
\label{sec:results:model}

As discussed in Section~\ref{sec:sparta:rsp}, we have made improvements to the \sparta code that change the splashback radii and masses by a few percent. Thus, we recalibrate the \citetalias{diemer_17_rsp} model including the following changes. We calculate mass accretion rates using \moria as described in Section~\ref{sec:moria:accrate}. Including halos that were subhalos or did not exist one dynamical time ago helps to extend the high-$z$ samples, especially at high accretion rates. We have confirmed that excluding backsplash halos makes a small difference and thus leave them in the sample. Since we are constraining $\rsp/\rtom$ and $\msp / \mtom$, we should use only halos for which $\rtom$ and $\mtom$ are reliably measured. In particular, we wish to avoid halos that include a significant contribution from a neighboring halo, which is relatively common at low masses (Sections~\ref{sec:results:completeness} and \ref{sec:results:so}). We could sidestep this issue by using bound-only masses, but there are good reasons to calibrate the relationship against all-particle quantities. First, bound-only masses cannot be measured observationally and depend on the unbinding algorithm; second, particle orbits (and thus $\rsp$) react to the total mass inside their radius rather than the bound mass; and third, all-particle masses are more clearly related to the halo density profile and the steepening feature. Thus, we use all-particle radii and masses but exclude halos for which $M_{\rm 200m,all}$ is more than 20\% greater than $M_{\rm 200m,bnd}$. While this cut may seem strict, including halos with mass increases of up to 50\% visibly increases the scatter in the lowest peak height bins, indicating that the splashback--SO relation is erratic in such halos. The higher peak height bins are barely affected by the cut. As in \citetalias{diemer_17_rsp}, we use only halos with $\ntom > 1000$ and with successful $\rsp$ and $\msp$ measurements from \sparta (as opposed to interpolated values).

With this halo sample in hand, we repeat the fitting procedure of \citetalias{diemer_17_rsp}. In particular, we fit the splashback results from both the \wmap and \planck cosmologies at redshifts $0.13$, $0.3$, $0.5$, $1$, $2$, $4$, and $8$, avoiding $z = 0$ owing to the corrections at the final snapshots (Section~\ref{sec:sparta:rsp}). We bin $\rsp$ and $\msp$ as a function of mass accretion rate ($0 < \gammadyn < 12$) and peak height (with bin edges at $0.5$, $1$, $1.5$, $2$, and $3$). We use a least-squares fit to minimize the difference between the binned relations and our fit. We combine the statistical uncertainty in each bin (due to the finite number of halos) with an additional 1\% systematic error in quadrature. This systematic error prevents high-occupation bins from dominating the fit and captures a number of inaccuracies in our methodology. The fitting function remains the same as in Equations~5-7 in \citetalias{diemer_17_rsp}. In particular, the relationships between $\gammadyn$ and the splashback quantities $\rsp$ and $\msp$ (which we summarily call $\xsp$) are fit with the general function
\begin{equation}
\label{eq:fit1}
\xsp = A_0 + B e^{-\gammadyn/C} \,, 
\end{equation}
where $A_0$, $B$, and $C$ are free parameters. The latter two are functions of peak height and $\Omega_{\rm m}(z)$,
\begin{align}
\label{eq:fit2}
B & = (B_0 + B_{\Omega} \Omega_{\rm m}) \times (1 + B_{\nu} \nu) \nonumber \\
C & = (c_0 + C_{\Omega} \Omega_{\rm m} + C_{\Omega 2} \Omega_{\rm m}^2) \times (1 + C_{\nu} \nu + C_{\nu 2} \nu^2) \,.
\end{align}
We separately fit $\rsp$ and $\msp$, as well as the mean-based definition and percentiles, leading to four sets of best-fit parameters. The dependence of the parameters on the percentile is parameterized in terms of $p$, the percentile divided by $100$,
\begin{align}
\label{eq:fit3}
A_0 &= a_{0} + a_{\rm p} \times p \nonumber \\
B_0 &= b_{0} + b_{\rm p} \times p \nonumber \\
B_{\Omega} &= b_{\Omega} + b_{\Omega \rm p} \times \exp( b_{\Omega \rm p2} \times p) \nonumber \\
B_{\nu} &= b_{\nu} + b_{\nu \rm p} \times p	 \nonumber \\
C_{\Omega} &= c_{\Omega} + c_{\Omega \rm p} \times \exp(c_{\Omega \rm p2} \times p)		 \nonumber \\
C_{\Omega 2} &= c_{\Omega 2} + c_{\Omega \rm 2p} \times \exp(c_{\Omega \rm 2p2} \times p) \nonumber \\	
C_{\nu} &= c_{\nu} + c_{\nu \rm p} \times p \nonumber \\
C_{\nu 2} &= c_{\nu 2} + c_{\nu \rm 2p} \times p	\,.
\end{align}
We also fit for the logarithmic scatter in the relation,
\begin{equation}
\label{eq:sigma}
\sigma_{\rm sp} = \sigma_0 + \sigma_{\Gamma} \gammadyn + \sigma_{\nu} \nu + \sigma_{\rm p} p \,.
\end{equation}
The new best-fit parameters are given in Table~\ref{table:fits}; some parameters are not necessary in all four fits as indicated by zeros. The new model is implemented in the publicly available \colossus code. The fit quality is very similar to that reported in \citetalias{diemer_17_rsp}, namely, better than 5\% essentially everywhere in the $\Gamma$--$\nu$--$\Omega_{\rm m}$ parameter space where we have data. $\deltasp$ is derived from $\rsp$ and $\msp$, with a combined uncertainty of 15\% or better. For a visual impression of these accuracies, we refer the reader to Figure~3 of \citetalias{diemer_17_rsp}. The model is valid for $0 < \gammadyn < 12$ and any percentile between $50$ and $90$ (higher percentiles become extremely noisy and were thus not computed by \sparta). We have checked the model against both the \wmap and \planck cosmologies and find similarly good agreement. The model does not describe self-similar simulations, although it comes close for slopes of $n \approx -2.5$ that resemble \LCDM \citepalias{diemer_17_rsp}.

\begin{deluxetable}{lcccc}
\tablecaption{Best-fit parameters for the Splashback--SO model
\label{table:fits}}
\tablewidth{0.47\textwidth}
\tablehead{
\colhead{Parameter} &
\colhead{$R_{\rm sp,mn}$} &
\colhead{$R_{\rm sp,\%}$} &
\colhead{$M_{\rm sp,mn}$} &
\colhead{$M_{\rm sp,\%}$}
}
\startdata
\multicolumn{5}{c}{\rule{0pt}{2ex} Parameters for $\rsp$ and $\msp$} \\
\hline
\rule{0pt}{3ex} $a_0$                & $   0.6597$ & $   0.3071$ & $   0.6962$ & $   0.2874$ \\
\rule{0pt}{0pt} $b_0$                & $   0.5562$ & $   0.2508$ & $   0.3736$ & $   0.6616$ \\
\rule{0pt}{0pt} $b_{\Omega}$         & $   0.1141$ & $   0.1527$ & $   0.3005$ & $   0.1321$ \\
\rule{0pt}{0pt} $b_{\nu}$            & $   0.0698$ & $   0.1956$ & $0$         & $0$         \\
\rule{0pt}{0pt} $c_0$                & $  -0.8508$ & $  -1.2214$ & $   3.3445$ & $   4.5913$ \\
\rule{0pt}{0pt} $c_{\Omega}$         & $  18.4464$ & $  17.5374$ & $   1.3718$ & $   3.0928$ \\
\rule{0pt}{0pt} $c_{\nu}$            & $  -0.3332$ & $0$         & $  -0.0825$ & $  -0.1155$ \\
\rule{0pt}{0pt} $c_{\Omega 2}$       & $ -10.0596$ & $ -10.3158$ & $0$         & $0$         \\
\rule{0pt}{0pt} $c_{\nu 2}$          & $   0.0474$ & $  -0.0189$ & $0$         & $0$         \\
\multicolumn{5}{c}{\rule{0pt}{3ex} Meta-parameters for Dependence on Percentile} \\
\hline
\rule{0pt}{3ex} $a_{\rm p}$          & $0$         & $   0.6428$ & $0$         & $   0.8228$ \\
\rule{0pt}{0pt} $b_{\rm p}$          & $0$         & $   0.5074$ & $0$         & $  -0.6567$ \\
\rule{0pt}{0pt} $b_{\Omega \rm p}$   & $0$         & $0$         & $0$         & $   0.0032$ \\
\rule{0pt}{0pt} $b_{\Omega \rm p2}$  & $0$         & $0$         & $0$         & $   4.9536$ \\
\rule{0pt}{0pt} $b_{\nu \rm p}$      & $0$         & $  -0.2128$ & $0$         & $   0.2882$ \\
\rule{0pt}{0pt} $c_{\Omega \rm p}$   & $0$         & $   0.0024$ & $0$         & $  -0.6609$ \\
\rule{0pt}{0pt} $c_{\Omega \rm p2}$  & $0$         & $   9.7115$ & $0$         & $  -1.1050$ \\
\rule{0pt}{0pt} $c_{\Omega \rm 2p}$  & $0$         & $  -0.0005$ & $0$         & $0$         \\
\rule{0pt}{0pt} $c_{\Omega \rm 2p2}$ & $0$         & $  10.7626$ & $0$         & $0$         \\
\rule{0pt}{0pt} $c_{\nu \rm p}$      & $0$         & $  -0.4735$ & $0$         & $  -0.3761$ \\
\rule{0pt}{0pt} $c_{\nu \rm 2p}$     & $0$         & $   0.0940$ & $0$         & $   0.0784$ \\
\multicolumn{5}{c}{\rule{0pt}{3ex} Parameters for 68\% Scatter (in dex)} \\
\hline
\rule{0pt}{3ex} $\sigma_0$           & $   0.0501$ & $   0.0419$ & $   0.0456$ & $   0.0224$ \\
\rule{0pt}{0pt} $\sigma_\Gamma$      & $   0.0035$ & $   0.0043$ & $   0.0017$ & $   0.0009$ \\
\rule{0pt}{0pt} $\sigma_\nu$         & $  -0.0108$ & $  -0.0141$ & $  -0.0079$ & $  -0.0091$ \\
\rule{0pt}{0pt} $\sigma_{\rm p}$     & $0$         & $   0.0235$ & $0$         & $   0.0454$
\enddata
\end{deluxetable}

\begin{figure}
\centering
\includegraphics[trim =  3mm 5mm 3mm 3mm, clip, scale=0.62]{\figdir/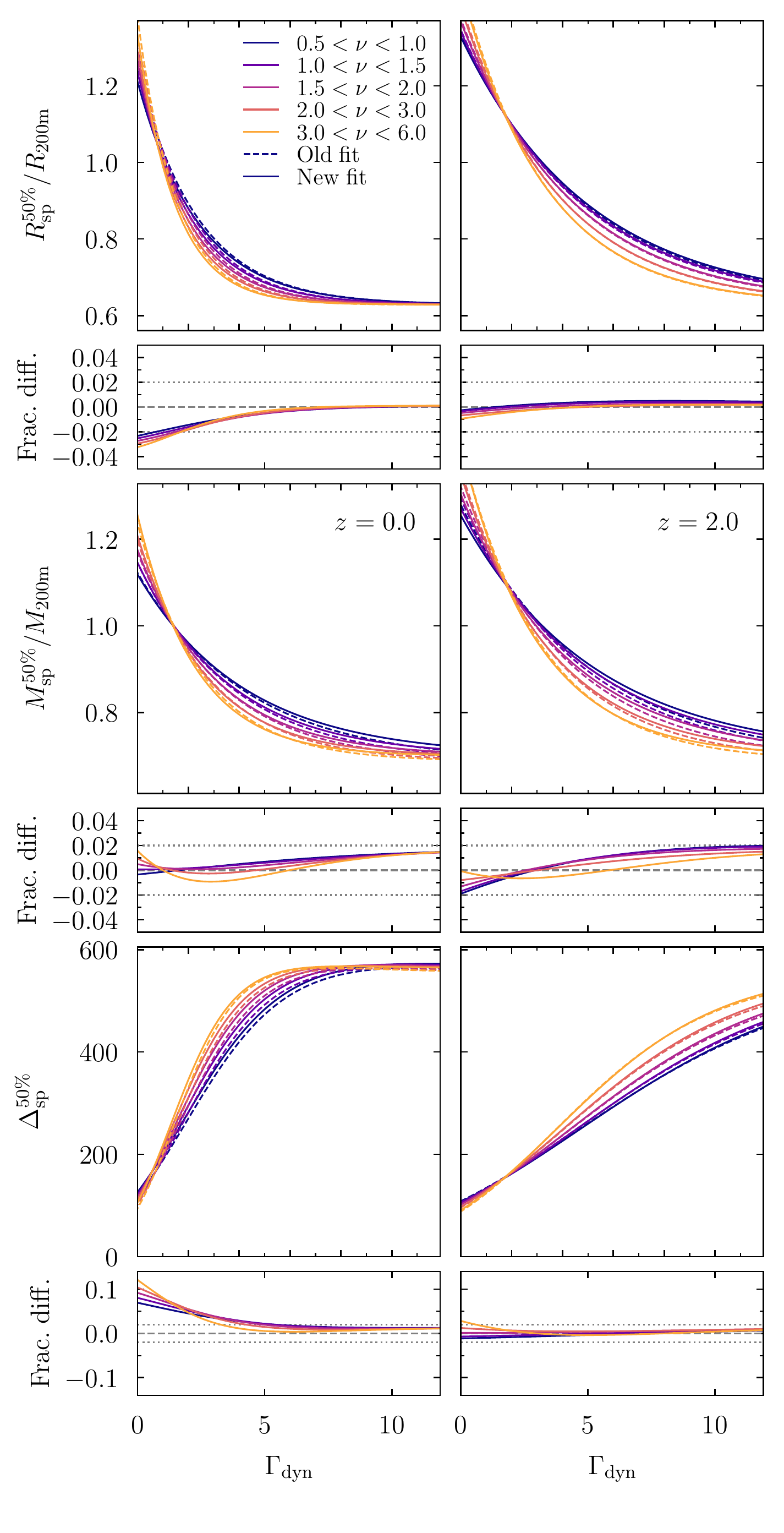}
\caption{Comparison of our new fitting function for the splashback--SO relation to the fit from \citetalias{diemer_17_rsp}. The large panels show the model predictions as a function of mass accretion rate for $\rsp/\rtom$, $\msp/\mtom$, and $\deltasp$ (from top to bottom); the small panels show the fractional difference between the new and old models. The left column shows $z = 0$; the right column, $z = 2$. We have chosen the median definition of $\rsp$ because it shows the largest variations between models. Nevertheless, the differences in $\rsp$ and $\msp$ are smaller than 5\%.}
\label{fig:fitcomp}
\end{figure}

Figure~\ref{fig:fitcomp} shows a visual comparison of the old and new fitting functions. The differences in $\rsp/\rtom$ and $\msp/\mtom$ are below 5\% for all redshifts, masses, and cosmologies that we tested and below 2\% for most of the parameter space. The differences in $\deltasp$ (which is derived from $\rsp$ and $\msp$) multiply to a maximum of 15\% but are typically around 5\%. There is no systematic upward or downward trend in the results. These small differences confirm that, despite numerous changes and bug fixes, \sparta's algorithm to determine $\rsp$ is robust.

We have also updated our fit to the mass accretion rate as a function of peak height and redshift. This formula is convenient when computing splashback properties without any knowledge of the mass accretion rate,
\begin{equation}
\gammadyn = A \nu + B \nu^{3/2} \,,
\end{equation}
where 
\begin{align}
\label{eq:gamma_fit}
A &= 1.1721 + 0.3255 z \nonumber \\
B &= -0.2565 + 0.0932 z - 0.0571 z^2 + 0.0042 z^3 \,.
\end{align}
The differences to the fit from \citetalias{diemer_17_rsp} are miniscule, but the new fit is more consistent with the parameters in Table~\ref{table:fits}.

\subsection{Splashback and the Radius of Steepest Slope}
\label{sec:results:steepest}

\begin{figure}
\centering
\includegraphics[trim =  8mm 5mm 4mm 3mm, clip, scale=0.58]{\figdir/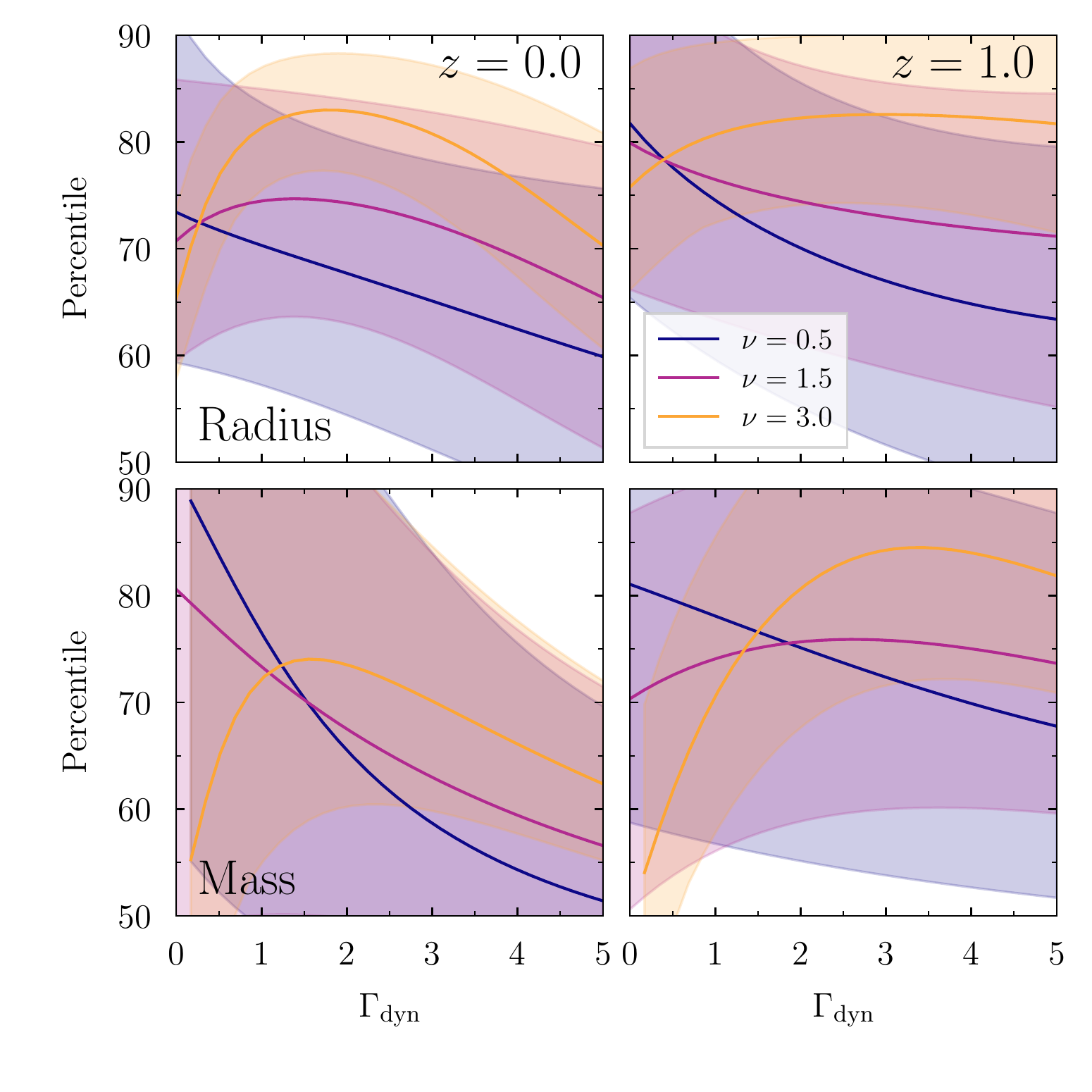}
\caption{Percentiles of particle splashback events that best match the radius of steepest slope according to the formula of \citet{more_15}. Results are shown for $z = 0$ (left) and $z = 1$ (right), as well as for radius (top) and mass (bottom). The dependencies on mass accretion rate, halo mass, and redshift are complicated, highlighting that there is no one percentile that corresponds to the radius of steepest slope in general.}
\label{fig:percentiles}
\end{figure}

While we have extensively investigated the relationship between \sparta's dynamically determined $\rsp$ and SO radii, we have not yet established a connection to the most commonly used definition of the splashback radius: the radius where the logarithmic slope of the spherically averaged density profile is steepest, $R_{\rm steep}$ (Figure~\ref{fig:viz_radii}). The splashback radius was first discovered as a result of the sharp drop in density \citep{diemer_14}, first quantified according to this definition \citep{more_15}, and all observational detections have hitherto relied on the density profile \citep[][and following work]{more_16}. Moreover, in theoretical models, there is no ambiguity between $R_{\rm steep}$ and the radius where the most recently accreted particles reach their first apocenter \citep{adhikari_14, shi_16_rsp}. In realistic halos, however, $\rsp$ and $R_{\rm steep}$ are not equivalent because the apocenters are spread out over a significant radial range owing to nonsphericity, complex accretion histories, and interactions with other halos \citepalias{diemer_17_sparta}. Similarly, the density drop represents a trade-off between the radially decreasing density of orbiting particles (or galaxies) and the stream of infalling material at large radii. As a result, it is not clear that there is a one-to-one correspondence between features measured in density profiles (or correlation functions) and the dynamically determined splashback radii from \sparta \citep{xhakaj_20, aung_20_phasespace, garcia_20}.

We attempt to quantify the connection in Figure~\ref{fig:percentiles}, where we show the percentiles that most closely match $R_{\rm steep}$ as approximated by the formula of \citet{more_15}. This fitting function was derived from the averaged (median) density profiles of halos binned in accretion rate and peak height. To be comparable to observations, the profiles included subhalo particles. \citet{more_15} used the \citet{diemer_14} definition of accretion rate, $\gammadk$, rather than $\gammadyn$. There, the time interval was defined by fixed redshift intervals instead of the dynamical time. To evaluate the fitting function, we crudely convert $\gammadyn$ to $\gammadk$ based on linear fits to the median relation in the \wmap sample, which is well approximated as
\begin{equation}
\gammadk = \Gamma_0 + \gamma \gammadyn \,.
\end{equation}
At redshifts $0$ and $0.5$, the accretion rates are virtually identical because both methods choose the same redshift to compare the current masses to. For the higher redshifts defined in \citet{diemer_14}, $z = [1, 2, 4]$, we find $\Gamma_0 = [0.51, 1.43, 0.88]$ and $\gamma = [0.74, 0.53, 0.78]$, respectively. We now obtain $R_{\rm steep} / \rtom$ and $M_{\rm steep} / \mtom$ from the \citet{more_15} fit at each $z$ and $\gammadyn$ and find the percentile where the fitting function of Equation~\ref{eq:fit1} most closely matches these values (if such a value exists between the 50th and 90th percentiles). We also evaluate the scatter using Equation~\ref{eq:sigma} and find the range of percentiles where $R_{\rm steep}$ and $M_{\rm steep}$ lie within the $1\sigma$ scatter (shaded areas in Figure~\ref{fig:percentiles}).

Clearly, the relationship between dynamical $\rsp$ and $R_{\rm steep}$ depends on accretion rate, mass, and redshift in a complex fashion. For example, Figure~\ref{fig:percentiles} explains the trend we noticed in the example of Figure~\ref{fig:viz_radii}, where higher $\gammadyn$ drives $R_{\rm steep}$ toward a lower percentile. The mass dependence is driven solely by \sparta's measurements because the \citet{more_15} formula does not depend on mass, a conclusion that should be revisited with improved simulation data. The ranges where the scatter around the medians matches $R_{\rm steep}$ are large, which reflects the fact that the differences between the percentiles are comparable to the scatter in the relation. Nevertheless, the median relations are statistically well defined.

We conclude that there is no one percentile value that can be used to mimic $R_{\rm steep}$ and $M_{\rm steep}$. The exact conversion will depend not only on accretion rate, mass, and redshift but also on the exact fitting procedure used to derive $R_{\rm steep}$ and on whether the profile of dark matter particles or subhalos is considered \citep{xhakaj_20}. We leave a more detailed calibration of the conversion for future work. For practical purposes, percentiles such as those in Figure~\ref{fig:percentiles} can be obtained by numerically matching the fitting functions of \citet{more_15} and Section~\ref{sec:results:model} for a given accretion rate, mass, and redshift, for example, using the implementations in \colossus.


\section{Discussion: Opportunities and Limitations}
\label{sec:discussion}

We have presented publicly available halo catalogs and merger trees for the \erebos simulations with the purpose of enabling the community to explore the impact of mass and radius definitions. In this section, we discuss the advantages and limitations of this dataset.

Our catalogs present new opportunities because they are, to the best of our knowledge, the first to contain SO definitions with multiple thresholds, bound-only and all-particle definitions, splashback radii and masses, and the new tracer mass definition for subhalos. Previously, exploring the impact of radius definition often meant running a halo finder with multiple density thresholds and manually matching the resulting catalogs \citep[e.g.,][]{villareal_17}. Our catalogs obviate the need for such complex procedures, as they contain multiple host--subhalo relations side by side.

Similarly, our new tree format makes it easy to analyze the evolution of halos. For instance, a few simple lines of Python code can pick out all epochs where subhalos fall into their hosts and plot the evolution of their masses before or after such events. The uniform catalog format and nature of the \erebos simulations make them perfect for combining data from different box sizes, as demonstrated in Section~\ref{sec:results} and in previous works. With box sizes spanning from $62.5$ to $2000 \mpch$ at $z = 0$, even a strict limit of $1000$ particles per halo results in a resolved mass range of more than five orders of magnitude \citep[see also][]{mansfield_20_resolution}. The self-similar universes do not represent realistic cosmologies but allow the user to explore the behavior of structure in extreme, yet easy-to-understand, limits of the power spectrum shape \citep[e.g.,][]{diemer_19_cm}.

However, no set of simulations is appropriate for all types of investigation. The most obvious limitation of the \erebos simulations is their particle number. With $1024^3 \approx 1$ billion particles per simulation, they yield more than sufficient number statistics at any given mass, but they are dwarfed by recent ultralarge simulations such as Millennium-XXL, DarkSky, Multidark-Planck, or Outer Rim \citep{angulo_12, skillman_14_darksky, klypin_16, heitmann_19}. Increasing the particle number does not necessarily add much to the already sufficient statistics on host halos, but larger simulations can resolve a wider range of subhalo-to-host mass ratios. For the particular use case of small subhalos in large hosts, combining simulations of different box sizes does not help. On the other hand, the more moderate particle numbers of \erebos mean that we can store 100 snapshots per simulation, enabling the kind of dynamical analysis performed by \sparta. Another limitation is that the \erebos suite contains only two \LCDM cosmologies, meaning that it cannot be used for many cosmological analyses such as forecasts on parameter constraints. Large suites of simulations with different cosmologies exist for such purposes, but generally only a few snapshots are stored \citep[e.g.,][]{heitmann_10, derose_19, nishimichi_19}.

Finally, we caution the user regarding certain limitations of our catalogs. We have tried to exclude poorly resolved halos with our cut of $N_{\rm 200m,peak} \geq 200$, but different definitions converge at different resolutions. For example, splashback radii and masses are reliable for halos with at least 500, or ideally 1000, particles. The cutoff can also lead to selection effects at the low-mass end. We recommend to always compare simulations with different resolutions at fixed mass; any nonconvergence typically shows up as disagreements between the boxes \citep[see, e.g.,][]{diemer_20_mfunc, diemer_20_subs, mansfield_20_resolution}. Similarly, not all definitions make sense for all types of halos: all-particle SO and splashback masses and radii should be trusted for hosts only, while bound-only and tracer masses are applicable to subhalos. Due to a correction applied to splashback radii in the final snapshots of a simulation, the splashback data should ideally be used at $z \geq 0.13$ in the \LCDM simulations, or about $0.2$ dynamical times before the end of the simulation in general. We also note that the snapshot spacing in the self-similar simulations corresponds to only about four snapshots per dynamical time. Such sparse coverage can lead to biases of a few percent in the splashback radii \citepalias{diemer_17_sparta}. Finally, due to a bug, the positions of ghost halos are missing for the final snapshot of each simulation. This error will be fixed in future iterations of the catalogs.


\section{Conclusions}
\label{sec:conclusion}

We have introduced \moria, an extension to the \sparta code framework that creates catalogs and merger trees with an arbitrary number of mass and radius definitions, as well as the corresponding host--subhalo relations. We present catalogs and merger trees that are based on \rockstar calculations but also include SO definitions with all and bound particles, splashback definitions, and subhalo masses computed using a novel particle tracking scheme. The \sparta code and our catalogs are publicly available at \href{http://www.benediktdiemer.com}{benediktdiemer.com}. Our main conclusions are as follows:
\begin{enumerate}
\item We have introduced a merger tree format based on 2D arrays in compressed hdf5 files, which facilitates the extraction of information at fixed time and for a particular halo.
\item Our catalogs are almost entirely complete for all sensible SO masses and mostly complete for splashback definitions, partly due to \moria's algorithms for interpolation and model estimation.
\item The halo boundary definition profoundly affects all aspects of halo masses, radii, and subhalo assignments. This statement holds both for the difference between SO and splashback definitions and for different SO thresholds.
\item We confirm that gravitational unbinding is not important for host halos but critical for subhalos. We compare different unbinding algorithms and conclude that the results depend strongly on a prior determination of halo membership.
\item We update our previous model of the splashback--SO connection. Despite numerous improvements to the splashback calculations, the results change by only a few percent, highlighting the robustness of the \sparta algorithm.
\item We show that there is a complex relation between the dynamically determined splashback radii from \sparta and the steepest slope in the density profile.
\end{enumerate}
The main purpose of this paper is to introduce new algorithms and data products; we have deferred most scientific questions to future work. In two companion papers, we investigate the impact of the halo boundary definition on mass functions  \citep{diemer_20_mfunc} and subhalo abundances \citep{diemer_20_subs}. We will also present our algorithms for subhalo particle tracking and ghost halos (Diemer \& Behroozi 2020, in preparation) and analyze the correlation between splashback and other halo properties (Shin \& Diemer 2020, in preparation). We anticipate further investigations of the galaxy--halo connection, semianalytic models of galaxy formation, and assembly bias. Most importantly, however, we encourage the community to use our catalogs to probe the impact of mass definition on any area of structure formation.


\vspace{0.5cm}

I am deeply indebted to Peter Behroozi for making \rockstar publicly available, for generously allowing me to adapt his kd-tree algorithm in \sparta, and for many enlightening discussions. I am grateful to Han Aung, Andrew Hearin, Philip Mansfield, Daisuke Nagai, and Enia Xhakaj for valuable feedback on a draft and to Alexie Leauthaud, R\"udiger Pakmor, and Tae-Hyeon Shin for helpful conversations. This work was partially completed during the coronavirus lockdown and would not have been possible without the essential workers who did not enjoy the privilege of working from the safety of their homes. All computations were run on the \textsc{Midway} computing cluster provided by the University of Chicago Research Computing Center. This research made extensive use of the Python packages \textsc{NumPy} \citep{code_numpy2}, \textsc{SciPy} \citep{code_scipy}, \textsc{Matplotlib} \citep{code_matplotlib}, and \colossus \citep{diemer_18_colossus}. This research was supported in part by the National Science Foundation under grant No. NSF PHY-1748958. Support for program No. HST-HF2-51406.001-A was provided by NASA through a grant from the Space Telescope Science Institute, which is operated by the Association of Universities for Research in Astronomy, Incorporated, under NASA contract NAS5-26555. 


\bibliographystyle{aasjournal}
\bibliography{\includedir/bib_mine.bib,\includedir/bib_general.bib,\includedir/bib_structure.bib,\includedir/bib_galaxies.bib,\includedir/bib_clusters.bib}

\begin{thebibliography}{}
\expandafter\ifx\csname natexlab\endcsname\relax\def\natexlab#1{#1}\fi
\providecommand{\url}[1]{\href{#1}{#1}}

\bibitem[{{Abel} {et~al.}(2012){Abel}, {Hahn}, \& {Kaehler}}]{abel_12}
{Abel}, T., {Hahn}, O., \& {Kaehler}, R. 2012, \mnras, 427, 61

\bibitem[{{Adhikari} {et~al.}(2014){Adhikari}, {Dalal}, \&
  {Chamberlain}}]{adhikari_14}
{Adhikari}, S., {Dalal}, N., \& {Chamberlain}, R.~T. 2014, JCAP, 11, 19

\bibitem[{{Angulo} {et~al.}(2012){Angulo}, {Springel}, {White}, {Jenkins},
  {Baugh}, \& {Frenk}}]{angulo_12}
{Angulo}, R.~E., {Springel}, V., {White}, S.~D.~M., {et~al.} 2012, \mnras, 426,
  2046

\bibitem[{{Aung} {et~al.}(2020){Aung}, {Nagai}, {Rozo}, \&
  {Garcia}}]{aung_20_phasespace}
{Aung}, H., {Nagai}, D., {Rozo}, E., \& {Garcia}, R. 2020, arXiv e-prints,
  arXiv:2003.11557

\bibitem[{{Balogh} {et~al.}(2000){Balogh}, {Navarro}, \& {Morris}}]{balogh_00}
{Balogh}, M.~L., {Navarro}, J.~F., \& {Morris}, S.~L. 2000, \apj, 540, 113

\bibitem[{{Banerjee} {et~al.}(2020){Banerjee}, {Adhikari}, {Dalal}, {More}, \&
  {Kravtsov}}]{banerjee_20}
{Banerjee}, A., {Adhikari}, S., {Dalal}, N., {More}, S., \& {Kravtsov}, A.
  2020, \jcap, 2020, 024

\bibitem[{{Baxter} {et~al.}(2017){Baxter}, {Chang}, {Jain}, {Adhikari},
  {Dalal}, {Kravtsov}, {More}, {Rozo}, {Rykoff}, \& {Sheth}}]{baxter_17}
{Baxter}, E., {Chang}, C., {Jain}, B., {et~al.} 2017, \apj, 841, 18

\bibitem[{{Behroozi} {et~al.}(2014){Behroozi}, {Wechsler}, {Lu}, {Hahn},
  {Busha}, {Klypin}, \& {Primack}}]{behroozi_14}
{Behroozi}, P.~S., {Wechsler}, R.~H., {Lu}, Y., {et~al.} 2014, \apj, 787, 156

\bibitem[{{Behroozi} {et~al.}(2013{\natexlab{a}}){Behroozi}, {Wechsler}, \&
  {Wu}}]{behroozi_13_rockstar}
{Behroozi}, P.~S., {Wechsler}, R.~H., \& {Wu}, H.-Y. 2013{\natexlab{a}}, \apj,
  762, 109

\bibitem[{{Behroozi} {et~al.}(2013{\natexlab{b}}){Behroozi}, {Wechsler}, {Wu},
  {Busha}, {Klypin}, \& {Primack}}]{behroozi_13_trees}
{Behroozi}, P.~S., {Wechsler}, R.~H., {Wu}, H.-Y., {et~al.} 2013{\natexlab{b}},
  \apj, 763, 18

\bibitem[{{Benson}(2017)}]{benson_17}
{Benson}, A.~J. 2017, \mnras, 467, 3454

\bibitem[{{Bertschinger}(1985)}]{bertschinger_85}
{Bertschinger}, E. 1985, \apjs, 58, 39

\bibitem[{{Bond} {et~al.}(1991){Bond}, {Cole}, {Efstathiou}, \&
  {Kaiser}}]{bond_91}
{Bond}, J.~R., {Cole}, S., {Efstathiou}, G., \& {Kaiser}, N. 1991, \apj, 379,
  440

\bibitem[{{Bryan} \& {Norman}(1998)}]{bryan_98}
{Bryan}, G.~L., \& {Norman}, M.~L. 1998, \apj, 495, 80

\bibitem[{{Busch} \& {White}(2017)}]{busch_17}
{Busch}, P., \& {White}, S.~D.~M. 2017, \mnras, 470, 4767

\bibitem[{{Chang} {et~al.}(2018){Chang}, {Baxter}, {Jain}, {S{\'a}nchez},
  {Adhikari}, {Varga}, {Fang}, {Rozo}, {Rykoff}, {Kravtsov}, {Gruen},
  {Hartley}, {Huff}, {Jarvis}, {Kim}, {Prat}, {MacCrann}, {McClintock},
  {Palmese}, {Rapetti}, {Rollins}, {Samuroff}, {Sheldon}, {Troxel}, {Wechsler},
  {Zhang}, {Zuntz}, {Abbott}, {Abdalla}, {Allam}, {Annis}, {Bechtol},
  {Benoit-L{\'e}vy}, {Bernstein}, {Brooks}, {Buckley-Geer}, {Carnero Rosell},
  {Carrasco Kind}, {Carretero}, {D{\textquoteright}Andrea}, {da Costa},
  {Davis}, {Desai}, {Diehl}, {Dietrich}, {Drlica-Wagner}, {Eifler}, {Flaugher},
  {Fosalba}, {Frieman}, {Garc{\'\i}a-Bellido}, {Gaztanaga}, {Gerdes},
  {Gruendl}, {Gschwend}, {Gutierrez}, {Honscheid}, {James}, {Jeltema},
  {Krause}, {Kuehn}, {Lahav}, {Lima}, {March}, {Marshall}, {Martini},
  {Melchior}, {Menanteau}, {Miquel}, {Mohr}, {Nord}, {Ogando}, {Plazas},
  {Sanchez}, {Scarpine}, {Schindler}, {Schubnell}, {Sevilla-Noarbe}, {Smith},
  {Smith}, {Soares-Santos}, {Sobreira}, {Suchyta}, {Swanson}, {Tarle},
  {Weller}, \& {DES Collaboration}}]{chang_18}
{Chang}, C., {Baxter}, E., {Jain}, B., {et~al.} 2018, \apj, 864, 83

\bibitem[{{Contigiani} {et~al.}(2019){Contigiani}, {Hoekstra}, \&
  {Bah{\'e}}}]{contigiani_19_wl}
{Contigiani}, O., {Hoekstra}, H., \& {Bah{\'e}}, Y.~M. 2019, \mnras, 485, 408

\bibitem[{{Crocce} {et~al.}(2006){Crocce}, {Pueblas}, \&
  {Scoccimarro}}]{crocce_06}
{Crocce}, M., {Pueblas}, S., \& {Scoccimarro}, R. 2006, \mnras, 373, 369

\bibitem[{{Croton}(2013)}]{croton_13_h}
{Croton}, D.~J. 2013, \pasa, 30, e052

\bibitem[{{Cuesta} {et~al.}(2008){Cuesta}, {Prada}, {Klypin}, \&
  {Moles}}]{cuesta_08}
{Cuesta}, A.~J., {Prada}, F., {Klypin}, A., \& {Moles}, M. 2008, \mnras, 389,
  385

\bibitem[{{Dalal} {et~al.}(2010){Dalal}, {Lithwick}, \& {Kuhlen}}]{dalal_10}
{Dalal}, N., {Lithwick}, Y., \& {Kuhlen}, M. 2010, arXiv:1010.2539,
  arXiv:1010.2539

\bibitem[{{DeRose} {et~al.}(2019){DeRose}, {Wechsler}, {Tinker}, {Becker},
  {Mao}, {McClintock}, {McLaughlin}, {Rozo}, \& {Zhai}}]{derose_19}
{DeRose}, J., {Wechsler}, R.~H., {Tinker}, J.~L., {et~al.} 2019, \apj, 875, 69

\bibitem[{{Diemand} {et~al.}(2005){Diemand}, {Moore}, \& {Stadel}}]{diemand_05}
{Diemand}, J., {Moore}, B., \& {Stadel}, J. 2005, \nat, 433, 389

\bibitem[{{Diemer}(2017)}]{diemer_17_sparta}
{Diemer}, B. 2017, \apjs, 231, 5

\bibitem[{{Diemer}(2018)}]{diemer_18_colossus}
---. 2018, The Astrophysical Journal Supplement Series, 239, 35

\bibitem[{{Diemer}(2020{\natexlab{a}})}]{diemer_20_subs}
---. 2020{\natexlab{a}}, arXiv e-prints, arXiv:2007.10992

\bibitem[{{Diemer}(2020{\natexlab{b}})}]{diemer_20_mfunc}
---. 2020{\natexlab{b}}, \apj, 903, 87

\bibitem[{{Diemer} \& {Joyce}(2019)}]{diemer_19_cm}
{Diemer}, B., \& {Joyce}, M. 2019, \apj, 871, 168

\bibitem[{{Diemer} \& {Kravtsov}(2014)}]{diemer_14}
{Diemer}, B., \& {Kravtsov}, A.~V. 2014, \apj, 789, 1

\bibitem[{{Diemer} \& {Kravtsov}(2015)}]{diemer_15}
---. 2015, \apj, 799, 108

\bibitem[{{Diemer} {et~al.}(2013{\natexlab{a}}){Diemer}, {Kravtsov}, \&
  {More}}]{diemer_13_scalingrel}
{Diemer}, B., {Kravtsov}, A.~V., \& {More}, S. 2013{\natexlab{a}}, \apj, 779,
  159

\bibitem[{{Diemer} {et~al.}(2017){Diemer}, {Mansfield}, {Kravtsov}, \&
  {More}}]{diemer_17_rsp}
{Diemer}, B., {Mansfield}, P., {Kravtsov}, A.~V., \& {More}, S. 2017, \apj,
  843, 140

\bibitem[{{Diemer} {et~al.}(2013{\natexlab{b}}){Diemer}, {More}, \&
  {Kravtsov}}]{diemer_13_pe}
{Diemer}, B., {More}, S., \& {Kravtsov}, A.~V. 2013{\natexlab{b}}, \apj, 766,
  25

\bibitem[{{Eisenstein} \& {Hu}(1998)}]{eisenstein_98}
{Eisenstein}, D.~J., \& {Hu}, W. 1998, \apj, 496, 605

\bibitem[{{Elahi} {et~al.}(2019){Elahi}, {Poulton}, {Tobar}, {Ca{\~n}as},
  {Lagos}, {Power}, \& {Robotham}}]{elahi_19_treefrog}
{Elahi}, P.~J., {Poulton}, R. J.~J., {Tobar}, R.~J., {et~al.} 2019, \pasa, 36,
  e028

\bibitem[{{Fillmore} \& {Goldreich}(1984)}]{fillmore_84}
{Fillmore}, J.~A., \& {Goldreich}, P. 1984, \apj, 281, 1

\bibitem[{{Fong} {et~al.}(2018){Fong}, {Bowyer}, {Whitehead}, {Lee}, {King},
  {Applegate}, \& {McCarthy}}]{fong_18}
{Fong}, M., {Bowyer}, R., {Whitehead}, A., {et~al.} 2018, \mnras, 478, 5366

\bibitem[{{Garc{\'\i}a} \& {Rozo}(2019)}]{garcia_19}
{Garc{\'\i}a}, R., \& {Rozo}, E. 2019, \mnras, 489, 4170

\bibitem[{{Garcia} {et~al.}(2020){Garcia}, {Rozo}, {Becker}, \&
  {More}}]{garcia_20}
{Garcia}, R., {Rozo}, E., {Becker}, M.~R., \& {More}, S. 2020, arXiv e-prints,
  arXiv:2006.12751

\bibitem[{{Gill} {et~al.}(2005){Gill}, {Knebe}, \& {Gibson}}]{gill_05}
{Gill}, S.~P.~D., {Knebe}, A., \& {Gibson}, B.~K. 2005, \mnras, 356, 1327

\bibitem[{{Green} {et~al.}(2020){Green}, {Aung}, {Nagai}, \& {van den
  Bosch}}]{green_20}
{Green}, S.~B., {Aung}, H., {Nagai}, D., \& {van den Bosch}, F.~C. 2020, arXiv
  e-prints, arXiv:2002.01934

\bibitem[{{Gunn} \& {Gott}(1972)}]{gunn_72}
{Gunn}, J.~E., \& {Gott}, III, J.~R. 1972, \apj, 176, 1

\bibitem[{{Guo} {et~al.}(2010){Guo}, {White}, {Li}, \&
  {Boylan-Kolchin}}]{guo_10}
{Guo}, Q., {White}, S., {Li}, C., \& {Boylan-Kolchin}, M. 2010, \mnras, 404,
  1111

\bibitem[{{Hahn} {et~al.}(2013){Hahn}, {Abel}, \& {Kaehler}}]{hahn_13}
{Hahn}, O., {Abel}, T., \& {Kaehler}, R. 2013, \mnras, 434, 1171

\bibitem[{{Han} {et~al.}(2018){Han}, {Cole}, {Frenk}, {Benitez-Llambay}, \&
  {Helly}}]{han_18_hbt}
{Han}, J., {Cole}, S., {Frenk}, C.~S., {Benitez-Llambay}, A., \& {Helly}, J.
  2018, \mnras, 474, 604

\bibitem[{{Han} {et~al.}(2012){Han}, {Jing}, {Wang}, \& {Wang}}]{han_12_hbt}
{Han}, J., {Jing}, Y.~P., {Wang}, H., \& {Wang}, W. 2012, \mnras, 427, 2437

\bibitem[{{Harris} {et~al.}(2020){Harris}, {Jarrod Millman}, {van der Walt},
  {Gommers}, {Virtanen}, {Cournapeau}, {Wieser}, {Taylor}, {Berg}, {Smith},
  {Kern}, {Picus}, {Hoyer}, {van Kerkwijk}, {Brett}, {Haldane}, {Fern{\'a}ndez
  del R{\'\i}o}, {Wiebe}, {Peterson}, {G{\'e}rard-Marchant}, {Sheppard},
  {Reddy}, {Weckesser}, {Abbasi}, {Gohlke}, \& {Oliphant}}]{code_numpy2}
{Harris}, C.~R., {Jarrod Millman}, K., {van der Walt}, S.~J., {et~al.} 2020,
  arXiv e-prints, arXiv:2006.10256

\bibitem[{{Heitmann} {et~al.}(2010){Heitmann}, {White}, {Wagner}, {Habib}, \&
  {Higdon}}]{heitmann_10}
{Heitmann}, K., {White}, M., {Wagner}, C., {Habib}, S., \& {Higdon}, D. 2010,
  \apj, 715, 104

\bibitem[{{Heitmann} {et~al.}(2019){Heitmann}, {Finkel}, {Pope}, {Morozov},
  {Frontiere}, {Habib}, {Rangel}, {Uram}, {Korytov}, {Child}, {Flender},
  {Insley}, \& {Rizzi}}]{heitmann_19}
{Heitmann}, K., {Finkel}, H., {Pope}, A., {et~al.} 2019, \apjs, 245, 16

\bibitem[{{Heitmann} {et~al.}(2020){Heitmann}, {Frontiere}, {Rangel}, {Larsen},
  {Pope}, {Sultan}, {Uram}, {Habib}, {Finkel}, {Korytov}, {Kovacs}, {Rizzi}, \&
  {Insley}}]{heitmann_20}
{Heitmann}, K., {Frontiere}, N., {Rangel}, E., {et~al.} 2020, arXiv e-prints,
  arXiv:2006.01697

\bibitem[{{Hu} \& {Kravtsov}(2003)}]{hu_03}
{Hu}, W., \& {Kravtsov}, A.~V. 2003, \apj, 584, 702

\bibitem[{Hunter(2007)}]{code_matplotlib}
Hunter, J.~D. 2007, Computing in Science Engineering, 9, 90

\bibitem[{{Joyce} {et~al.}(2020){Joyce}, {Garrison}, \&
  {Eisenstein}}]{joyce_20}
{Joyce}, M., {Garrison}, L., \& {Eisenstein}, D. 2020, arXiv e-prints,
  arXiv:2004.07256

\bibitem[{{Kaehler} {et~al.}(2012){Kaehler}, {Hahn}, \& {Abel}}]{kaehler_12}
{Kaehler}, R., {Hahn}, O., \& {Abel}, T. 2012, ArXiv e-prints, arXiv:1208.3206

\bibitem[{{Klypin} {et~al.}(1999){Klypin}, {Gottl{\"o}ber}, {Kravtsov}, \&
  {Khokhlov}}]{klypin_99_overmerging}
{Klypin}, A., {Gottl{\"o}ber}, S., {Kravtsov}, A.~V., \& {Khokhlov}, A.~M.
  1999, \apj, 516, 530

\bibitem[{{Klypin} {et~al.}(2016){Klypin}, {Yepes}, {Gottl{\"o}ber}, {Prada},
  \& {He{\ss}}}]{klypin_16}
{Klypin}, A., {Yepes}, G., {Gottl{\"o}ber}, S., {Prada}, F., \& {He{\ss}}, S.
  2016, \mnras, 457, 4340

\bibitem[{{Klypin} {et~al.}(2011){Klypin}, {Trujillo-Gomez}, \&
  {Primack}}]{klypin_11}
{Klypin}, A.~A., {Trujillo-Gomez}, S., \& {Primack}, J. 2011, \apj, 740, 102

\bibitem[{{Knollmann} {et~al.}(2008){Knollmann}, {Power}, \&
  {Knebe}}]{knollmann_08}
{Knollmann}, S.~R., {Power}, C., \& {Knebe}, A. 2008, \mnras, 385, 545

\bibitem[{{Komatsu} {et~al.}(2011){Komatsu}, {Smith}, {Dunkley}, {Bennett},
  {Gold}, {Hinshaw}, {Jarosik}, {Larson}, {Nolta}, {Page}, {Spergel},
  {Halpern}, {Hill}, {Kogut}, {Limon}, {Meyer}, {Odegard}, {Tucker}, {Weiland},
  {Wollack}, \& {Wright}}]{komatsu_11}
{Komatsu}, E., {Smith}, K.~M., {Dunkley}, J., {et~al.} 2011, \apjs, 192, 18

\bibitem[{{Lacey} \& {Cole}(1993)}]{lacey_93}
{Lacey}, C., \& {Cole}, S. 1993, \mnras, 262, 627

\bibitem[{{Lacey} \& {Cole}(1994)}]{lacey_94}
---. 1994, \mnras, 271, 676

\bibitem[{{Lahav} {et~al.}(1991){Lahav}, {Lilje}, {Primack}, \&
  {Rees}}]{lahav_91}
{Lahav}, O., {Lilje}, P.~B., {Primack}, J.~R., \& {Rees}, M.~J. 1991, \mnras,
  251, 128

\bibitem[{{Lau} {et~al.}(2015){Lau}, {Nagai}, {Avestruz}, {Nelson}, \&
  {Vikhlinin}}]{lau_15}
{Lau}, E.~T., {Nagai}, D., {Avestruz}, C., {Nelson}, K., \& {Vikhlinin}, A.
  2015, \apj, 806, 68

\bibitem[{{Leroy} {et~al.}(2020){Leroy}, {Garrison}, {Eisenstein}, {Joyce}, \&
  {Maleubre}}]{leroy_20}
{Leroy}, M., {Garrison}, L., {Eisenstein}, D., {Joyce}, M., \& {Maleubre}, S.
  2020, arXiv e-prints, arXiv:2004.08406

\bibitem[{{Lewis} {et~al.}(2000){Lewis}, {Challinor}, \& {Lasenby}}]{lewis_00}
{Lewis}, A., {Challinor}, A., \& {Lasenby}, A. 2000, \apj, 538, 473

\bibitem[{{Lithwick} \& {Dalal}(2011)}]{lithwick_11}
{Lithwick}, Y., \& {Dalal}, N. 2011, \apj, 734, 100

\bibitem[{{Luki{\'c}} {et~al.}(2009){Luki{\'c}}, {Reed}, {Habib}, \&
  {Heitmann}}]{lukic_09}
{Luki{\'c}}, Z., {Reed}, D., {Habib}, S., \& {Heitmann}, K. 2009, \apj, 692,
  217

\bibitem[{{Mamon} {et~al.}(2004){Mamon}, {Sanchis}, {Salvador-Sol{\'e}}, \&
  {Solanes}}]{mamon_04}
{Mamon}, G.~A., {Sanchis}, T., {Salvador-Sol{\'e}}, E., \& {Solanes}, J.~M.
  2004, \aap, 414, 445

\bibitem[{{Mansfield} \& {Avestruz}(2020)}]{mansfield_20_resolution}
{Mansfield}, P., \& {Avestruz}, C. 2020, arXiv e-prints, arXiv:2008.08591

\bibitem[{{Mansfield} \& {Kravtsov}(2020)}]{mansfield_20_ab}
{Mansfield}, P., \& {Kravtsov}, A.~V. 2020, \mnras, 493, 4763

\bibitem[{{Mansfield} {et~al.}(2017){Mansfield}, {Kravtsov}, \&
  {Diemer}}]{mansfield_17}
{Mansfield}, P., {Kravtsov}, A.~V., \& {Diemer}, B. 2017, \apj, 841, 34

\bibitem[{{Moore} {et~al.}(1996){Moore}, {Katz}, \& {Lake}}]{moore_96}
{Moore}, B., {Katz}, N., \& {Lake}, G. 1996, \apj, 457, 455

\bibitem[{{More} {et~al.}(2015){More}, {Diemer}, \& {Kravtsov}}]{more_15}
{More}, S., {Diemer}, B., \& {Kravtsov}, A.~V. 2015, \apj, 810, 36

\bibitem[{{More} {et~al.}(2016){More}, {Miyatake}, {Takada}, {Diemer},
  {Kravtsov}, {Dalal}, {More}, {Murata}, {Mandelbaum}, {Rozo}, {Rykoff},
  {Oguri}, \& {Spergel}}]{more_16}
{More}, S., {Miyatake}, H., {Takada}, M., {et~al.} 2016, \apj, 825, 39

\bibitem[{{Murata} {et~al.}(2020){Murata}, {Sunayama}, {Oguri}, {More},
  {Nishizawa}, {Nishimichi}, \& {Osato}}]{murata_20}
{Murata}, R., {Sunayama}, T., {Oguri}, M., {et~al.} 2020, arXiv e-prints,
  arXiv:2001.01160

\bibitem[{{Nishimichi} {et~al.}(2019){Nishimichi}, {Takada}, {Takahashi},
  {Osato}, {Shirasaki}, {Oogi}, {Miyatake}, {Oguri}, {Murata}, {Kobayashi}, \&
  {Yoshida}}]{nishimichi_19}
{Nishimichi}, T., {Takada}, M., {Takahashi}, R., {et~al.} 2019, \apj, 884, 29

\bibitem[{{Nishizawa} {et~al.}(2018){Nishizawa}, {Oguri}, {Oogi}, {More},
  {Nishimichi}, {Nagashima}, {Lin}, {Mandelbaum}, {Takada}, {Bahcall},
  {Coupon}, {Huang}, {Jian}, {Komiyama}, {Leauthaud}, {Lin}, {Miyatake},
  {Miyazaki}, \& {Tanaka}}]{nishizawa_18}
{Nishizawa}, A.~J., {Oguri}, M., {Oogi}, T., {et~al.} 2018, \pasj, 70, S24

\bibitem[{{Okumura} {et~al.}(2018){Okumura}, {Nishimichi}, {Umetsu}, \&
  {Osato}}]{okumura_18}
{Okumura}, T., {Nishimichi}, T., {Umetsu}, K., \& {Osato}, K. 2018, \prd, 98,
  023523

\bibitem[{{Patej} \& {Loeb}(2016)}]{patej_16}
{Patej}, A., \& {Loeb}, A. 2016, \apj, 824, 69

\bibitem[{{Peebles}(1980)}]{peebles_80}
{Peebles}, P.~J.~E. 1980, {The large-scale structure of the universe}
  (Princeton University Press)

\bibitem[{{Planck Collaboration} {et~al.}(2014){Planck Collaboration}, {Ade},
  {Aghanim}, {Armitage-Caplan}, {Arnaud}, {Ashdown}, {Atrio-Barandela},
  {Aumont}, {Baccigalupi}, {Banday}, \& et~al.}]{planck_14}
{Planck Collaboration}, {Ade}, P.~A.~R., {Aghanim}, N., {et~al.} 2014, \aap,
  571, A16

\bibitem[{{Press} \& {Schechter}(1974)}]{press_74}
{Press}, W.~H., \& {Schechter}, P. 1974, \apj, 187, 425

\bibitem[{{Reddick} {et~al.}(2013){Reddick}, {Wechsler}, {Tinker}, \&
  {Behroozi}}]{reddick_13}
{Reddick}, R.~M., {Wechsler}, R.~H., {Tinker}, J.~L., \& {Behroozi}, P.~S.
  2013, \apj, 771, 30

\bibitem[{{Rees} \& {Ostriker}(1977)}]{rees_77}
{Rees}, M.~J., \& {Ostriker}, J.~P. 1977, \mnras, 179, 541

\bibitem[{{Rodriguez-Gomez} {et~al.}(2015){Rodriguez-Gomez}, {Genel},
  {Vogelsberger}, {Sijacki}, {Pillepich}, {Sales}, {Torrey}, {Snyder},
  {Nelson}, {Springel}, {Ma}, \& {Hernquist}}]{rodriguezgomez_15}
{Rodriguez-Gomez}, V., {Genel}, S., {Vogelsberger}, M., {et~al.} 2015, \mnras,
  449, 49

\bibitem[{{Shi}(2016)}]{shi_16_rsp}
{Shi}, X. 2016, \mnras, 459, 3711

\bibitem[{{Shin} {et~al.}(2019){Shin}, {Adhikari}, {Baxter}, {Chang}, {Jain},
  {Battaglia}, {Bleem}, {Bocquet}, {DeRose}, {Gruen}, {Hilton}, {Kravtsov},
  {McClintock}, {Rozo}, {Rykoff}, {Varga}, {Wechsler}, {Wu}, {Zhang}, {Aiola},
  {Allam}, {Bechtol}, {Benson}, {Bertin}, {Bond}, {Brodwin}, {Brooks},
  {Buckley-Geer}, {Burke}, {Carlstrom}, {Carnero Rosell}, {Carrasco Kind},
  {Carretero}, {Castander}, {Choi}, {Cunha}, {Crawford}, {da Costa}, {De
  Vicente}, {Desai}, {Devlin}, {Dietrich}, {Doel}, {Dunkley}, {Eifler},
  {Evrard}, {Flaugher}, {Fosalba}, {Gallardo}, {Garc{\'\i}a-Bellido},
  {Gaztanaga}, {Gerdes}, {Gralla}, {Gruendl}, {Gschwend}, {Gupta}, {Gutierrez},
  {Hartley}, {Hill}, {Ho}, {Hollowood}, {Honscheid}, {Hoyle}, {Huffenberger},
  {Hughes}, {James}, {Jeltema}, {Kim}, {Krause}, {Kuehn}, {Lahav}, {Lima},
  {Madhavacheril}, {Maia}, {Marshall}, {Maurin}, {McMahon}, {Menanteau},
  {Miller}, {Miquel}, {Mohr}, {Naess}, {Nati}, {Newburgh}, {Niemack}, {Ogando},
  {Page}, {Partridge}, {Patil}, {Plazas}, {Rapetti}, {Reichardt}, {Romer},
  {Sanchez}, {Scarpine}, {Schindler}, {Serrano}, {Smith}, {Smith},
  {Soares-Santos}, {Sobreira}, {Staggs}, {Stark}, {Stein}, {Suchyta},
  {Swanson}, {Tarle}, {Thomas}, {van Engelen}, {Wollack}, \&
  {Xu}}]{shin_19_rsp}
{Shin}, T., {Adhikari}, S., {Baxter}, E.~J., {et~al.} 2019, \mnras, 487, 2900

\bibitem[{{Silk}(1977)}]{silk_77}
{Silk}, J. 1977, \apj, 211, 638

\bibitem[{{Skillman} {et~al.}(2014){Skillman}, {Warren}, {Turk}, {Wechsler},
  {Holz}, \& {Sutter}}]{skillman_14_darksky}
{Skillman}, S.~W., {Warren}, M.~S., {Turk}, M.~J., {et~al.} 2014, arXiv
  e-prints, arXiv:1407.2600

\bibitem[{{Springel}(2005)}]{springel_05_gadget2}
{Springel}, V. 2005, \mnras, 364, 1105

\bibitem[{{Springel} {et~al.}(2001){Springel}, {White}, {Tormen}, \&
  {Kauffmann}}]{springel_01_subfind}
{Springel}, V., {White}, S.~D.~M., {Tormen}, G., \& {Kauffmann}, G. 2001,
  \mnras, 328, 726

\bibitem[{{Srisawat} {et~al.}(2013){Srisawat}, {Knebe}, {Pearce}, {Schneider},
  {Thomas}, {Behroozi}, {Dolag}, {Elahi}, {Han}, {Helly}, {Jing}, {Jung},
  {Lee}, {Mao}, {Onions}, {Rodriguez-Gomez}, {Tweed}, \& {Yi}}]{srisawat_13}
{Srisawat}, C., {Knebe}, A., {Pearce}, F.~R., {et~al.} 2013, \mnras, 436, 150

\bibitem[{{Tomooka} {et~al.}(2020){Tomooka}, {Rozo}, {Wagoner}, {Aung},
  {Nagai}, \& {Safonova}}]{tomooka_20}
{Tomooka}, P., {Rozo}, E., {Wagoner}, E.~L., {et~al.} 2020, arXiv e-prints,
  arXiv:2003.11555

\bibitem[{{Trenti} {et~al.}(2010){Trenti}, {Smith}, {Hallman}, {Skillman}, \&
  {Shull}}]{trenti_10}
{Trenti}, M., {Smith}, B.~D., {Hallman}, E.~J., {Skillman}, S.~W., \& {Shull},
  J.~M. 2010, \apj, 711, 1198

\bibitem[{{Tully}(2015)}]{tully_15}
{Tully}, R.~B. 2015, \aj, 149, 54

\bibitem[{{Umetsu} \& {Diemer}(2017)}]{umetsu_17}
{Umetsu}, K., \& {Diemer}, B. 2017, \apj, 836, 231

\bibitem[{{van den Bosch}(2017)}]{vandenbosch_17}
{van den Bosch}, F.~C. 2017, \mnras, 468, 885

\bibitem[{{van Kampen}(1995)}]{vankampen_95}
{van Kampen}, E. 1995, \mnras, 273, 295

\bibitem[{{Villarreal} {et~al.}(2017){Villarreal}, {Zentner}, {Mao}, {Purcell},
  {van den Bosch}, {Diemer}, {Lange}, {Wang}, \& {Campbell}}]{villareal_17}
{Villarreal}, A.~S., {Zentner}, A.~R., {Mao}, Y.-Y., {et~al.} 2017, \mnras,
  472, 1088

\bibitem[{{Virtanen} {et~al.}(2019){Virtanen}, {Gommers}, {Oliphant},
  {Haberland}, {Reddy}, {Cournapeau}, {Burovski}, {Peterson}, {Weckesser},
  {Bright}, {van der Walt}, {Brett}, {Wilson}, {Jarrod Millman}, {Mayorov},
  {Nelson}, {Jones}, {Kern}, {Larson}, {Carey}, {Polat}, {Feng}, {Moore}, {Vand
  erPlas}, {Laxalde}, {Perktold}, {Cimrman}, {Henriksen}, {Quintero}, {Harris},
  {Archibald}, {Ribeiro}, {Pedregosa}, {van Mulbregt}, \&
  {Contributors}}]{code_scipy}
{Virtanen}, P., {Gommers}, R., {Oliphant}, T.~E., {et~al.} 2019, arXiv
  e-prints, arXiv:1907.10121

\bibitem[{{Vogelsberger} {et~al.}(2011){Vogelsberger}, {Mohayaee}, \&
  {White}}]{vogelsberger_11_similarity}
{Vogelsberger}, M., {Mohayaee}, R., \& {White}, S.~D.~M. 2011, \mnras, 414,
  3044

\bibitem[{{Wang} {et~al.}(2006){Wang}, {Li}, {Kauffmann}, \& {De
  Lucia}}]{wang_06_orphans}
{Wang}, L., {Li}, C., {Kauffmann}, G., \& {De Lucia}, G. 2006, \mnras, 371, 537

\bibitem[{{Wechsler} \& {Tinker}(2018)}]{wechsler_18}
{Wechsler}, R.~H., \& {Tinker}, J.~L. 2018, ArXiv e-prints, arXiv:1804.03097

\bibitem[{{White}(2001)}]{white_01_mass}
{White}, M. 2001, \aap, 367, 27

\bibitem[{{White}(2002)}]{white_02}
---. 2002, \apjs, 143, 241

\bibitem[{{White} \& {Rees}(1978)}]{white_78}
{White}, S.~D.~M., \& {Rees}, M.~J. 1978, \mnras, 183, 341

\bibitem[{{Xhakaj} {et~al.}(2020){Xhakaj}, {Diemer}, {Leauthaud}, {Wasserman},
  {Huang}, {Luo}, {Adhikari}, \& {Singh}}]{xhakaj_20}
{Xhakaj}, E., {Diemer}, B., {Leauthaud}, A., {et~al.} 2020, \mnras, 499, 3534

\bibitem[{{Xhakaj} {et~al.}(2019){Xhakaj}, {Leauthaud}, {Diemer}, \&
  {Behroozi}}]{xhakaj_19_accrate}
{Xhakaj}, E., {Leauthaud}, A., {Diemer}, B., \& {Behroozi}, P. 2019, Research
  Notes of the American Astronomical Society, 3, 169

\bibitem[{{Zemp}(2014)}]{zemp_14}
{Zemp}, M. 2014, \apj, 792, 124

\bibitem[{{Zu} {et~al.}(2016){Zu}, {Mandelbaum}, {Simet}, {Rozo}, \&
  {Rykoff}}]{zu_17}
{Zu}, Y., {Mandelbaum}, R., {Simet}, M., {Rozo}, E., \& {Rykoff}, E.~S. 2016,
  arXiv:1611.00366, arXiv:1611.00366

\bibitem[{{Z{\"u}rcher} \& {More}(2019)}]{zuercher_19}
{Z{\"u}rcher}, D., \& {More}, S. 2019, \apj, 874, 184

\end{thebibliography}

\end{document}